\documentclass[12pt]{article}

\usepackage{amsmath}
\usepackage{graphicx,psfrag,epsf}
\usepackage{enumerate}
\usepackage{natbib}
\usepackage{url} % not crucial - just used below for the URL 
\usepackage[total={170mm,230mm},
 left=25mm,
 top=20mm]{geometry}
\usepackage{amssymb}
\usepackage[colorlinks=true, allcolors=blue]{hyperref}
\usepackage[table,xcdraw]{xcolor}
\usepackage{booktabs}
\usepackage{array}
\usepackage{bbm}
\newcolumntype{P}[1]{>{\centering\arraybackslash}p{#1}}
\newcolumntype{M}[1]{>{\centering\arraybackslash}m{#1}}
\newcolumntype{L}[1]{>{\raggedright\let\newline\\\arraybackslash\hspace{0pt}}m{#1}}
\newcolumntype{C}[1]{>{\centering\let\newline\\\arraybackslash\hspace{0pt}}m{#1}}
\newcolumntype{R}[1]{>{\raggedleft\let\newline\\\arraybackslash\hspace{0pt}}m{#1}}
\usepackage{adjustbox}

\usepackage{amsthm}
\usepackage{amsfonts}
\usepackage{bbm}

\usepackage{xpatch}
\usepackage{multirow}
\usepackage{placeins}
\xpatchcmd{\proof}{\itshape}{\normalfont\proofnamefont}{}{}
\newcommand{\proofnamefont}{\bfseries}
 \usepackage{float}

\usepackage{hyperref}
\usepackage{appendix}
\newcommand{\blind}{1}

\definecolor{ch}{RGB}{30,136,229}
\definecolor{chbs}{RGB}{34,139,34}

\newcommand\extrafootertext[1]{%
    \bgroup
    \renewcommand\thefootnote{\fnsymbol{footnote}}%
    \renewcommand\thempfootnote{\fnsymbol{mpfootnote}}%
    \footnotetext[0]{#1}%
    \egroup
}

\newcommand\blfootnote[1]{%
  \begingroup
  \renewcommand\thefootnote{}\footnote{#1}%
  \addtocounter{footnote}{-1}%
  \endgroup
}

\begin{document}
\def\spacingset#1{\renewcommand{\baselinestretch}%
{#1}\small\normalsize} \spacingset{1}

\if1\blind
{
  \title{\bf XGBoost meets INLA: a two-stage
spatio-temporal forecasting of wildfires in
Portugal}}
  \author{Chenglei Hu$^1$* \blfootnote{Corresponding author. Email address: chenglei.hu@glasgow.ac.uk}, Regina Baltazar Bispo$^{2,5}$,  Håvard Rue$^3$, \\Carlos C. DaCamara$^4$, Ben Swallow$^2$,  Daniela Castro-Camilo$^1$ }
  
\footnotetext[1]{
\baselineskip=10pt School of Mathematics and Statistics,  University of Glasgow, UK}
\footnotetext[2]{ 
\baselineskip=10pt  School of Mathematics and Statistics, Centre for Research into Ecological and Environmental Modelling, University of St Andrews, UK}
\footnotetext[3]{ 
\baselineskip=10pt King Abdullah University of Science and Technology, Saudi Arabia }
\footnotetext[4]{ 
\baselineskip=10pt Instituto Dom Luiz, Faculdade de Ci\^encias, Universidade de Lisboa, 1749-016 Lisboa, Portugal}
\footnotetext[5]{ 
\baselineskip=10pt Center for Mathematics and Applications (NOVA Math), NOVA School of Science and Technology (NOVA FCT), Portugal}
  \maketitle
 \fi

\if0\blind
{
  \bigskip
  \bigskip
  \bigskip
  \begin{center}
    {\LARGE\bf Title}
\end{center}
  \medskip
} \fi

\begin{abstract}

{Wildfires pose a major threat to Portugal, with over 115,000 hectares burned annually on average during 1980-2024, and the country has faced devastating mega-fires such as those in 2017. 
Accurate forecasts of wildfire occurrence and burned area are therefore essential for firefighting resource allocation and emergency preparedness. 
In this study, we propose a novel two-stage ensemble that extends the widely used latent Gaussian modelling framework with integrated nested Laplace approximation (INLA) for spatio-temporal wildfire forecasting. 
Stage 1 applies a gradient boosting model (XGBoost) to environmental covariates and historical fire records to produce one-month-ahead point forecasts of fire counts and burned area. 
Stage 2 uses these predictions as external covariates in a latent Gaussian model with additional spatiotemporal random effects to generate probabilistic forecasts of monthly total fire counts and burned area at the council level. 
To capture both moderate and extreme events, we implement the extended generalised Pareto (eGP) likelihood (a sub-asymptotic distribution) within INLA, develop Penalised Complexity (PC) priors for its parameters, and compare the eGP likelihood with common alternatives (e.g., Gamma and Weibull). 
Our framework tackles the unavailability of future environmental covariates at prediction time and performs strongly for one-month-ahead forecasts.}

\end{abstract}

\noindent%
\vfill

\newpage
\spacingset{1.45} % DON'T change the spacing!
\section{Introduction}
\label{sec:intro}
Wildfires affect a vast portion of the vegetated surfaces of the Earth and may be defined as unplanned and uncontrolled fires that quickly spread over the terrain.
Wildfires are favoured by prolonged drought, heat waves and by hot, dry and windy weather, and they may be triggered by natural phenomena, such as lightning strikes, or by human activities, including negligence, arson, and various forms of accidental ignition. 
% Fuelled by dry vegetation and wind, wildfires can spread rapidly across vast areas, devastating everything in their path.

Portugal is among the countries most severely affected by wildfires, due to its mild and humid winters followed by hot and dry summers, strong winds, and extensive areas with forests and shrublands.
According to official records, in 2017 alone, more than 21,000 fire ignitions were recorded, resulting in over 540,000 hectares burned \citep{dacamara2024signature}.
In addition to the high number of fire events and extent of burnt area, Portugal regularly experiences mega-fires, i.e., individual wildfires with extreme consequences. 
For instance, the deadly wildfires of 17 June 2017 claimed at least 66 lives and affected more than 220,000 hectares in 24 hours. 
These types of events result not only in economic losses and human casualties but also cause substantial environmental damage, including widespread deforestation. 
Consequently, the development of an effective wildfire forecasting system is crucial for enabling early warnings and better allocation of firefighting resources \citep{dacamara2018ceasefire}.

The increasing demand for accurate wildfire modelling has led to a wide range of studies employing statistical and machine learning methods. These approaches can be broadly categorised based on how they represent wildfires.
One common approach models wildfire as an event occurring at an ignition point, optionally with an associated burnt area.
This perspective motivates point process models for fire ignitions \citep{Xu2011PointPM, gabriel2017detecting, opitz2020point, woolford2021development} and marked point processes where the burnt area serves as the mark \citep{tonini2017evolution, xi2019statistical, koh2023spatiotemporal, de2024spatio, Duvstenwildfire2025}.
Alternatively, ignition events and associated burnt areas can be aggregated over spatial partitions, with models targeting either the total burnt area or both fire count and burnt area per unit, leading to areal modelling approaches \citep{opitz2023eva, cisneros2024deep, lawler2024anthropogenic}.

To capture the risk of extremely large burnt areas, extreme value theory (EVT) is often employed to estimate high quantiles of the burnt area distribution. 
The Generalised Pareto Distribution (GPD) is a widely used EVT-based model for peak-over-threshold methods and has been applied to model exceedance probabilities \citep{pimont2021prediction, richards2023insights, koh2023spatiotemporal}. 
When modelling the entire distribution of burnt areas, however, GPD-based methods require an auxiliary distribution for values below the threshold, which can lead to discontinuities in the likelihood.
Recent advances have introduced sub-asymptotic distributions, which offer continuous density, flexible tail behaviour, and theoretical justification within EVT \citep{papastathopoulos2013extended, naveau2016modeling}.
These distributions have seen successful applications in domains such as precipitation \citep{naveau2016modeling}, landslides \citep{yadav2021spatial}, and wildfires \citep{cisneros2024deep, lawler2024anthropogenic}.

Bayesian hierarchical models with latent Gaussian fields are a popular framework for high-dimensional spatial and spatio-temporal modelling \citep{opitz2017latent}. 
While inference is traditionally performed via Markov Chain Monte Carlo (MCMC), the integrated nested Laplace approximation (INLA) offers a faster and accurate alternative for posterior approximation, especially in space-time applications \citep{rue2009approximate}. 
Several studies have applied INLA-based frameworks to wildfire modelling \citep{gabriel2017detecting, opitz2020point, koh2023spatiotemporal}. 
Machine learning methods have also been explored, including tree-based models \citep{koh2023gradient, cisneros2023combined} and neural networks \citep{richards2022regression, richards2023insights, cisneros2024deep}.

Despite the breadth of existing research, practical wildfire forecasting methods remain limited. 
The development and spread of wildfires are strongly influenced by environmental factors such as humidity, temperature, wind, and vegetation type. Incorporating such information is essential for improving predictive accuracy. However, most current spatio-temporal frameworks \citep[e.g.][]{koh2023spatiotemporal, cisneros2024deep} are designed for retrospective analysis, using covariates observed at time $t+1$ to predict wildfires also occurring at time $t+1$.
This setup limits their use in real-time forecasting, as it assumes access to future covariates that would not be available at the times for which predictions are to be made.
Furthermore, within the popular latent Gaussian modelling framework, the additive structure and the practical constraints on the number of hyperparameters restrict the inclusion of multiple covariates in INLA-based models. 
As a result, studies often rely on a small number of representative variables, such as the Fire Weather Index (FWI), an index developed by the Canadian Forestry Service that has proven to be an especially suitable indicator of meteorological fire danger in Mediterranean ecosystems \citep{dacamara2014calibration,pinto2018fire, nunes2023assessing}

% rely on a simultaneous mapping of covariates and wildfire occurrence, using covariates at time $t$ to predict wildfires at time $t$.
% This approach requires access to future covariates, which is often infeasible \dcc{not really infeasible, right? Pretty much the only choice here is to use NWP outputs, which don't come without issues, such as bias. So we could say something like: which limits its forecasting ability to the use of numerical weather prediction (NWP) model outputs, which often exhibit systematic biases, coarse spatial resolution, and limited accuracy for extreme events, posing challenges for downstream applications such as wildfire modelling \citep{veraart2024biases}.}
% , particularly for models with coarse temporal resolution (e.g. monthly). 
% Forecasts from Numerical Weather Prediction systems generally have fewer variables, lower spatial resolution, and are limited to short time horizons.

In this work, we aim to address the twin challenges of acquiring future covariates and the limited capacity of the INLA framework to accommodate numerous predictors. 
We propose a two-stage, interpretable modelling framework that relies entirely on readily available reanalysis data for probabilistic forecasting, leveraging both machine learning and INLA. 
In the first stage, we train a tree-based ensemble model, specifically, XGBoost \citep{chen2016xgboost}, on a window-based dataset, incorporating environmental covariates and historical wildfire data up to time $t$, with the target being wildfire activity at time $t+1$. 
This model learns patterns from historical data to produce a point forecast for the next time step. 
In the second stage, the XGBoost forecast is used as a synthetic future covariate, combined with spatial and temporal Gaussian effects in a latent Gaussian model estimated via INLA, yielding posterior predictive distributions.

The XGBoost model effectively encodes the information from all available covariates into a single, most informative predictor for Portuguese wildfires.
{This strategy reduces the reliance on future environmental covariates, circumvents the limitations of INLA in handling a large number of predictors, and enhances the INLA model’s ability to capture complex interactions among covariates.}
Meanwhile, INLA provides a principled Bayesian framework for uncertainty quantification through the posterior predictive distribution.
{This hybrid modelling framework can be viewed as a stacking strategy \citep{wolpert1992stacked}, which has been shown to yield substantial improvements in predictive performance \citep{van2007super}.
In contrast to other hybrid approaches for spatial and spatio-temporal modelling that require specialised mappings between spatial processes or covariance structures \citep{wikle2023statistical}, our approach minimises modifications to the underlying spatial structure, thereby preserving interpretability and facilitating its application to other spatial or spatio-temporal modelling tasks within the INLA framework.
Our proposed method is also closely related to residual-learning approaches \citep{macbride2025spatial}, but does not rely on iterative updates between the first-stage and second-stage models. 
In our setting, such iterative training would be computationally expensive due to the inclusion of complex spatio-temporal random effects in the INLA model, but the potential gains from iterative residual learning are likely to be marginal.}

Additionally, we contribute to the integration of the extended Generalised Pareto (eGP) likelihood \citep{naveau2016modeling} within the INLA framework. 
The eGP distribution belongs to the family of sub-asymptotic models capable of jointly modelling the bulk and tail of the data. 
We use this distribution to model burnt area data, thereby capturing the moderate and extreme fires simultaneously in a continuous manner. 
As part of our implementation, we derive and incorporate penalised complexity (PC) priors with a closed-form expression for the two shape parameters of the eGP distribution.
% We also derive a PC prior for the other shape parameter, which, while lacking a closed-form solution, is provided in a usable form.

The remainder of the paper is structured as follows. Section \ref{sec: data} introduces the Portuguese wildfire dataset and the environmental covariates used. Section \ref{sec: method} presents the two-stage modelling framework, including the choice of priors for the eGP parameters. Section \ref{sec: results} reports forecasting results and model interpretation. Section \ref{sec: discussion} discusses the use of the eGP likelihood and considerations for longer-horizon forecasting. Finally, Section \ref{sec: conclusion} concludes the study.

% \dcc{General comment: font sizes of figures are very small compared to those of the main text. Not a deal breaker, but if at some point you have to/can update figures, would be good to try to match the size in the main text.}

% 1. Leverage machine learning techniques to incorporate additional covariates (mainly to improve the interaction between covariates and avoid making the model overcomplex) to enhance the forecasting capability.

% 2. interpretability of the model while still having high performance

% 3. extended GPD for full likelihood estimation.

% 4. bayesian probability prediction, easy for policy maker

% 5. Additional covariates do not have enough forecasting horizon, and could be oversmoothed and not imformed. Machine learning technique to provide a forecasted local feature

\section{Data Preparation}
\label{sec: data}
\subsection{Wildfire Data Scope}
The wildfire dataset used in this study is sourced from the Portuguese Institute for Nature Conservation and Forests (ICNF, \url{https://www.icnf.pt/florestas/gfr/gfrgestaoinformacao/estatisticas}), a governmental agency responsible for forest and conservation policy. 
The dataset includes detailed information on wildfire events, such as ignition time (year, month, day, hour), fire duration, geographical coordinates (longitude and latitude), burnt area by land type (urban, bush, or agricultural), and additional attributes including cause and FWI.

This study focuses on wildfire records from 2011 to 2023, a 13-year period that captures both typical wildfire activity and extreme events such as the 2017 mega-fires, while keeping the temporal dimension manageable for computational modelling. 
To ensure that the analysis excludes intentional land management fires (e.g. crop residue burning), only fire events with a total burnt area exceeding 1 hectare and a duration longer than 3 hours are retained.

Administrative metadata, including council and district information, is mapped to each fire record. 
Fires are then aggregated to the council-month level to facilitate areal modelling. 
This choice is motivated by two factors: 
(1) The recorded ignition coordinates lack high spatial precision, and repeated wildfires are often observed at the same location, making a regional modelling unit such as the council more appropriate. 
(2) Council-level predictions are more interpretable and actionable for policymakers than point-level estimates. As such, modelling directly on aggregated data is preferred over integrating point process models post hoc.
The temporal aggregation to a monthly resolution is a deliberate balance between computational feasibility and practical relevance.
Modelling at higher temporal resolution (e.g. daily or hourly) over a 13-year span would be computationally intensive and likely require oversimplified spatio-temporal structures, potentially compromising model accuracy.
Monthly aggregation allows for a more complex spatio-temporal modelling without excessive computational burden \citep{krainski2018advanced}.

\begin{figure}
\begin{center}
\includegraphics[width=0.49\textwidth]{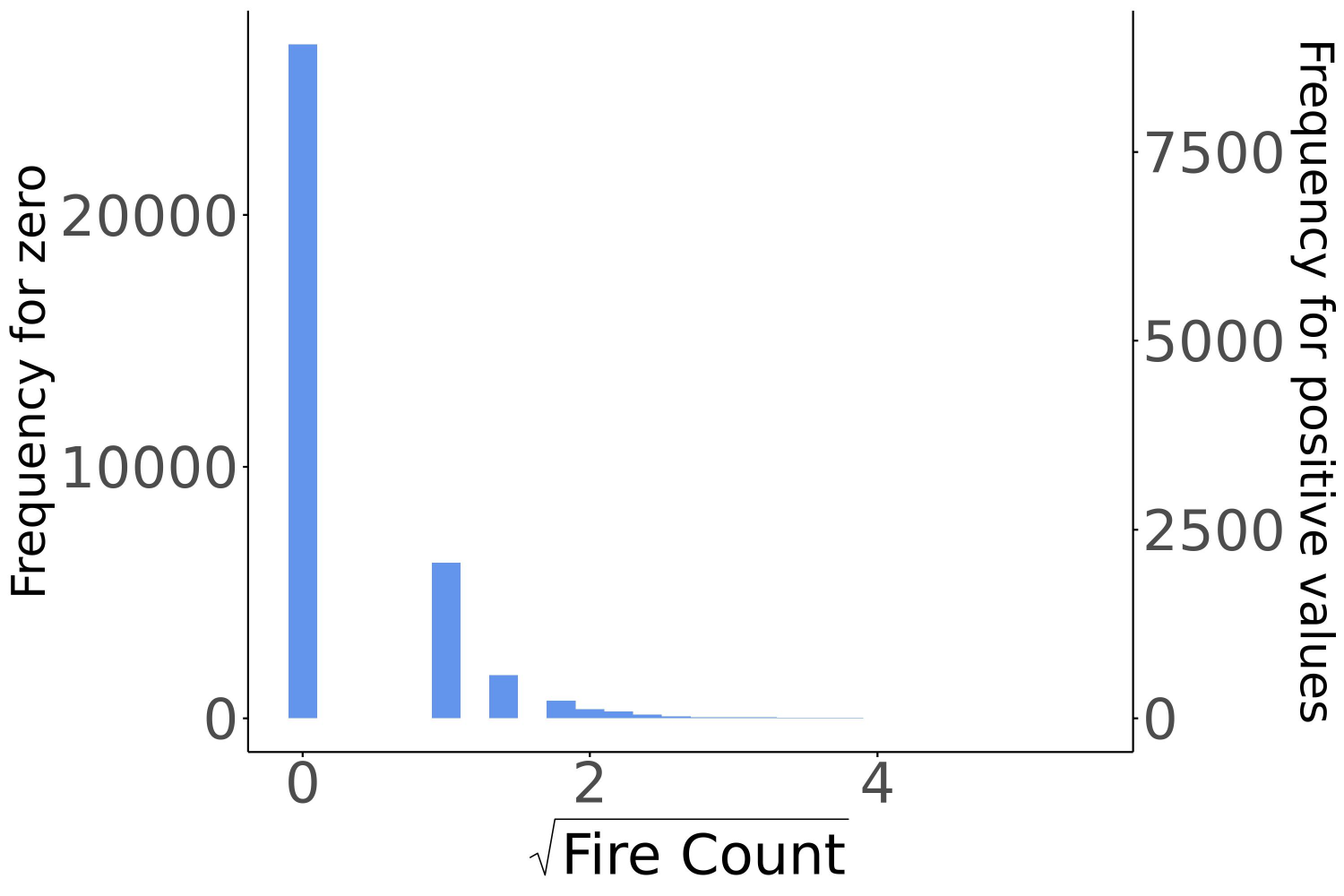}
\includegraphics[width=0.49\textwidth]
{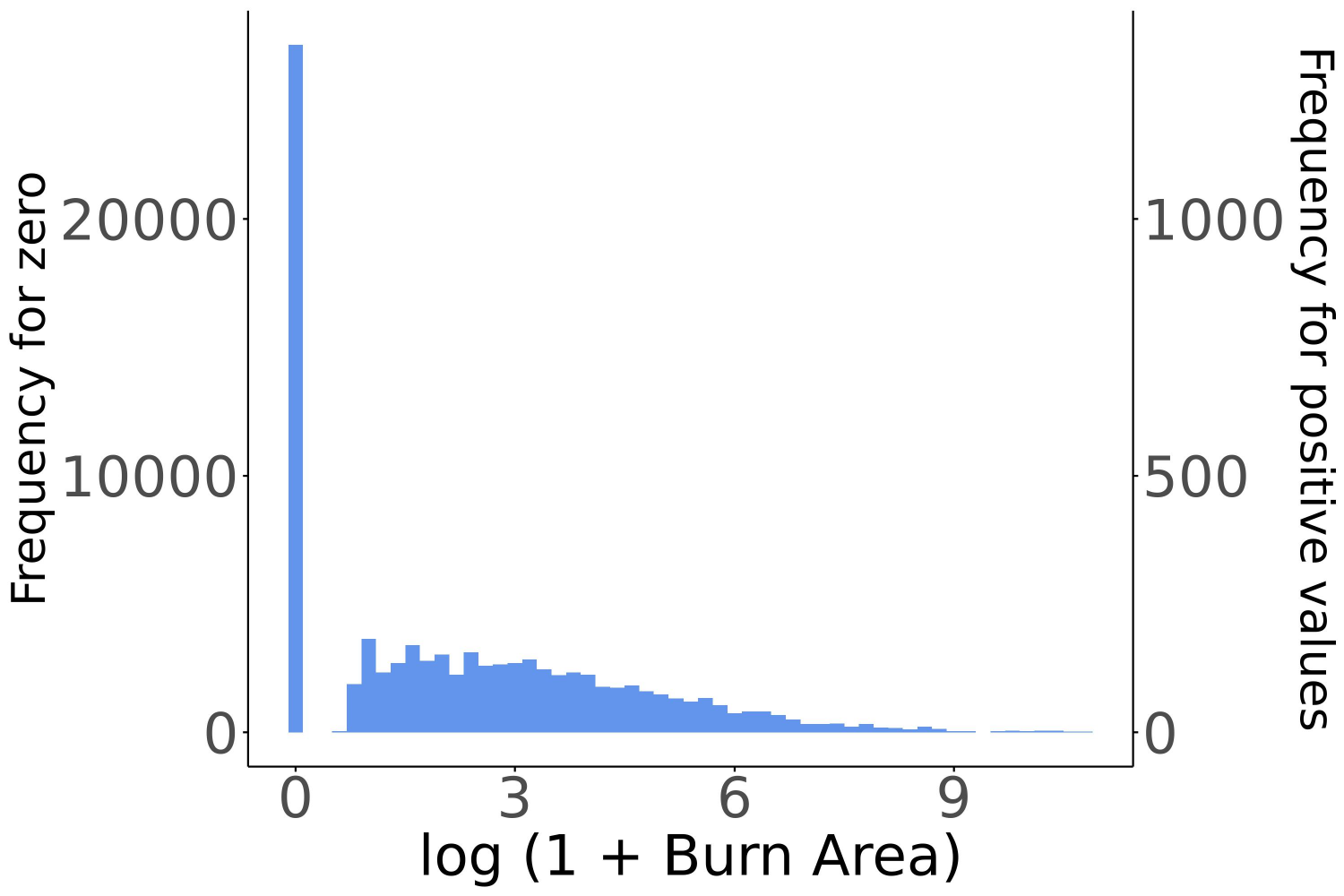}
\end{center}
% \vspace{-1cm}
\caption{\footnotesize{Histograms of fire count and burnt area at the council-month level, highlighting the prevalence of zeros.
Quantities are rescaled for visualisation purposes using a square-root transformation for fire counts and a logarithmic transformation for burned area.}}
\label{fig: exploratory_plot}
\end{figure}

After preprocessing, the dataset comprises 278 councils over 156 months, resulting in 43,368 council-month observations. 
For each observation, the total fire count and total burnt area are computed and serve as the primary response variables in the modelling framework. 
Observations with no recorded fire activity are assigned zeros for both responses. 
Unless otherwise specified, ``fire count" and ``burnt area" hereafter refer to these council-monthly totals.
This aggregation process naturally introduces a large number of zeros into the data, as illustrated in Figure \ref{fig: exploratory_plot}, resulting in zero-inflation in both response variables. 
Additionally, both the fire count and burnt area distributions exhibit strong skewness, even after transformation (square root for fire count, logarithm for burnt area).
Due to the pronounced skewness and heavy-tailed nature of the burnt area distribution, we model its square root to mitigate these effects.
Alternative transformations for burnt area are discussed in Section \ref{sec: discussion}.

\begin{figure}[ht]
\begin{center}
\includegraphics[width=0.3\textwidth]{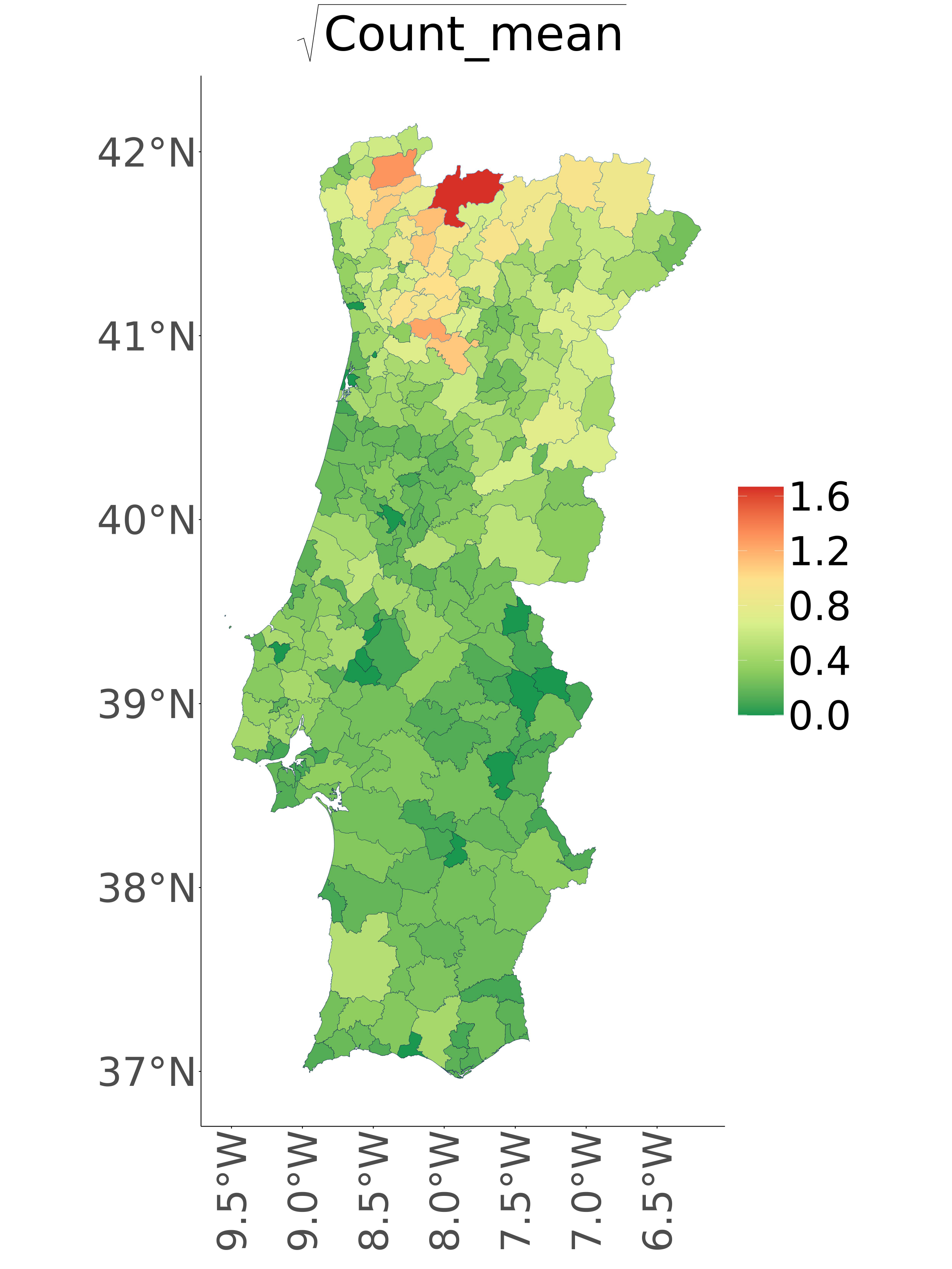}
\includegraphics[width=0.3\textwidth]
{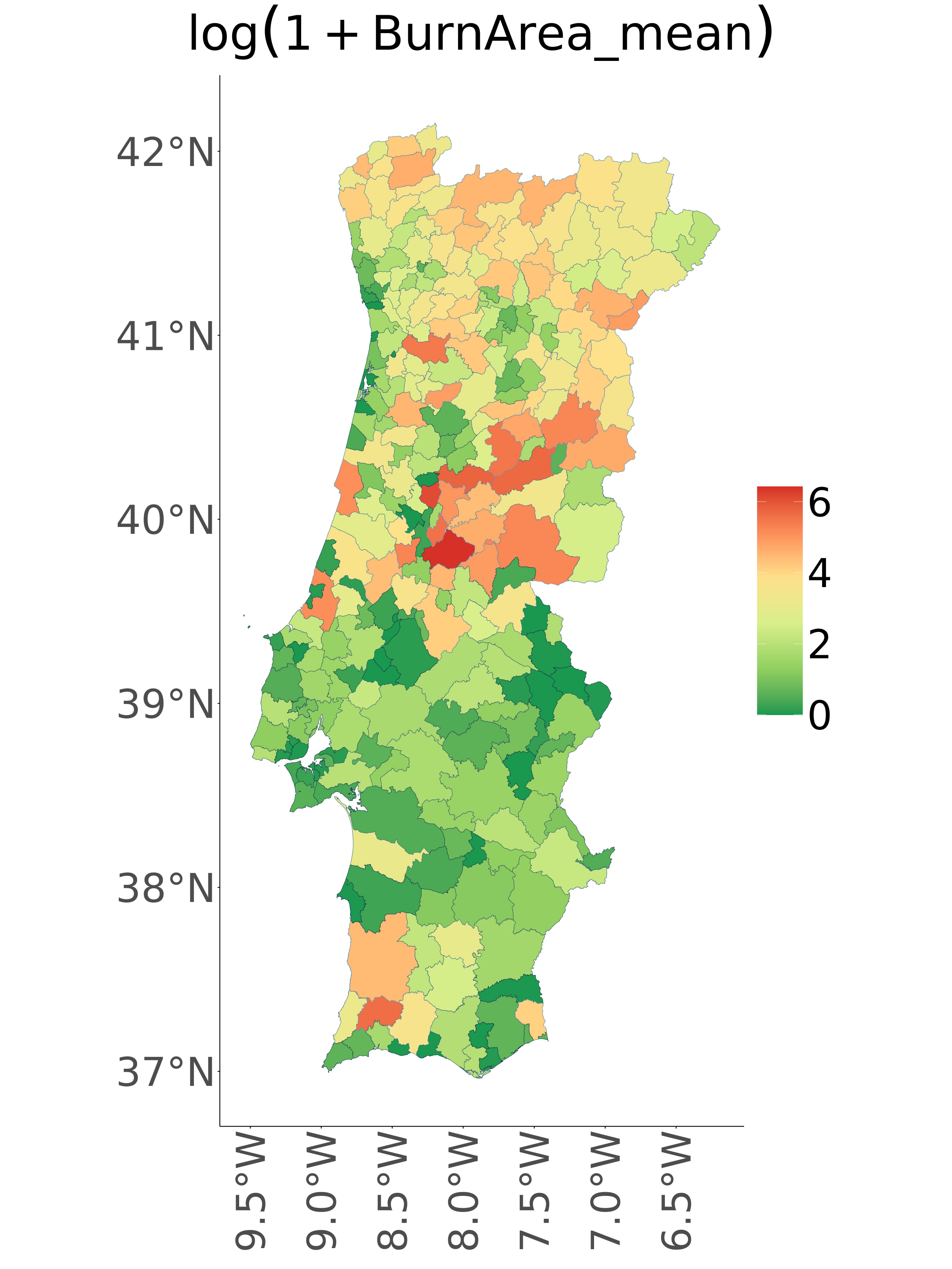}
\includegraphics[width=0.6\textwidth]{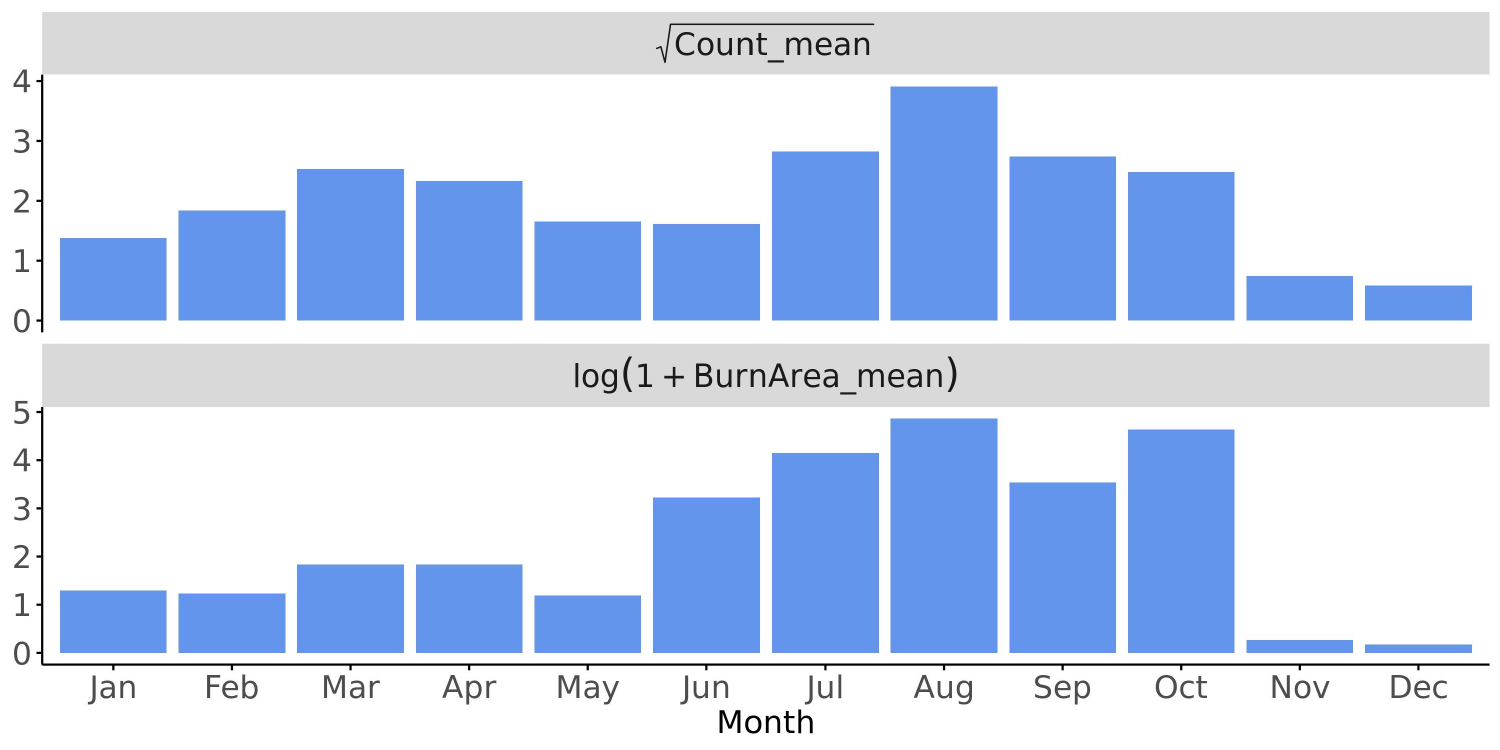}
\end{center}
% \vspace{-1cm}
\caption{\footnotesize{Top: Average council-level fire count and burnt area. Bottom: Monthly average fire count and burnt area across all councils. Spatial and seasonal variation is evident.
Quantities are rescaled for visualisation purposes using a square-root transformation for fire counts and a logarithmic transformation for burned area.}}
\label{fig: raw_fire_map}
\end{figure}

Figure \ref{fig: raw_fire_map} presents the spatial and seasonal patterns of wildfires across Portugal.
Spatial heterogeneity is evident: while fire count tends to increase with latitude and displays spatial autocorrelation among neighbouring councils, extreme burnt areas are more concentrated in central-north Portugal, with notable variability even between adjacent councils. 
Strong seasonal patterns are also observed, with peak activity occurring during summer and autumn.

\subsection{Environmental Covariates}
Wildfire behaviour is strongly influenced by environmental conditions, particularly, climate related including, e.g., wind speed and direction, air temperature, and humidity.
To enhance the predictive capabilities of our model, we incorporate 11 environmental covariates derived from reanalysis datasets spanning 2011–2023.

Five meteorological (air temperature, precipitation, wind speed and direction, dew point ) and two vegetation-related covariates (green leaf area for two vegetation types) are obtained from ERA5-Land, which offers a fine spatial resolution ($0.1^\circ \times 0.1^\circ$).
Four additional vegetation covariates, related to land cover types and their coverage percentages, are sourced from the ERA5, which has a relatively coarse resolution ($0.25^\circ \times 0.25^\circ$) and is time-invariant.
In addition, two derived covariates computed from the daily meteorological data are included: relative humidity and FWI.

All covariates are spatially mapped to the nearest council unit by assigning each grid cell to the council containing its centroid. 
Temporal aggregation is performed at the monthly level.
For continuous variables, the aggregated mean is used; for categorical variables, the mode is applied.
A complete list and description of all covariates are provided in Table \ref{tab:covariates_appendix} in the Supplementary Materials.

\section{Methods}
\label{sec: method}
\subsection{Overview}

\begin{figure}[ht]
\begin{center}
\includegraphics[width=0.9\textwidth]{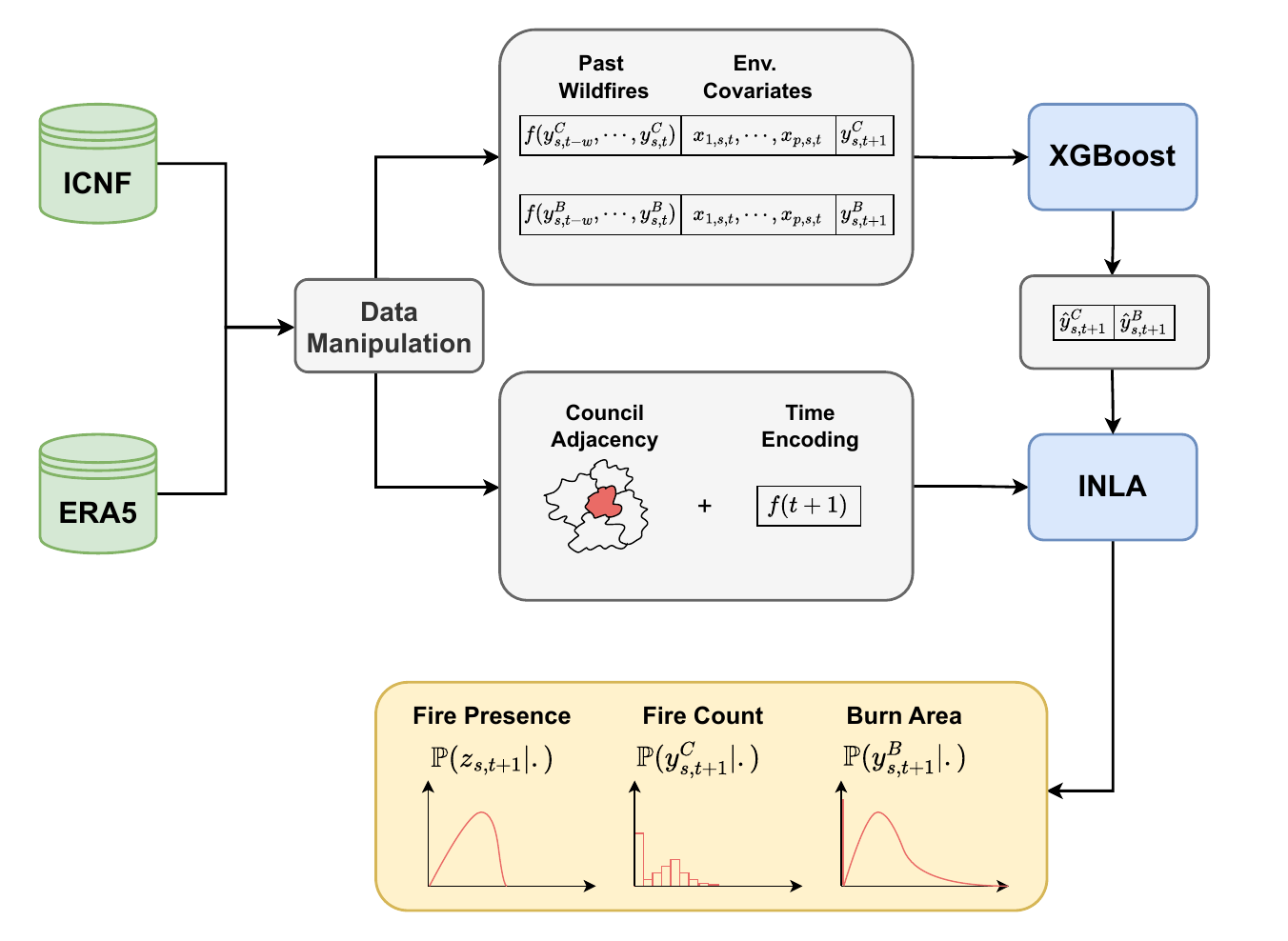} 
\end{center}
% \vspace{-1cm}
\caption{\footnotesize{Diagram of the proposed two-stage wildfire forecasting framework combining XGBoost and INLA.}}
\label{fig: model_framework}
\end{figure}

We propose a two-stage modelling framework to provide probabilistic forecasts of wildfires at the monthly council level. 
In the first stage, we employ the XGBoost algorithm to integrate historical wildfire activity data and current environmental covariates to generate one-month-ahead forecasts of wildfire count and burnt area for each council. 
These forecasts are then passed to a latent Gaussian model estimated via INLA, which incorporates spatio-temporal dependencies through council adjacency and temporal encoding.
The INLA framework outputs full posterior predictive distributions for fire presence, fire count, and burnt area.
To model the heavy-tailed behaviour of burnt areas, we implement the extended generalised Pareto (eGP) likelihood in INLA, which blends a Gamma-like left tail with a Pareto-like right tail, and is characterised by two shape parameters and a scale parameter (see Section~\ref{sec:likelihood-spec} for details).
This two-stage approach can be viewed as a form of model stacking \citep{wolpert1992stacked}, a type of ensemble learning where the predictions of one model are used as input features for another.
Figure \ref{fig: model_framework} provides a high-level overview of the proposed modelling pipeline.

The combination of XGBoost and INLA utilises the strengths of each framework while mitigating their individual limitations. 
XGBoost is particularly effective in modelling complex, nonlinear interactions among environmental and historical variables. 
By incorporating past wildfire activity in an autoregressive manner, the two XGBoost models, one for fire count and one for burnt area, can partially recover information that may be lost due to the spatial smoothing inherent in gridded environmental data. 
However, a key drawback of XGBoost is its lack of native uncertainty quantification. 
This is addressed in the second stage of our framework, where the INLA-based Bayesian hierarchical model provides full posterior distributions, allowing for straightforward uncertainty quantification.

\subsection{Stage I: XGBoost}
XGBoost is a scalable and efficient gradient boosting framework \citep{friedman2001greedy} designed for structured data. 
As an ensemble method, it constructs a strong predictive model by sequentially combining multiple tree models, each of which serves as a weak learner. 
The fundamental idea is to improve model performance by learning residual patterns from previous models in an additive and iterative fashion.

In a regression task, each regression tree $f(\boldsymbol{x})$, with input $\boldsymbol{x}\in \mathbb{R}^d$, partitions the feature space into $L$ disjoint regions $R_1, R_2,\cdots, R_L$ using  a series of binary splits.
For any input $\boldsymbol{x}_i,\;i=1,\ldots,n$, the model assigns a prediction $w_l$, which is the average response within the region $R_l$ containing $\boldsymbol{x}_i$:
\begin{align*}
f(\boldsymbol{x}_i) = \sum_{l=1}^{L}w_l\mathbbm{1}\{{\boldsymbol{x}_i\in R_l}\},
\end{align*}
\noindent with $\mathbbm{1}\{\cdot\}$ denoting an indicator function. The final prediction of an XGBoost model is the sum of $M$ such regression trees: 
\begin{align*}
\widehat{y}_i = \sum_{m=1}^{M}f_m(\boldsymbol{x}_i), \quad f_m \in \mathcal{F},
\end{align*}
where $\mathcal{F}$ denotes the space of all possible trees.
Each tree is fitted in a forward stagewise manner, with tree $f_m$ at iteration $m$ trained to minimise the regularised objective function:
\begin{align*}
\mathcal{L}^{(m)} = \sum_{i=1}^n\ell(y_i,\widehat{y}_i^{(m-1)} + f_m(\boldsymbol{x}_i)) + \Omega(f_m),
\end{align*}
where $\ell(\cdot)$ is a differentiable convex loss function, $\widehat{y}_{i}^{(m-1)} = \sum_{j=1}^{m-1}f_j(\boldsymbol{x}_i)$ is the prediction from the ensemble up to iteration $m-1$, and $\Omega(f_m) = \gamma L + \frac{1}{2}\lambda \|\boldsymbol{w}\|^2$ penalises the model complexity through the number of leaves $L$ and the leaf weights $\boldsymbol{w}$.
To improve the optimisation efficiency, the loss is approximated using a second-order Taylor expansion \citep{chen2016xgboost}:
\begin{align*}
    \mathcal{L}^{(m)} \approx \sum_{i=1}^n\left[\ell(y_i,\widehat{y}_i^{(m-1)}) + g_if_m(\boldsymbol{x}_i) + \frac{1}{2}h_i f_m^2(\boldsymbol{x}_i) \right]+ \gamma L + \frac{1}{2}\lambda \sum_{j=1}^L w_j^2,
    \label{eq: xgb_obj}
\end{align*}
where $g_i$ and $h_i$ are the first and second derivatives of the loss function with respect to $\widehat{y}^{m-1}_i$.
Given a fixed tree structure, the optimal leaf weights can be derived analytically in terms of $g_i$, $h_i$ and $\lambda$.

The choice of loss function $\ell$ is central to the performance of XGBoost and should align with the distributional properties of the response variable.
In our case, fire count is a non-negative integer and is naturally modelled using a Poisson loss.
By contrast, the burnt area is of a mixed type: it is continuous and positive when fires occur, but has a point mass at zero when no fire is observed.
For this, we adopt the Tweedie deviance loss \citep{jorgensen1987exponential}, which corresponds to a compound Poisson-Gamma distribution.
This distribution models a sum of Gamma-distributed fire sizes conditional on a Poisson-distributed number of occurrences:
\begin{equation*}
    Y = \begin{cases}
        0 \quad &\text{if } N=0,\\
        \sum_{i=1}^N X_i  \quad &\text{if } N=1,2,\ldots
    \end{cases}
\end{equation*}
where $N$ is a Poisson random variable, and $X_i$ are i.i.d. Gamma random variables.
Here, $N$ and $X_i$ can be interpreted as the council-month level fire count and the burnt area in each fire ignition, respectively, and $Y$ represents the total burnt area at the council-month level.
The corresponding Tweedie deviance loss function of true value $y$ and prediction (mean of the Tweedie) $\widehat{y}$ is 
\begin{equation*}
    \ell(y,\widehat{y},k) = 2\left(\frac{\max\{y,0\}^{2-k}}{(1-k)(2-k)}- \frac{y\widehat{y}^{1-k}}{1-k}+\frac{\widehat{y}^{2-k}}{2-k}\right), \quad 1<k<2,
\end{equation*}
where the index $k$ governs the shape of the distribution: values closer to 1 approximate a Poisson, and those nearer 2 approach a Gamma.

\subsubsection{Window-Based Modelling}
Tree-based models, including XGBoost, do not inherently account for sequential dependencies in time-series data.
A naive implementation would treat each time point independently, using covariates $\boldsymbol{x}_{t+1}$ to predict the target $y_{t+1}$, thereby neglecting temporal autocorrelation and requiring future covariates for forecasting.
Given our 1-month modelling granularity, it is challenging to obtain the future environmental covariates over such a large horizon.

\begin{figure}
\begin{center}
\includegraphics[width=1\textwidth]{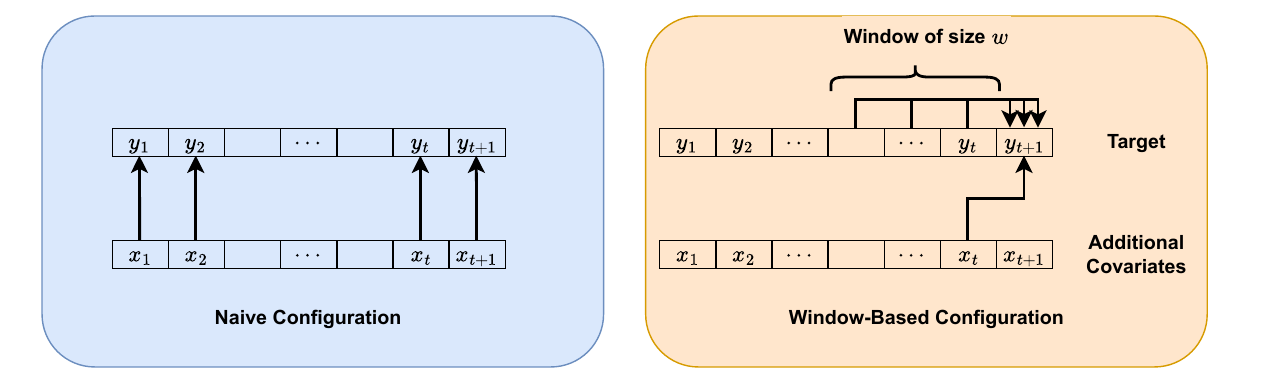} 
\end{center}
% \vspace{-1cm}
\caption{\footnotesize{Comparison of naive and window-based modelling configurations. Arrows indicate the direction of information flow from predictors to targets.}}
\label{fig: window_based_format}
\end{figure}

To address this limitation, we adopt a window-based approach, a common practice in time-series forecasting for non-sequential models \citep{elsayed2021reallyneeddeeplearning}. 
For a given forecasting horizon of one month, we reformulate the data into a lagged autoregressive structure: each target $y_{t+1}$ is modelled as a function of covariates and wildfire records from previous time steps, such as $(\boldsymbol{x}_{t},y_{t},y_{t-1},\ldots,y_{t-w+1})$, where $w$ is the time window size.
This configuration enables the model to learn temporal dependencies and make forecasts based on available historical wildfire data and covariates.
It also helps mitigate the issue of smoothed covariate values, which are uninformative for local fire prediction,  by incorporating recent wildfire history.
Figure \ref{fig: window_based_format} contrasts the window-based modelling with the naive approach.

Autocorrelation plots (ACF) of the monthly fire count and burnt area in Figure \ref{fig: ACF_plots} in the Supplementary Materials reveal short-term dependencies and seasonal peaks at lags 12, 24, and 36, suggesting strong annual cycles. 
Based on this,  we set $w=36$, and include both short and long-term lag features.
For short-term features, past fire count and burnt area up to lag 9 are included.
Long-term and periodic patterns are captured by feature-engineered covariates.
A full list of autoregressive features is provided in Table \ref{tab: autoregressive_covariates} in the Supplementary Materials.

Let $\widetilde{\boldsymbol{x}}_{s,t}^C$ and $\widetilde{\boldsymbol{x}}_{s,t}^B$ denote the complete feature vectors used to forecast fire count ($C$) and burnt area ($B$), respectively, at council $s$ and time $t$.
Both vectors share the same covariates listed in Table \ref{tab:covariates_appendix} and \ref{tab: autoregressive_covariates}, except $\widetilde{\boldsymbol{x}}_{s,t}^C$ includes only fire count-based autoregressive covariates, while $\widetilde{\boldsymbol{x}}_{s,t}^B$ includes only those based on burnt area.
The forecasts for fire count and burnt area at council $s$ and time $t+1$ are then given by:
\begin{align}
\widehat{y}_{s,t+1}^C = \sum_{m_1}f_{m_1}^C(\widetilde{\boldsymbol{x}}_{s,t}^C),\label{eq: xgb_cnt}\\
\sqrt{\widehat{y}_{s,t+1}^B} = \sum_{m_2}f_{m_2}^B(\widetilde{\boldsymbol{x}}_{s,t}^B),\label{eq: xgb_ba}
\end{align}
where $f_{m_1}^C$ and $f_{m_2}^B$ are regression trees trained to minimise Poisson and Tweedie deviance losses, respectively.

\subsubsection{Output Generation}
\label{sec: cv_score}
{We adopt a time-adjusted Super Learner cross-validation (CV) scheme \citep{van2007super} to tune hyperparameters and generate one-month-ahead forecasts of fire counts and burnt area.
Standard $k$-fold CV is unsuitable for time series, as random splits may allow training on observations that occur after the validation set, thereby introducing look-ahead bias.
To avoid this, we employ a chronological CV strategy, in which the model is trained exclusively on past folds and evaluated on subsequent future folds.}

\begin{figure}
\begin{center}
\includegraphics[width=1\textwidth]{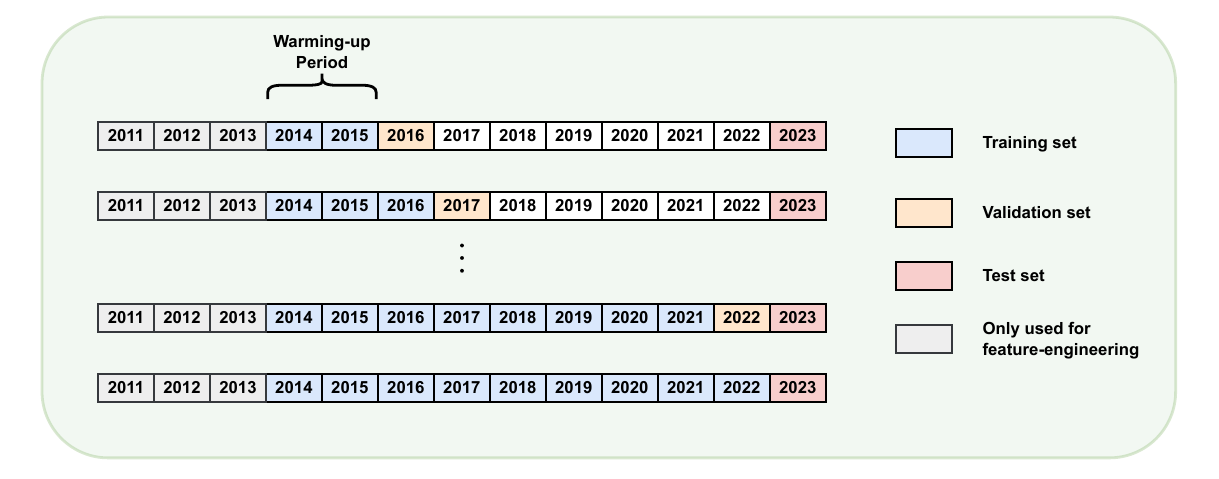} 
\end{center}
% \vspace{-1cm}
\caption{\footnotesize{Time-series–aware cross-validation is implemented using yearly blocks for both hyperparameter tuning and the generation of one-month-ahead forecasts.
The first three years are excluded from the training set because feature construction based on a 36-month rolling window results in missing values during this period. }}
\label{fig: cross_validation}
\end{figure}

{Formally, let $D_k, k=1,\ldots,12$ denote the subsets of records corresponding to calendar years 2011–2022 in the training set, and let $k_0$ mark the end of the warm-up period required to accumulate sufficient history for stable forecasting in the first validation fold.
For each validation year $k_{\mathrm{val}} > k_0$, we fit the model on $\bigcup_{k<k_{\mathrm{val}}} D_k$ and evaluate it on $D_{k_{\mathrm{val}}}$.
Hyperparameters are selected by optimising metrics computed on pooled out-of-fold predictions, i.e. aggregating predictions from all validation years ${k_{\mathrm{val}} > k_0}$ and evaluating the metric once on this combined set.
After selection, we (i) refit on $\bigcup_{k<k_{\mathrm{val}}} D_k$ to generate within-training one-month-ahead forecasts for $D_{k_{\mathrm{val}}}$, and (ii) refit on $\bigcup_{k=1}^{12} D_k$ to forecast the test period (2023).
This procedure prevents information leakage and provides performance estimates that reflect out-of-time generalisation, as illustrated in Figure \ref{fig: cross_validation}.}

\subsection{Stage II: Bayesian Latent Gaussian Modelling}
We incorporate the one-month-ahead forecasts of fire count and burnt area from equations \eqref{eq: xgb_cnt} and \eqref{eq: xgb_ba} as external covariates in a Bayesian latent Gaussian model, which yields the final probabilistic predictions.
This two-stage setup reflects the principles of stacked generalisation \citep{wolpert1992stacked}, where the predictions from one model serve as informative inputs to a second, more interpretable model to enhance overall performance.

The XGBoost forecasts compensate for the structural inflexibility of latent Gaussian models in handling many covariates, particularly when those covariates have nonlinear interactions. 
Conversely, the Bayesian latent Gaussian model addresses a key limitation of XGBoost: the lack of native uncertainty quantification. 
The Bayesian latent Gaussian model provides full posterior distributions for quantities of interest, thus combining flexibility with interpretability and probabilistic reasoning.

\subsubsection{Model Structure and INLA}
% \dcc{needs to be updated, this is the old INLA (now there is no nested Laplace). See \url{https://www.sciencedirect.com/science/article/pii/S0167947323000038?casa_token=JQ2h7XC4c1MAAAAA:fgWil0ywGGMGM0xHAYdfwkifN5TJtwBlGCZha2cXOOtFtJmc5a8Kr-V9wK8cTjY-vS0mG63VNg}}

In a Bayesian latent Gaussian model, each observation $y_i$ is assumed to be conditionally independent given the linear predictor $\eta_i$ and hyperparameters $\boldsymbol{\theta}_1$ of the observation model.
The response’s mean or quantiles are linked to $\eta_i$ as in the generalised linear model framework, and $\eta_i$ comprises random and fixed effects that describe the data in an additive way:
\begin{equation*}
    \eta_i = \beta_0 + \sum_{j}\beta_jv_{i,j} + \sum_k \omega_{i,k},
\end{equation*}
where $v_{i,j}$ are the fixed effects, $\omega_{i,k}$ are latent Gaussian random effects, and $\beta_0,\beta_j$ are linear coefficients.
Using $\boldsymbol{\theta}_2$ to denote all hyperparameters in $\beta_j$, $\omega_{i,k}$,
the latent field $\boldsymbol{u} = (\beta_0,\beta_1,\dots,\omega_{1,1},\cdots)$ is assumed to have a Gaussian prior 
\begin{equation*}
\boldsymbol{u}\mid\boldsymbol{\theta}_2 \sim \mathcal{N}(\boldsymbol{0}, Q^{-1}(\boldsymbol{\theta}_2)),
\end{equation*}
where $Q(\boldsymbol{\theta}_2)$ is the precision matrix.
The linear predictor $\boldsymbol{\eta}=(\eta_1,\cdots,\eta_n)$ can be expressed by $\boldsymbol{u}$ and a sparse design matrix $\boldsymbol{A}$ by
\begin{equation*}
    \boldsymbol{\eta} = \boldsymbol{Au}.
\end{equation*}

% Due to the additive Gaussian structure, $\eta_i$ is itself Gaussian.
% Using $\boldsymbol{\theta}_2$ to denote all hyperparameters in $\beta_j$, $\omega_{i,k}$,
% the latent field $\boldsymbol{u} = (\beta_0,\beta_1,\dots,\omega_{1,1},\cdots)$ has a Gaussian prior 
% \begin{equation*}
% \boldsymbol{u}\mid\boldsymbol{\theta}_2 \sim \mathcal{N}(\mu(\boldsymbol{\theta_2}), Q^{-1}(\boldsymbol{\theta}_2)),
% \end{equation*}
% where $\mu(\boldsymbol{\theta_2})$ is the mean vector and $Q(\boldsymbol{\theta}_2)$ is the precision matrix.

Let $\boldsymbol{\theta}=(\boldsymbol{\theta}_1,\boldsymbol{\theta}_2)$ and its prior as $\pi(\boldsymbol{\theta})$, the full posterior distribution is
\begin{align*}
\pi(\boldsymbol{u},\boldsymbol{\theta}\mid\boldsymbol{y}) &\propto \pi(\boldsymbol{y}\mid \boldsymbol{A}\boldsymbol{u},\boldsymbol{\theta})\pi(\boldsymbol{u}\mid\boldsymbol{\theta})\pi(\boldsymbol{\theta})\\
&\propto \pi(\boldsymbol{u}\mid\boldsymbol{\theta})\pi(\boldsymbol{\theta})\prod_i\pi(y_i\mid (\boldsymbol{Au})_i,\boldsymbol{\theta}),
\end{align*}
and this joint posterior could be approximated using MCMC. 
% However, under the latent Gaussian model framework, it suffices to compute the marginal posterior distributions $\pi(\eta_i\mid\boldsymbol{y})$ and $\pi(\theta_j\mid\boldsymbol{y})$  for further inference of $\boldsymbol{y}$, thanks to the conditional independence assumption. 
{However, in many spatial applications, the marginal posterior distributions $\pi(u_i \mid \boldsymbol{y})$ and $\pi(\theta_j \mid \boldsymbol{y})$ are of greater practical interest than the full joint posterior, and knowledge of these marginals is sufficient for subsequent inference and analysis.}
\citet{rue2009approximate} proposed a fast approximation algorithm for marginal posteriors in latent Gaussian models using the integrated nested Laplace approximation (INLA).
This algorithm was later reformulated to accommodate big data settings, enhancing numerical scalability and enabling fast inference \citep{van2023new}. 
These developments form the foundation of the R-INLA package \citep{martins2013bayesian}.

A key idea in INLA is to approximate $\pi(\boldsymbol{u}\mid\boldsymbol{\theta},\boldsymbol{y})$ by the Laplace approximation $\widetilde{\pi}(\boldsymbol{u}\mid\boldsymbol{\theta},\boldsymbol{y})$ so that the joint posterior $\pi(\boldsymbol{\theta\mid\boldsymbol{y}})$ can be approximated by
\begin{align*}
    \left.\widetilde{\pi}(\boldsymbol{\theta}\mid\boldsymbol{y})\propto \frac{\pi(\boldsymbol{u},\boldsymbol{\theta},\boldsymbol{y})}{\widetilde{\pi}(\boldsymbol{u}\mid\boldsymbol{\theta},\boldsymbol{y})} \right\vert_{\boldsymbol{u}=\boldsymbol{u}^*(\boldsymbol{\theta})},
\end{align*}
where $\boldsymbol{u}^*(\boldsymbol{\theta})$ is the mode of ${\pi}(\boldsymbol{u}\mid\boldsymbol{\theta},\boldsymbol{y})$.
Then, the marginal posterior $\pi(\theta_j\mid\boldsymbol{y})$ can be obtained by numerically integrating out the nuisance parameters $\boldsymbol{\theta}_{-j}$:
\begin{align*}
   \pi(\theta_j\mid\boldsymbol{y}) = \int \widetilde{\pi}(\boldsymbol{\theta}\mid\boldsymbol{y})\text{d}\boldsymbol{\theta}_{-j}.
\end{align*}
Next, taking the Gaussian margins $\widetilde{\pi}\left(u_i \mid \boldsymbol{\theta}, \mathbf{y}\right)$ from $\widetilde{\pi}(\boldsymbol{u}|\boldsymbol{\theta},\boldsymbol{y})$, the marginal posterior $\pi(u_i\mid\boldsymbol{y})$ is approximated by
\begin{equation*}
    \widetilde{\pi}\left(u_i \mid \boldsymbol{y}\right) \approx \sum_k \widetilde{\pi}\left(u_i \mid \boldsymbol{\theta}_k, \boldsymbol{y}\right) \widetilde{\pi}\left(\boldsymbol{\theta}_k \mid \boldsymbol{y}\right) \Delta_k,
\end{equation*}
with integration points $\boldsymbol{\theta}_k$ and weights $ \Delta_k$.
The marginal posterior of the linear predictors $\pi(\eta_i \mid \boldsymbol{y})$ is derived in a similar manner. 
Staring with the Gaussian approximation $\widetilde{\pi}(\boldsymbol{u}\mid\boldsymbol{\theta},\boldsymbol{y})$, the conditional posterior $\pi(\eta_i|\boldsymbol{\theta},\boldsymbol{y})$ is then also approximated as Gaussian.
Its mean and variance can be efficiently computed leveraging the linear relationship $\boldsymbol{\eta}=\boldsymbol{Au}$ and some tricks on computing the $i$-th diagonal element of the inverse of the precision matrix associated with $\widetilde{\pi}(\boldsymbol{u}|\boldsymbol{\theta},\boldsymbol{y})$ \citep{van2023new}.
The marginal posterior $\pi(\eta_i|\boldsymbol{y})$ is then approximated  by:
\begin{equation*}
    \widetilde{\pi}\left(\eta_i \mid \boldsymbol{y}\right) \approx \sum_k \widetilde{\pi}\left(\eta_i \mid \boldsymbol{\theta}_k, \boldsymbol{y}\right) \widetilde{\pi}\left(\boldsymbol{\theta}_k \mid \boldsymbol{y}\right) \Delta_k.
\end{equation*}
Finally, \cite{van2023new, van2024low} proposed a low-rank correction to the mean of $\widetilde{\pi}(\boldsymbol{u} \mid \boldsymbol{\theta}, \boldsymbol{y})$ using variational Bayes, further improving the approximation of both $\widetilde{\pi}(u_i \mid \boldsymbol{y})$ and $\widetilde{\pi}(\eta_i \mid \boldsymbol{y})$.
% Gaussian, Laplace, and a simplified Laplace approximations are available for  $\widetilde{\pi}\left(u_i \mid \boldsymbol{\theta}_k, \mathbf{y}\right)$, as described in Rue's original paper.
% The above Bayesian latent Gaussian model fitting and inference are performed using the R-INLA package \citep{martins2013bayesian}.

\subsubsection{Likelihood Specification}\label{sec:likelihood-spec}
Although Poisson and Tweedie likelihoods perform well in modelling general wildfire trends via XGBoost, they are not optimal in the latent Gaussian framework. 
Specifically, the Tweedie distribution, being composed of i.i.d. Gamma components, features a light right tail and is therefore unsuitable for capturing extreme burnt areas.
Instead, we adopt the extended generalised Pareto (eGP) distribution \citep{naveau2016modeling} for modelling the burnt area.
This distribution combines features of Gamma and Pareto distributions by applying a power transformation to the standard Pareto distribution. 
It is defined on the positive real line and behaves like a Gamma for small values and like a Pareto for large values. 
The right tail behaviour is controlled by a shape parameter $\xi$, allowing flexibility in modelling heavy tails.

To handle zero-inflated data, we use a hurdle model, which separately models zero and positive outcomes. The hurdle model is defined as:
\begin{align}
    \pi(y)=
    \begin{cases}
    \mathbb{P}(Z=0), \quad  &y=0,\\
    \mathbb{P}(Z=1)\pi(y\mid Z=1), \quad  & y > 0,
    \end{cases}
    \label{eq: hurdle}
\end{align}
where $Z$ is a latent Bernoulli variable indicating the presence of a non-zero event.
The structure in \eqref{eq: hurdle} can be easily implemented in the latent Gaussian model framework for wildfire modelling by introducing an auxiliary Bernoulli variable $Z$ of the same length as the observations.
Conditional on $Z=1$, the response $y$ is modelled using a suitable distribution $\pi(y\mid Z=1)$ with positive support.

For the fire count,  the conditional distribution $\pi(y| Z=1)$ is modelled using a zero-truncated Poisson distribution with parameter $\lambda$:

\begin{align*}
    \mathbb{P}(Y=y;\lambda) = \frac{1}{1-\exp(-\lambda)}\frac{\lambda^y}{y!}\exp(-\lambda), \quad y = 1, 2, \ldots
\end{align*}
% It is derived from normal Poisson by truncating 0 and moving the probability mass on 0 to the positive integers.

For the burnt area, we model the square root of the area using the eGP distribution.
The cumulative distribution function of eGP is given by:
\begin{align}
F(y;\sigma,\xi,\kappa)= \begin{cases}\left[1-\left(1+\xi{y}/{\sigma}\right)^{-1/\xi}\right]^{\kappa},\; &\quad y>0, \; \xi \neq0, \\
\left[ 1-\exp(-{y}/{\sigma}) \right]^\kappa, &\quad y>0,\; \xi=0,
\end{cases}
\label{eq: egpd}
\end{align}
where  $\xi \in \mathbb{R} $ controls the rate of upper tail decay, $\sigma > 0$ is the scale, and $\kappa > 0$ governs the shape of the lower tail.
The linear predictor $\eta$ is linked to the $\alpha$-quantile $q_\alpha$ of eGP by $q_\alpha = \exp(\eta)$, 
where $\alpha$ is typically set to be 0.5.
By setting $\kappa$ and $\xi$ as hyperparameters, the scale parameter $\sigma$ can be expressed as a function of $\eta$, $\alpha$, $\kappa$ and $\xi$:
\begin{align*}
    \sigma(\eta)=\frac{\xi q_\alpha}{(1-\alpha^{1/\kappa})^{-\xi}-1}=\frac{\xi \exp(\eta)}{(1-\alpha^{1/\kappa})^{-\xi}-1}.
    % \label{eq: scale_parametration}
\end{align*}

We define three linear predictors for each council $s$ at time $t+1$: $\eta^Z_{s,t+1}$, $\eta^C_{s,t+1}$ and $\eta^B_{s,t+1}$, corresponding to fire presence ($Z$), fire count ($C$) and burnt area ($B$), respectively.
The full hierarchical model is described as:
\begin{subequations}
       \begin{align*}
        Z_{s,t+1}\mid\eta_{s,t+1}^Z &\sim \text{Bernoulli}\{\text{logit}^{-1}(\eta_{s,t+1}^Z)\} \\ 
        \{Y_{s,t+1}^C\mid \eta_{s,t+1}^C, Z_{s,t+1}=1\} &\sim \text{Trucated Poisson}\{\exp(\eta_{s,t+1}^C)\}  \\ 
        \left\{\sqrt{Y_{s,t+1}^B}\mid\eta_{s,t+1}^B, Z_{s,t+1}=1\right\} &\sim \text{eGP}\{\sigma\left(\eta_{s,t+1}^B\right), \xi, \kappa\} \\
        \xi, \kappa & \sim \text{Hyperpriors}.
        \end{align*}
\end{subequations}

\subsubsection{Effects in the Linear Predictor}
The latent effects in the linear predictors comprise Gaussian random effects derived from XGBoost predictions, as well as spatio-temporal Gaussian effects informed by adjacency structures and time.

Although the XGBoost predictions $\widehat{y}^C_{s,t+1}$ and $\widehat{y}^B_{s,t+1}$ could be included as fixed effects, {doing so would impose} a linear relationship with the linear predictor. Given the complex spatial and spatio-temporal {dependencies present in the data, such a restriction is overly limiting}.
Instead, {we treat the XGBoost outputs as smooth random effects, allowing the second-stage model to flexibly recalibrate the machine-learning predictions in a spatially coherent and uncertainty-aware manner. Specifically, }we model the effect of $\widehat{y}^{(\cdot)}_{s,t+1}$ using a first-order random walk (RW1) prior on 20 discretised bins of the prediction $\widehat{y}_{s,t+1}^{(\cdot)}$:
\begin{align}
    R_k - R_{k-1} \sim \mathcal{N}(0,\tau^{-1}_{R}),
    \label{eq: rw1}
\end{align}
where $R_k$ represents the effect of the $k$-th bin of the discretised covariate and $\tau_{R}$ is the precision parameter.
As the fire count and burnt area contribute to the Poisson and eGP likelihoods only when fire is present (i.e., when $y^C_{s,t+1}>0$) in the hurdle model training, we discretise $\widehat{y}_{s,t+1}^C$ and  $\widehat{y}_{s,t+1}^B$ conditional on ${y}_{s,t+1}^C>0$ when constructing the $R_k$ in their respective linear predictors $\eta_{s,t+1}^C$ and  $\eta_{s,t+1}^B$.
By contrast, $\widehat{y}_{s,t+1}^C$ and $\widehat{y}_{s,t+1}^B$ are discretised unconditionally when constructing the corresponding $R_k$ in $\eta_{s,t+1}^Z$.

We incorporate spatially structured and unstructured effects through the Besag–York–Mollié model, using the reparameterised BYM2 formulation \citep{riebler2016intuitive, simpson2017penalising}, as implemented in R-INLA.
A BYM2 effect $\boldsymbol{b}$ combines a scaled intrinsic conditional autoregressive (CAR) component $\boldsymbol{\delta}$ (with unit variance) and unstructured noise $\boldsymbol{\epsilon} \sim \mathcal{N}(\boldsymbol{0},\boldsymbol{I})$.
Here, $\boldsymbol{I}$ represents the identity matrix with dimension defined by the length of $\boldsymbol{\epsilon}$.
The effect is defined as 
\begin{align*}
    \boldsymbol{b}=\frac{1}{\sqrt{\tau}}(\sqrt{1-\phi}\boldsymbol{\epsilon} + \sqrt{\phi}\boldsymbol{\delta}),
\end{align*}
where $\tau$ is the precision parameter, and $\phi\in [0,1]$  controls the balance between spatial structure ($\phi=1$) and unstructured variation ($\phi=0$).
To reflect different spatial scales, we define two adjacency graphs based on Portugal's administrative boundaries: one at the council level (fine granularity) and the other at the district level (coarse granularity), resulting in two distinct BYM2 effects, termed $\boldsymbol{b}_c$ for the council-level effect and $\boldsymbol{b}_d$ for the district-level effect.
If we further denote the covariance of $\boldsymbol{b}_c$ as $\Sigma_c$ and the covariance of $\boldsymbol{b}_d$ as $\Sigma_d$, then
\begin{align*}
    \boldsymbol{b}_c \sim \mathcal{N}(\boldsymbol{0},\boldsymbol{\Sigma}_c), \quad
    \boldsymbol{b}_d \sim \mathcal{N}(\boldsymbol{0},\boldsymbol{\Sigma}_d).
\end{align*}

To introduce temporal dynamics, we group the spatial effects over time. 
This provides greater flexibility than additive spatial and temporal terms.
We consider two grouping schemes: Group 1 is based on calendar month (i.e., periodic across years), capturing seasonality.
On the other hand, Group 2 is based on unique time indices, intended to capture residual temporal patterns not explained by Group 1.
For each group, we assume independent Gaussian priors:
\begin{align*}
    \boldsymbol{t}_1 \sim \mathcal{N}(\boldsymbol{0},\boldsymbol{I}_1),\quad
    \boldsymbol{t}_2 \sim \mathcal{N}(\boldsymbol{0},\boldsymbol{I}_2),
\end{align*}
where $\boldsymbol{t}_1$ and $\boldsymbol{t}_2$ represent the effects of unique indices in Group 1 and Group 2, respectively, and $\boldsymbol{I}_1$ and $\boldsymbol{I}_2$ are identity matrices with dimensions matching the length of $\boldsymbol{t}_1$ and $\boldsymbol{t}_2$.

To manage model complexity, we group the council-level BYM2 effect $\boldsymbol{b}_c$ by Group 1 and the district-level BYM2 effect $\boldsymbol{b}_d$ by Group 2.
This yields a council-level spatio-temporal effect $G_c$ and a district-level spatio-temporal effect $G_d$:
\begin{align*}
    G_c \sim \mathcal{N}(\boldsymbol{0}, \boldsymbol{\Sigma}_c\otimes\boldsymbol{I}_1),\quad
    G_d \sim \mathcal{N}(\boldsymbol{0},\boldsymbol{\Sigma}_d\otimes\boldsymbol{I}_2),
\end{align*}
with $\otimes$ denoting the Kronecker product.

We also include a pure temporal effect for the year $T$ to capture annual variation, potentially driven by policy changes following major wildfire events.
$T$ is assigned a Gaussian prior with precision $\tau_T$:
\begin{align*}
    T \sim \mathcal{N}(0,\tau_T^{-1}).
\end{align*}

Bringing all components together, the linear predictors for fire presence $\eta^Z$, fire count $\eta^C$ and burnt area $\eta^B$ are expressed as:
{\footnotesize
\begin{equation}
\begin{alignedat}{4}
&\eta_{s,t+1}^Z &&= \beta_0^Z
  + \phantom{\beta_{1}^C}\, &&G_c(s,t+1;\tau_{G_c},\phi_{G_c})
  + \phantom{\beta_{2}^C}\, &&G_d(s,t+1;\tau_{G_d},\phi_{G_d})
  + T(t+1;\tau_{T}^{Z})
  + R(\widehat{y}_{s,t+1}^C;\tau_{R}^{Z^C})
  + R(\widehat{y}_{s,t+1}^B;\tau_{R}^{Z^B}) ,\\
&\eta_{s,t+1}^C &&= \beta_0^C
  + \beta_{1}^C &&G_c(s,t+1;\tau_{G_c},\phi_{G_c})
  + \beta_{2}^C &&G_d(s,t+1;\tau_{G_d},\phi_{G_d})
  + T(t+1;\tau_{T}^{C})
  + R(\widehat{y}_{s,t+1}^C; \tau_{R}^{C}),\\
&\eta_{s,t+1}^B &&= \beta_0^B
  + \beta_{1}^B &&G_c(s,t+1;\tau_{G_c},\phi_{G_c})
  + \beta_{2}^B &&G_d(s,t+1;\tau_{G_d},\phi_{G_d})
  + T(t+1;\tau_{T}^{B})
  + R(\widehat{y}_{s,t+1}^B;\tau_{R}^{B}).
\end{alignedat}
\label{eq: linear_predictor}
\end{equation}
}
Here, $\beta_0^{(\cdot)}$ are intercepts, and  $\beta_1^{(\cdot)},\beta_2^{(\cdot)}$ are scaling parameters controlling the contribution of the shared spatio-temporal effects to each predictor.
For parameters and effects that involve subscripts and superscripts, subscripts indicate the associated effect, while superscripts specify the predictor.
Note that we include both predicted fire count and burnt area in $\eta^Z$ since both positive fire count and burnt area indicate a fire presence.
In addition, the spatio-temporal effects $G_c$ and $G_d$ are shared across all three linear predictors. 
Sharing allows the spatio-temporal effects for fire count and burnt area to be informed by the full dataset rather than only the subset with fire occurrence.
This reduces the risk of overparameterisation and mitigates uncertainty arising from the sparse nature of fire events.

\subsubsection{Priors}
We now elaborate on the priors used for the eGP likelihood hyperparameters ($\kappa$ and $\xi$) as well as those associated with the linear predictor.
For $\kappa$ and $\xi$, we adopt Penalised Complexity (PC) priors following the framework of \citet{simpson2017penalising}, which provide a principled mechanism to control model complexity by penalising deviations from a simpler base model.
This deviation is measured via the Kullback–Leibler divergence (KLD; \citealt{kullback1951information}), and the corresponding distance is defined as $d = \sqrt{2\text{KLD}}$.
An exponential prior is then placed on this distance, yielding a memoryless penalty on increasing model complexity.
The PC prior for each parameter is obtained by transforming this exponential prior back to the original parameter scale.

In the special case of the standard GDP ($\kappa=1$), \citet{opitz2018inla} derived a PC prior for $\xi>0$, using the base model with $\xi=0$. 
They computed the KLD between a GPD density $f_{\xi}(y)=f_{\text{GPD}}(y;\xi)$ and the base model $f_{\xi_0}(y)=f_{\text{GPD}}(y;\xi = 0)$ as:
\begin{align}
    \text{KLD}\{f_{\xi} \| f_{\xi_0}\}= \frac{\xi^2}{1-\xi}, \quad{0\leq\xi<1.}
    \label{eq: KLD}
\end{align}
Two options were then considered for deriving the PC prior for $\xi$: (1) using the exact expression in \eqref{eq: KLD}, or (2) using the approximation $\text{KLD}\{f_{\xi} \| f_{\xi_0}\} \approx \xi^2$ as $\xi \rightarrow 0$. 
These yield the following prior formulations:
\begin{align*}
    &\text{Option 1: }\quad \pi_1(\xi)=\lambda \exp \left\{- \frac{\lambda\xi}{(1-\xi)^{1 / 2}}\right\}\left\{\frac{1-\xi / 2}{(1-\xi)^{3 / 2}}\right\},& \quad 0\leq\xi<1,\\
    &\text{Option 2: }\quad \pi_2(\xi)=\lambda \exp \left\{- \lambda\xi\right\}, &\quad 0\leq\xi<1,
\end{align*}
where $\lambda$ is the rate parameter of the exponential distribution, controlling the strength of penalisation. 
The two priors are nearly indistinguishable for large $\lambda$ (e.g., $\lambda > 3$), but diverge notably for smaller values, as illustrated in Figure 3 of \citet{opitz2018inla}.

For the eGP setting, we relax the constraint $\xi > 0$ to allow for negative values, which may be relevant in practice.
The eGP density exhibits a Pareto-type tail whose heaviness is governed by the shape parameter $\xi$, independent of $\kappa$.
For analytical tractability in deriving the $\text{KLD}$, we fix $\kappa=1$, which leads to a form of KLD same as that in \eqref{eq: KLD} except for the support.
Given the limited prior knowledge on the sign of $\xi$, we adopt a symmetric PC prior centred at zero.
Using a second-order expansion of  \eqref{eq: KLD} around $\xi = 0$, we obtain:
\begin{align}
    \pi(\xi)=\frac{\lambda \exp \left\{- \lambda|\xi|\right\}}{\int_{\xi_L}^{\xi_U} \lambda \exp\left\{ -\lambda |x| \right\}  \mathrm{d}x}, &\quad  \xi_L<\xi<\xi_U,
    \label{eq: pc_prior_xi}
\end{align}
where $\xi_L$ and $\xi_U$ are lower and upper bounds for $\xi$ selected to enforce desirable properties such as finite moments.
The resulting prior is symmetric around $\xi = 0$, matches $\pi_2(\xi)$ on the positive half-line, and incorporates truncations to preserve key theoretical properties of the eGP. 
In the R-INLA implementation, we use $(\xi_L, \xi_U) = (-0.5, 0.5)$, ensuring a finite mean and variance and desirable asymptotic properties for the maximum likelihood estimator of $\xi$.
% The prior $\pi(\xi)$ is symmetric for $|\xi| < \min(|\xi_L|, |\xi_U|)$ and coincides with $\pi_2(\xi)$ for $\xi > 0$. Meanwhile, a boundary constraint is imposed on $\pi(\xi)$ to ensure the eGP retains finite moments and other appealing properties. 
% The default values in the R-INLA package are $(\xi_L, \xi_U) = (-0.5, 0.5)$ to ensure a finite mean, variance of eGP and asymptotic properties of $\xi$.

% Unfortunately, a closed-form PC prior does not exist for $\kappa$. 
% In defining a PC prior for $\kappa$, we use the base model $\pi(\kappa = 1)$, corresponding to a generalised Pareto distribution.
% However, the KLD $\text{KLD}{\pi(\kappa) | \pi(\kappa = 1)}$ lacks a closed-form expression in this case. 
% Therefore, we adopt a log-Gamma prior: $\pi(\kappa) \sim \text{log-Gamma}(100, 100)$, which reflects a strong prior belief that $\kappa = 1$.

{The natural base model for constructing a PC prior for the parameter $\kappa$ in the eGP distribution is $\kappa = 1$, which corresponds to the standard GP distribution.
The KLD between $f_\kappa(y)=f_{\text{eGP}}(y;\kappa)$ and $f_{\kappa_1}(y)=f_{\text{eGP}}(y;\kappa=1)$ is given by
\begin{equation}
        \text{KLD}\{f_\kappa \| f_{\kappa_1}\}= \log \kappa - \frac{\kappa-1}{\kappa}, \quad{\kappa>0,}
        \label{eq: exact_kld_kappa}
\end{equation}
with derivation provided in the Supplementary Materials \ref{sec:appendix_pckappa}.
Following the approach for deriving a PC prior for the parameter $\xi$, we may either use the exact KLD in \eqref{eq: exact_kld_kappa}, or apply a second-order Taylor expansion around $\kappa=1$, which gives
\begin{equation}
            \text{KLD}\{f_\kappa \| f_{\kappa_1}\} \approx \frac{1}{2}({\kappa}-1)^2, \quad{\kappa>0.}
        \label{eq: approximate_kld_kappa}
\end{equation}
% to construct the PC prior for $\kappa$. 
Given a penalisation rate $\lambda$, the PC prior based on the exact KLD in \eqref{eq: exact_kld_kappa} takes the form
\begin{equation}
    \pi_1(\kappa) = \begin{cases}
        \frac{\lambda |\kappa-1|}{2\kappa^2\sqrt{2\log \kappa -2(\kappa-1)/\kappa}}\exp\left\{-\lambda \left( \sqrt{2 \log \kappa -{2(\kappa-1)}/{\kappa}} \right) \right\}, \qquad &\kappa>0, \;\kappa \neq1,\\
        \lambda/2, & \kappa=1,
    \end{cases}
    \label{eq: pc_prior_kappa_1}
\end{equation}
whereas the PC prior based on the approximated KLD in \eqref{eq: approximate_kld_kappa} is given by 
\begin{equation*}
    \pi_2(\kappa)\ = \frac{\lambda\exp(-\lambda|\kappa-1|)}{2-\exp(-\lambda)}, \qquad\kappa>0.
\end{equation*}
Figure \ref{fig: PC_priors_Kappa} displays  $\pi_1(\kappa)$ and $\pi_2(\kappa)$ under various values of $\lambda$.
When $\lambda$ is large (e.g. $\lambda>5$), both priors concentrates around $\kappa=1$, exhibiting similar behaviour.
As $\lambda$ decreases, the mode of $\pi_1(\kappa)$ shifts leftwards towards zero and diverge as $\kappa \rightarrow 0$, whereas $\pi_2(\kappa)$ remains locally symmetric around $\kappa=1$.
This suggests that $\pi_1(\kappa)$ always shrinks $\kappa$ towards a value in $(0,1]$, and it is less suitable for expressing weakly informative priors over $\kappa >1$ compared to $\pi_2(\kappa)$. 
To determine the more appropriate prior between $\pi_1(\kappa)$ and $\pi_2(\kappa)$, we consider the interpretation and functional role of $\kappa$.
According to \cite{naveau2016modeling}, $\kappa$ governs the lower-tail behaviour of the cumulative distribution function of an eGP random variable $Y$ via 
\begin{equation*}
    \mathbb{P}(Y<y) \approx \text{constant} \times y^{\kappa}, \qquad \text{as } y \rightarrow 0^+ .
\end{equation*}
Consequently, the corresponding density function $f_{\text{eGP}}(y)$ satisfies
\begin{equation*}
    f_{\text{eGP}}(y) \propto \kappa y^{\kappa-1},  \qquad \text{as } y \rightarrow 0^+.
\end{equation*}
This characterisation implies that $\kappa$ plays a role analogous to the shape parameter in the Gamma or Beta distribution.
Specifically, when $\kappa>1$, $f_{\text{eGP}}$ increases from zero to a mode, while for $0<\kappa<1$, the density exhibits a singularity at zero, sharply peaking near the origin. 
In environmental applications, data may be zero-inflated; however, the positive component is more frequently well modelled by an eGP distribution with $\kappa>1$ than with $0<\kappa<1$, as illustrated in Figure \ref{fig: exploratory_plot}.
From this perspective,  we seek a prior $\pi(\kappa)$ that does not favour the region $0<\kappa<1$, irrespective of $\lambda$.
Accordingly, we adopt $\pi(\kappa)=\pi_2(\kappa)$ for fitting the eGP model.
\begin{figure}[ht]
\begin{center}
\includegraphics[width=0.45\textwidth]{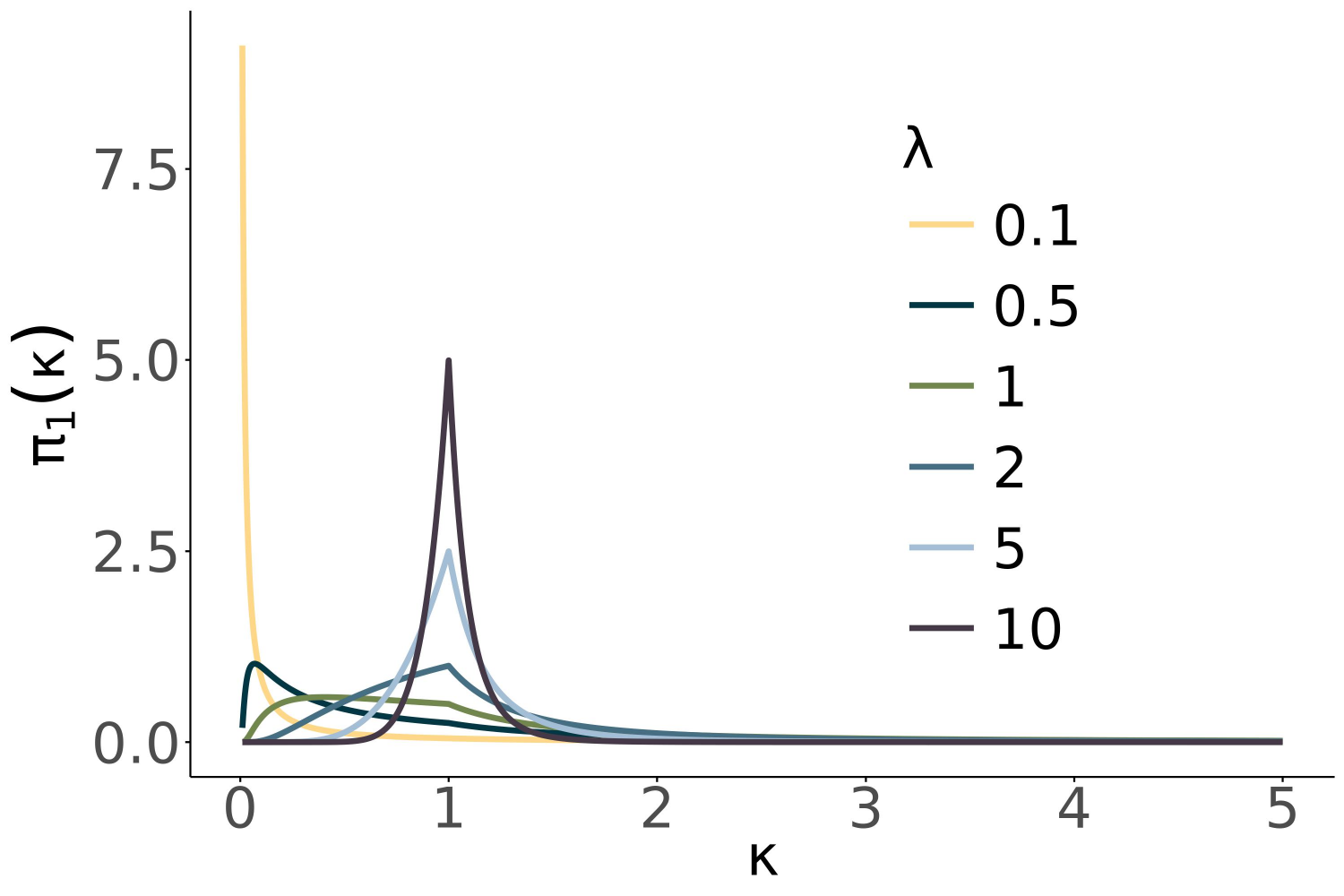}
\includegraphics[width=0.45\textwidth]{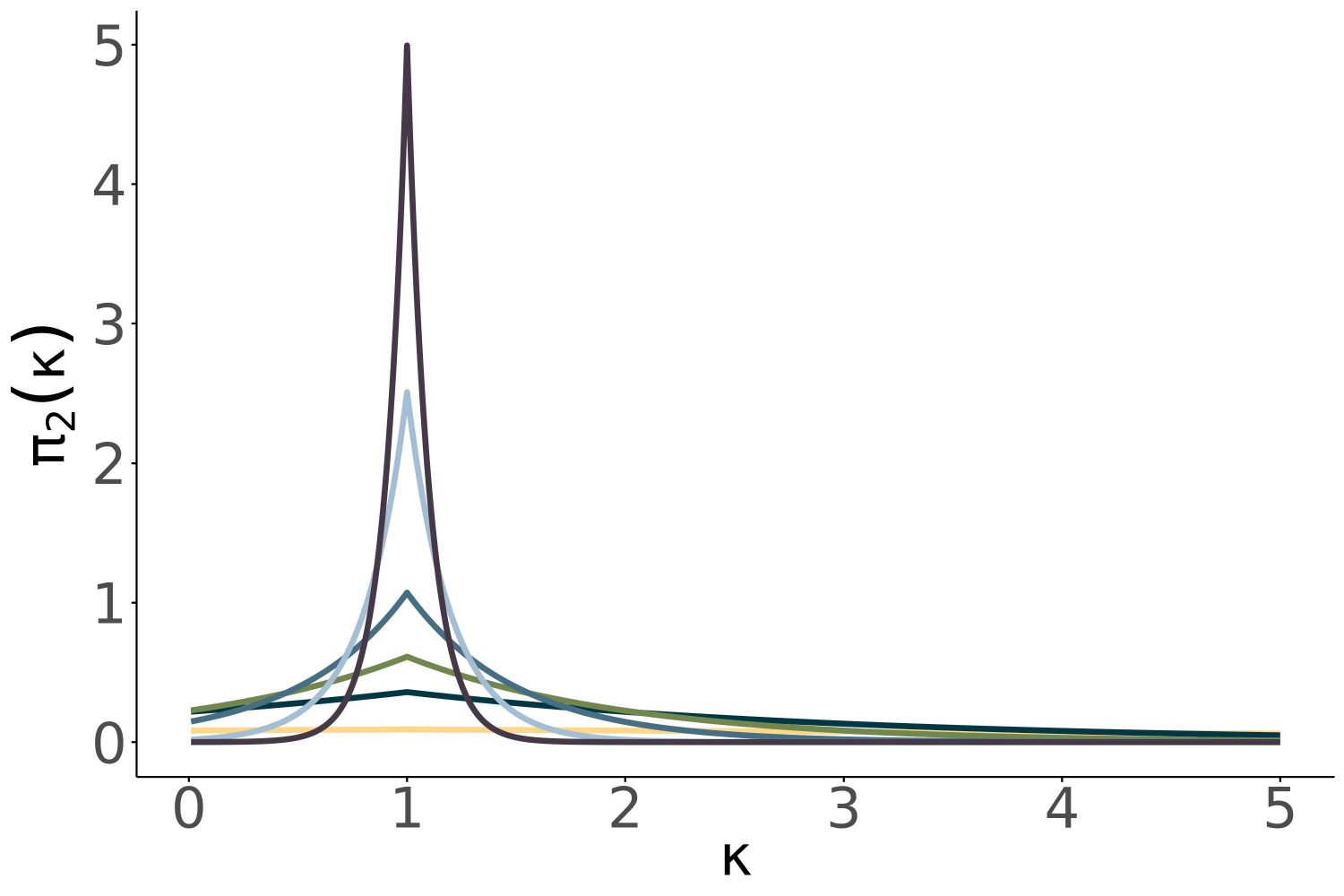}\\
\end{center}
\caption{\footnotesize{PC priors for $\kappa$ based on exact KLD in \eqref{eq: exact_kld_kappa} (left) and the approximated KLD around $\kappa=1$ \eqref{eq: approximate_kld_kappa} (right) under penalisation rates $\lambda\in\{0.1,0.5,1,2,5,10\}$.}}
\label{fig: PC_priors_Kappa}
\end{figure}
}

The priors for $\xi$, $\kappa$, and the remaining hyperparameters in the linear predictor in \eqref{eq: linear_predictor} are specified as follows:
\begin{enumerate}
    \item $\xi$ and $\kappa$ have priors in \eqref{eq: pc_prior_xi} and \eqref{eq: pc_prior_kappa_1}, respectively. In both cases, the penalisation rate parameter takes $\lambda=10$.
    \item Intercepts $\beta_0^{(\cdot)}$ are assigned weakly informative priors $\mathcal{N}(0, 1000)$;
    \item Scaling parameters $\beta_1^{(\cdot)}$ are given more informative priors $\mathcal{N}(0, 0.1)$;
    \item Precision parameters in $T(\cdot)$ and $R(\cdot)$ follow priors $\text{Gamma}(0.1, 0.1)$;
    \item PC priors are used for parameters in $G_c(\cdot)$ and $G_d(\cdot)$, with constraints $\mathbb{P}(1/\sqrt{\tau} > 1) = 0.01$ and $\mathbb{P}(\phi < 0.5) = 0.5$ to control the marginal standard deviation and spatial range, respectively.
\end{enumerate}
Here, we adopt Gamma priors rather than PC priors for the parameters in $T(\cdot)$ and $R(\cdot)$ because numerical instabilities were observed during model fitting across a wide range of PC prior specifications, from weakly informative to more informative choices. These instabilities were substantially mitigated after switching to Gamma priors.

\section{Results}
\label{sec: results}
\subsection{Model Comparison}
To evaluate the effectiveness of incorporating XGBoost-based wildfire forecasts into our modelling framework and to assess the appropriateness of the eGP likelihood for wildfire modelling, we compare several variations of the latent Gaussian model differing in their likelihood choices and linear predictor components.
Let M1 denote the full two-stage model introduced in Section \ref{sec: method}. 
To examine the role of the eGP likelihood, we substitute it with two alternative distributions commonly used in environmental applications: the Gamma and Weibull likelihoods, yielding models M2 and M3, respectively.
{Additionally, we construct M4 to emulate a setting in which future environmental covariates are unavailable. Specifically, the XGBoost-derived random effects are replaced by lag-1 values of the FWI and temperature. 
We retain FWI because it is one of the most widely used fire danger indicators, and temperature due to its high feature importance, as shown in Figure~\ref{fig: SHAP_values}.
All remaining random effects are kept identical to those in M1.}

We use wildfire data from 2011 to 2022 as the training set and data from 2023 as the test set. 
One-month-ahead forecasts of fire count and burnt area are generated as described in Section \ref{sec: cv_score}. 
The posterior predictive distributions of the fire presence $Z_{s,t+1}$, fire count $Y_{s,t+1}^C$ and square-root-transformed burnt area $\sqrt{Y_{s,t+1}^B}$ are obtained from 1000 posterior simulations.
To evaluate performance, we use Area Under Curve (AUC) to assess the predictive accuracy for fire presence, where the posterior mean of $\widehat{Z}_{s,t+1}$ serves as the estimate of $\mathbb{P}(Z_{s,t+1}=1)$.
The positive burnt area $\sqrt{\widehat{Y}_{s,t+1}^B}\mid Z_{s,t+1}=1$ is evaluated by continuous ranked probability score (CRPS), which is a proper scoring rule that measures the difference between a predictive distribution $F$ and a single observation $y$ by 
\begin{equation*}
\text{CRPS}(F,y) = \int_\mathbb{R}\left[F(v)-\mathbbm{1}(v\geq y)\right]^2 \mathrm{d}v,
\end{equation*}
where $\mathbbm{1}$ is the indicator function. 
CRPS is computed for each location–time pair and averaged over all instances.

Since forecasts are typically communicated as probabilities over categorised bins, with particular emphasis on large events, we also compute weighted binned scores following the structure proposed by \cite{opitz2023eva}.
Specifically, for a fire count threshold vector $\boldsymbol{h}^C=(0, 1,2,\ldots,10,15,20,25,30)$, the weighted scoring for fire count $r^C$ is the weighted sum of squared residuals between predicted and empirical probabilities across all observations and thresholds:
\begin{align*}
r^C = \sum_{s,t+1}\sum_{h \in \boldsymbol{h}^C} w^C(h)\left[\widehat{\mathbb{P}}(Y_{s,t+1}^C \leq h)- \mathbbm{1}(Y_{s,t+1}^C\leq h) \right]^2,
\end{align*}
where the normalised weight is given by $w^C(h) = \widetilde{w}^C(h)/\widetilde{w}^C(30)$, and the unnormalised weights are defined as $$ \widetilde{w}^C(h) = 1-(1+(h+1)^2/1000)^{-1/4},$$ which increases approximately linearly with $h$.
Here, $\mathbb{P}(Y_{s,t+1}^C \leq h)$ is the unconditional probability of $Y_{s,t+1}^C$ obtained by integrating out $Z_{s,t+1}$ from the conditional distribution $Y_{s,t+1}^C\mid Z_{s,t+1}=1$.

Similarly, for burnt area  (on the original scale),  the thresholds are defined as $\boldsymbol{h}^B = (0,20,40, \\60,80,100,200,300, 400, 500, 1000, 2000,5000,10000,20000,50000)$.
The corresponding weighted binned score $r^B$ is given by
\begin{align*}
r^B = \sum_{s,t+1}\sum_{h \in \boldsymbol{h}^B} w^B(h)\left[\widehat{\mathbb{P}}(Y_{s,t+1}^B \leq h)- \mathbbm{1}(Y_{s,t+1}^B\leq h) \right]^2,
\end{align*}
where the normalised weights are defined as $w^B(h) = \widetilde{w}^B(h)/\widetilde{w}^B(50000)$, $ \widetilde{w}^B(h) = 1-(1+(h+1)/1000)^{-1/4}$.
Similar to $w^C(h)$, $w^B(h)$ increases approximately linearly with the threshold $h$, placing more emphasis on larger fire events.
% \begin{table}
% \centering
% \caption{\footnotesize{Comparison of posterior predictive performance across model variants. 
% AUC evaluates fire presence predictions (higher is better). Lower values indicate better performance for CRPS and binned scores.
% All results are based on 1,000 posterior predictive samples. Bold values highlight relatively better performance.
% }}
% \vspace{0.3cm}
% \begin{tabular}{ccccc}
% \hline
% Metric & M1 (eGP) & M2 (Gamma) & M3 (Weibull) & M4 (eGP, no XGB) \\
% \hline
% AUC & \textbf{0.884} & 0.881 & \textbf{0.884} & 0.791 \\
% CRPS & 4.70 & 4.73 & \textbf{4.67} & 4.92 \\
% Unweighted $r^C$ & 347 & 346 & \textbf{345} & 371 \\
% Weighted $r^C$ & 26.9 & \textbf{25.9} & 26.1 & 27.0 \\
% Unweighted $r^B$ & \textbf{541} & 550 & 542 & 584 \\
% Weighted $r^B$ & 34.3 & 34.5 & \textbf{33.6} & 34.8 \\
% \hline
% \end{tabular}
% \label{tab: model_comparison}
% \end{table}

\begin{table}
\centering
\caption{\footnotesize{Comparison of posterior predictive performance across model variants on the test set. 
AUC evaluates fire presence predictions (higher is better). Lower values indicate better performance for CRPS and binned scores.
All results are based on 1,000 posterior predictive samples. Bold values highlight relatively better performance.
}}
\vspace{0.3cm}
\begin{tabular}{@{}ccccc@{}}
\toprule
Metric           & M1 (eGP)       & M2 (Gamma) & M3 (Weibull)  & M4 (eGP, lagged covariates) \\ \midrule
AUC              & \textbf{0.862} & 0.857      & 0.860         & 0.815            \\
CRPS             & \textbf{4.75}           & 4.83       & {4.78} & 4.89             \\
Unweighted $r^C$ & \textbf{362}            & 371        & {366}  & 382              \\
Weighted $r^C$   & \textbf{4.84}           & 4.93     & {4.88} & 5.14             \\
Unweighted $r^B$ & \textbf{565}   & 578        & 577           & 594              \\
Weighted $r^B$   & \textbf{14.2}           & 14.5       & {14.6} & 14.7             \\ \bottomrule
\end{tabular}
\label{tab: model_comparison}
\end{table}
Table \ref{tab: model_comparison} summarises the model performance on the test set in 2023 across six evaluation metrics. 
% Among models M1, M2, and M3, which differ only in the likelihood, no single likelihood dominates across all metrics.
% Model M1, using the eGP likelihood, achieves the highest AUC (0.883) and the lowest unweighted burnt area score (549).
% It is better than M2 (Gamma likelihood) for burnt area modelling, since the right tail of eGP has been adapted to allow for a heavy tail.
% However, its fire count performance lags slightly behind M2.
% Model M3 (Weibull likelihood), which can also accommodate heavy-tailed behaviour depending on its shape parameter, achieves the best CRPS, unweighted and weighted fire count score, and weighted burnt area score.
% M3 attains the best performance in four out of six metrics, whereas M1 leads in two. 
{Among models M1, M2, and M3, which differ only in the likelihood, the eGP likelihood achieves slightly better performance across all metrics.
M1 and M3 perform comparably, as both eGP and Weibull likelihoods can accommodate heavy-tailed data through their shape parameter.
In contrast, M2 employs a light-tailed distribution (Gamma) for burnt area and performs worst on five of the six metrics.
Notice that although the likelihood is modified only for burnt area, changes also propagate to metrics such as AUC and binned scores for counts, owing to the shared spatio-temporal effects in the joint modelling of fire presence, fire count, and burnt area.}
Nonetheless, performance differences between M1, M2, and M3 are marginal, and the reasons for this will be further discussed in Section \ref{sec: discussion_similar_perf}.
For consistency, we proceed with eGP (M1) in the remainder of the paper.

{
Model M4, which excludes XGBoost-derived covariates, relies only on lagged temperature and FWI covariates together with spatio-temporal effects in the latent Gaussian model to produce one-month-ahead forecasts. 
A substantial deterioration in performance is observed across all six evaluation metrics, with particularly pronounced declines in metrics that weight all events equally, such as AUC, CRPS, and the unweighted correlations $r^C$ and $r^B$. 
These results underscore the importance of incorporating dynamic, forecast-driven covariates that capture complex spatio-temporal structure for accurate wildfire forecasting.
}

\subsection{Posterior Predictions}
\begin{figure}[ht]
\begin{center}
\includegraphics[width=0.45\textwidth]{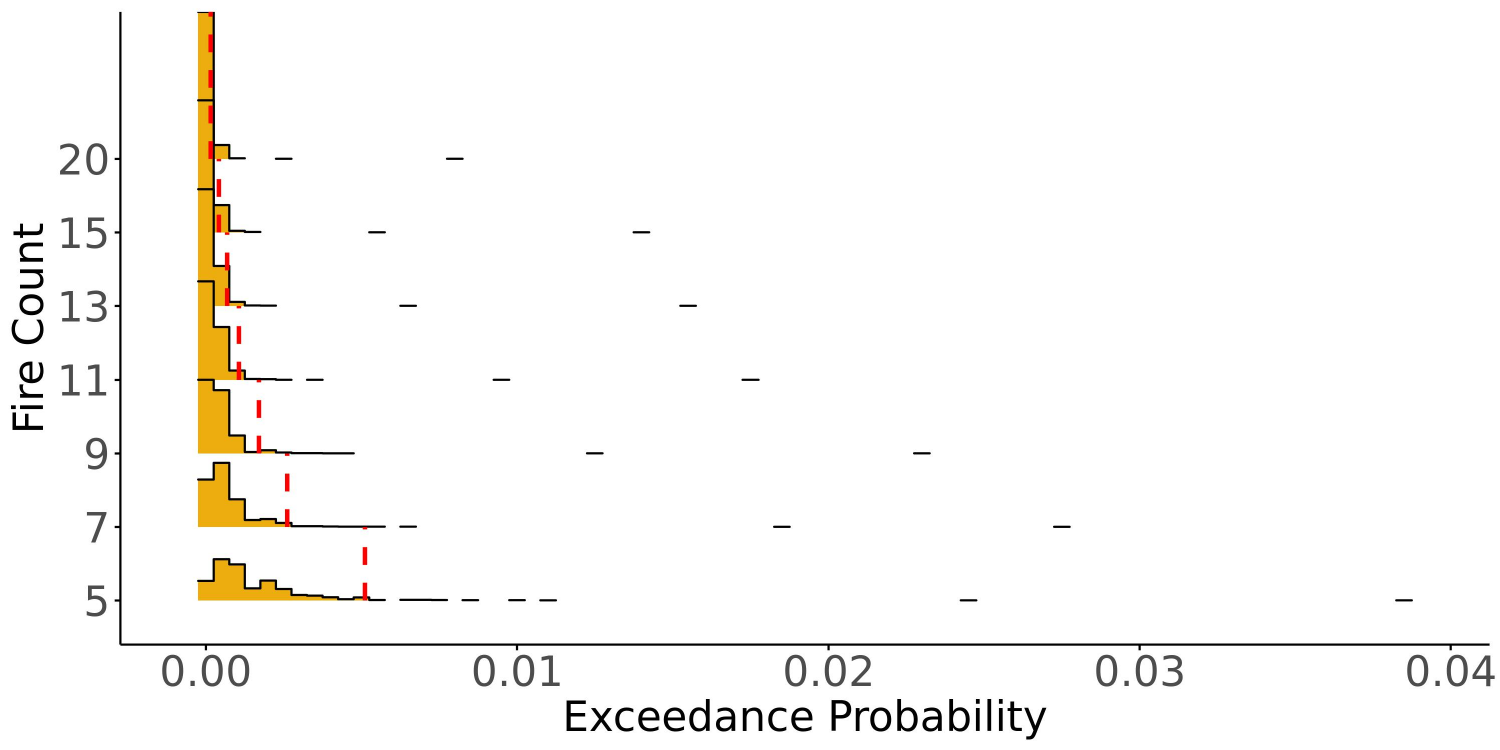}
\includegraphics[width=0.45\textwidth]{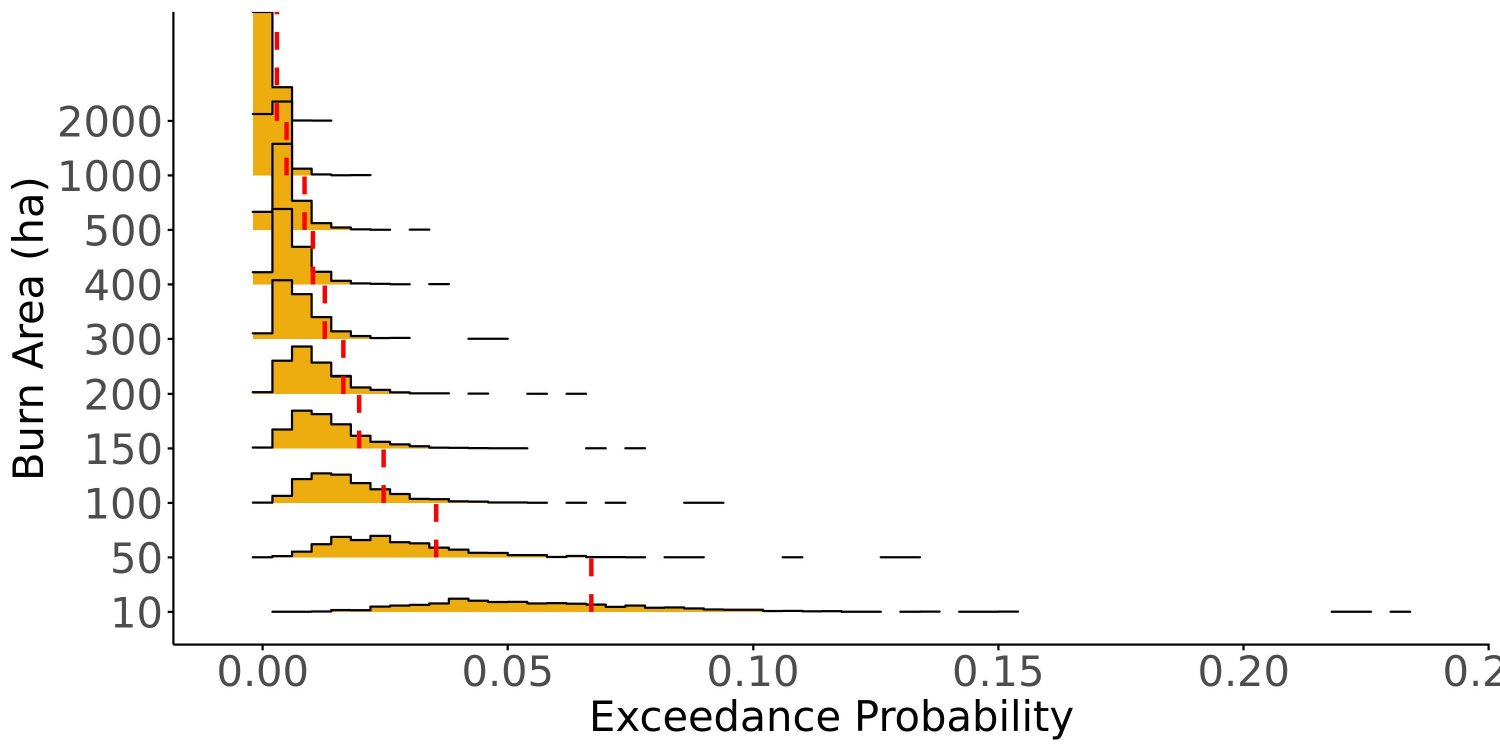}\\
\vspace{0.5cm}
\includegraphics[width=0.45\textwidth]{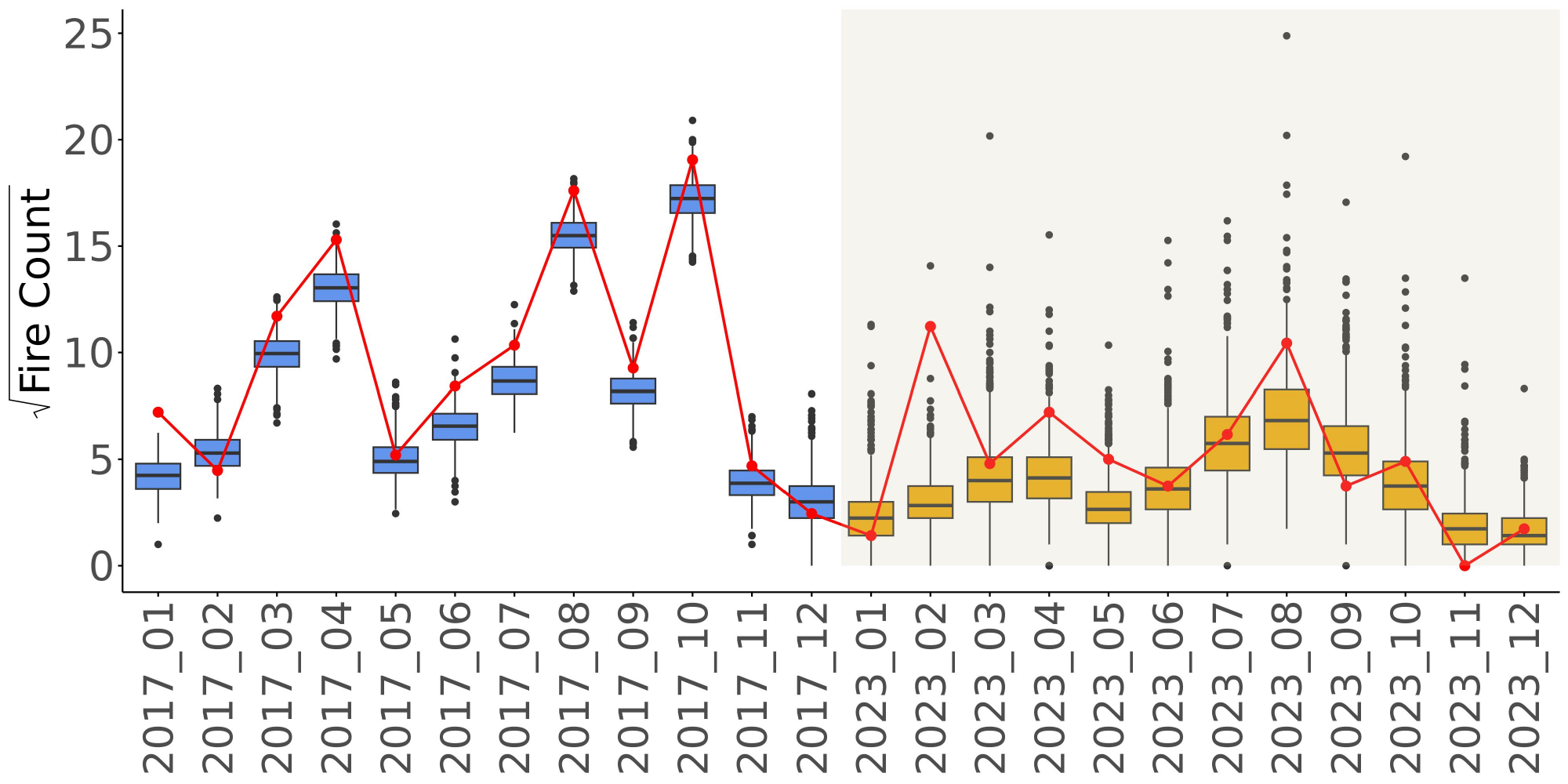}
\includegraphics[width=0.45\textwidth]{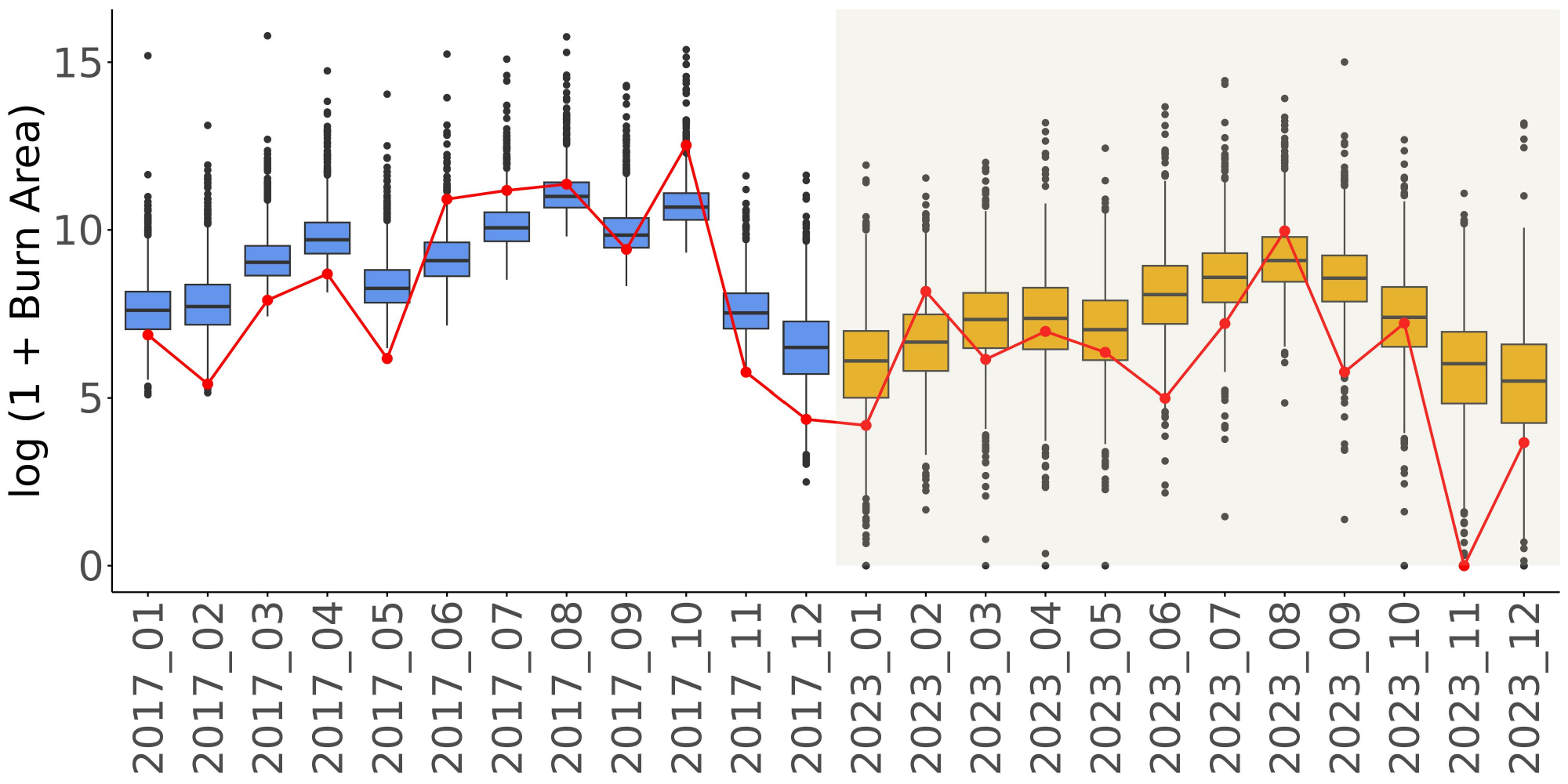}
\end{center}
\caption{\footnotesize{Posterior predictive checks of threshold exceedance probabilities for fire count and burnt area in the test set (top two panels), and posterior predictive distributions of total fire count and burnt area in Portugal for 2017 (training set) and 2023 (test set) (bottom two panels). 
In the top panels, red dashed lines indicate the empirical exceedance probabilities. 
In the bottom panels, red points denote the observed total values.
}}
\label{fig: posterior_prediction}
\end{figure}

Figure \ref{fig: posterior_prediction} presents a detailed view of the posterior predictive distributions, $\pi(\widehat{Y}^C_{s,t+1})$ and $\pi(\widehat{Y}^B_{s,t+1})$, obtained from the second stage model.
To assess predictive performance, we conduct posterior predictive checks based on the threshold exceedance probabilities of fire count and burnt area within the test set. 
The empirical exceedance probabilities at a given threshold $h$ are defined as
\begin{equation*}
\widehat{P}(Y^C > h) = \frac{1}{|\mathcal{I}|} \sum_{(s,t+1) \in \mathcal{I}} \mathbbm{1}(Y^C_{s,t+1} > h), \quad
\widehat{P}(Y^B  > h) = \frac{1}{|\mathcal{I}|} \sum_{(s,t+1) \in \mathcal{I}} \mathbbm{1}(Y^B_{s,t+1} > h),
\end{equation*}
where $\mathcal{I}$ denotes the set of all spatial and temporal indices in the test set, and $|\mathcal{I}|$ is its cardinality.
These quantities represent the overall proportions of exceedances in the test set, aggregated over space and time, and are not conditioned on specific locations or time points.

The upper panels of Figure \ref{fig: posterior_prediction} show the posterior predictive check of the exceedance probabilities over various thresholds based on 1000 posterior predictive replicates. 
% These illustrate the uncertainty in predicting whether fire count and burnt area exceed the given thresholds. 
Uncertainty is notably higher at lower thresholds (e.g., 5 fires or 10 hectares) and diminishes as the threshold increases.
{When compared with empirical exceedance probabilities, the model performs well for burnt area: the observed values (red dashed lines) consistently fall within the 50th–90th percentiles of the posterior predictive distribution.
For fire counts, some deviations occur at lower thresholds, but the observed statistics do not systematically fall outside the predictive envelopes.}
% Although the empirical exceedance probabilities (red dashed lines) may deviate from the centres of the predictive distributions, they consistently fall within the 50th to 90th percentiles, suggesting satisfactory model calibration.

The lower panels of Figure \ref{fig: posterior_prediction} display the posterior predictive distributions for the total fire count and burnt area across Portugal at selected time points.
We focus on one year from the training set (2017), during which several severe wildfires occurred, and a full year from the test set (2023), to evaluate the model’s ability to capture temporal dynamics.
Each box plot is generated from 1000 posterior samples, with aggregated totals over the entire Portugal and specified time periods.

Overall, the results demonstrate that the model effectively captures the temporal evolution trend of wildfire activity.
For most time points, the observed fire counts and burnt areas lie within 1.5 times the interquartile range (IQR) from the first and third quartiles of the posterior predictive distributions.
Notably, the model yields accurate predictions for October 2017, when Portugal experienced an exceptionally intense wildfire episode, with over 350 reported fires and a burnt area exceeding 250,000 hectares.

\subsection{Covariates and latent effects}
\subsubsection{XGBoost model interpretation}
The XGBoost model provides point forecasts of wildfire activity, while the latent Gaussian model primarily quantifies associated uncertainty. 
It is therefore essential to understand which covariates most strongly influence the predictions generated by XGBoost. 
To this end, we assess covariate importance using SHapley Additive exPlanations (SHAP) values \citep{lundberg2017unified}.

SHAP values offer a principled approach to attributing the marginal contribution of each covariate to the model output, drawing on the concept of Shapley values from cooperative game theory \citep{shapley:book1952}.
For a model $f$ fitted on a covariate set $M=\{\widetilde{x}_1,\widetilde{x}_2,\cdots,\widetilde{x}_m\}$, and a subset $S \subseteq M$, the SHAP value $\phi_j$ of covariate $\widetilde{x}_j$ is the average difference between the predictions $f(S \cup \{\widetilde{x}_j\})$ and $f(S)$ over all possible $S$.
Formally, $\phi_j$ is defined by
\begin{align*}
    \phi_j = \sum_{S\subseteq M \backslash \{\widetilde{x}_j\}}\frac{|S|!(|M| - |S|-1)!}{|M|!}\left[f(S \cup \{\widetilde{x}_j\}) - f(S)\right].
\end{align*}
Since $f$ typically requires the full covariate set $M$,
the output for a reduced set $S$ is approximated as $f(S) = \mathbb{E}[f(M)\mid S]$.
For tree-based models such as XGBoost, this conditional expectation can be efficiently computed using the algorithm proposed by \citet{lundberg2018consistent}.

A key property of SHAP values is that they enable an additive decomposition of the model prediction:
\begin{align}
 f({M}) = \mathbb{E}(f(M)) + \sum_{j=1}^m\phi_j.
 \label{eq: SHAP}
\end{align}
\cite{lundberg2017unified} showed that \eqref{eq: SHAP} is the unique additive representation that satisfies local accuracy, missingness, and consistency. 
This decomposition provides an intuitive and theoretically grounded measure of covariate influence, based on both the sign and magnitude of the SHAP values.

\begin{figure}[ht]
\begin{center}
\includegraphics[width=0.49\textwidth]{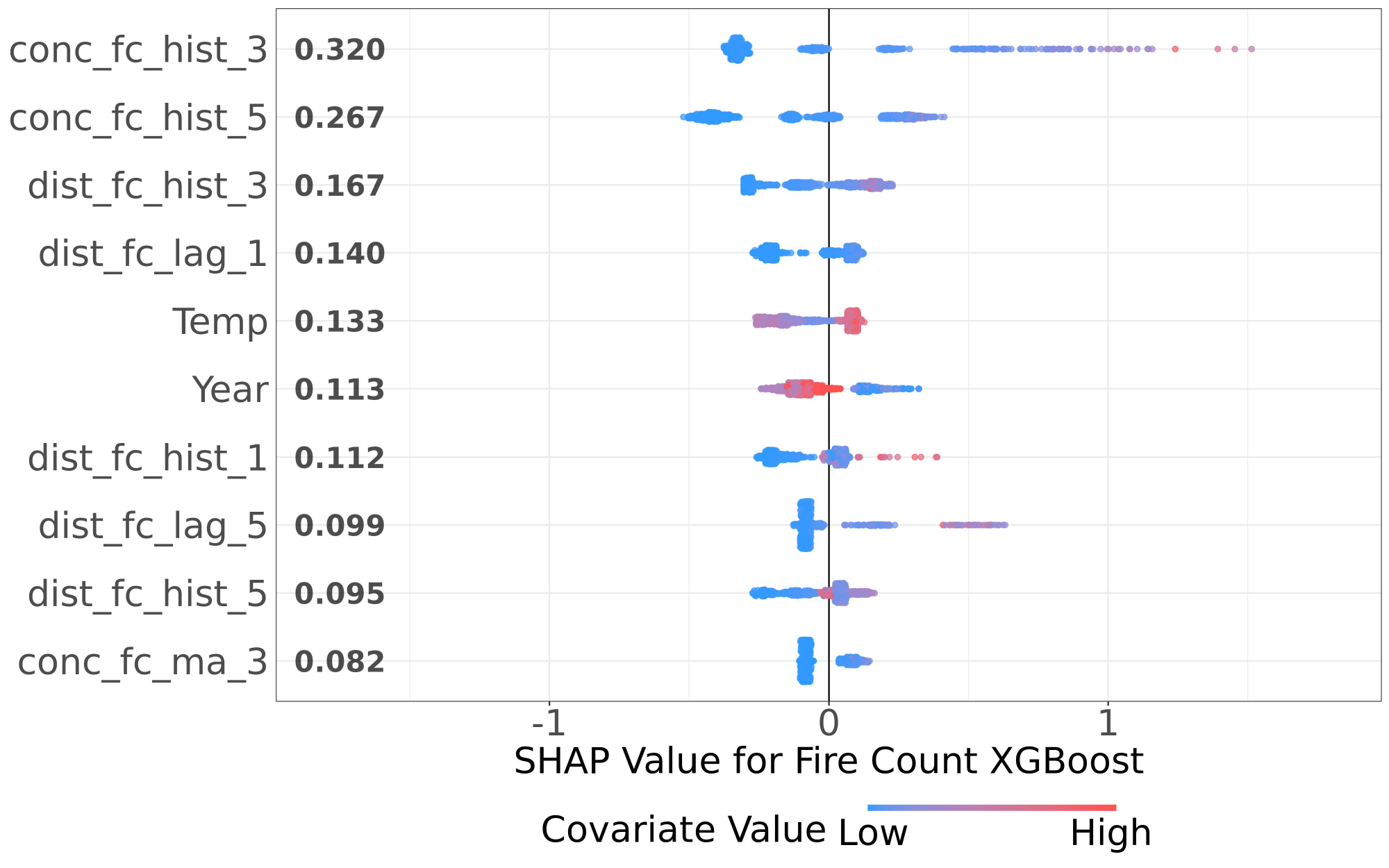}
\includegraphics[width=0.49\textwidth]{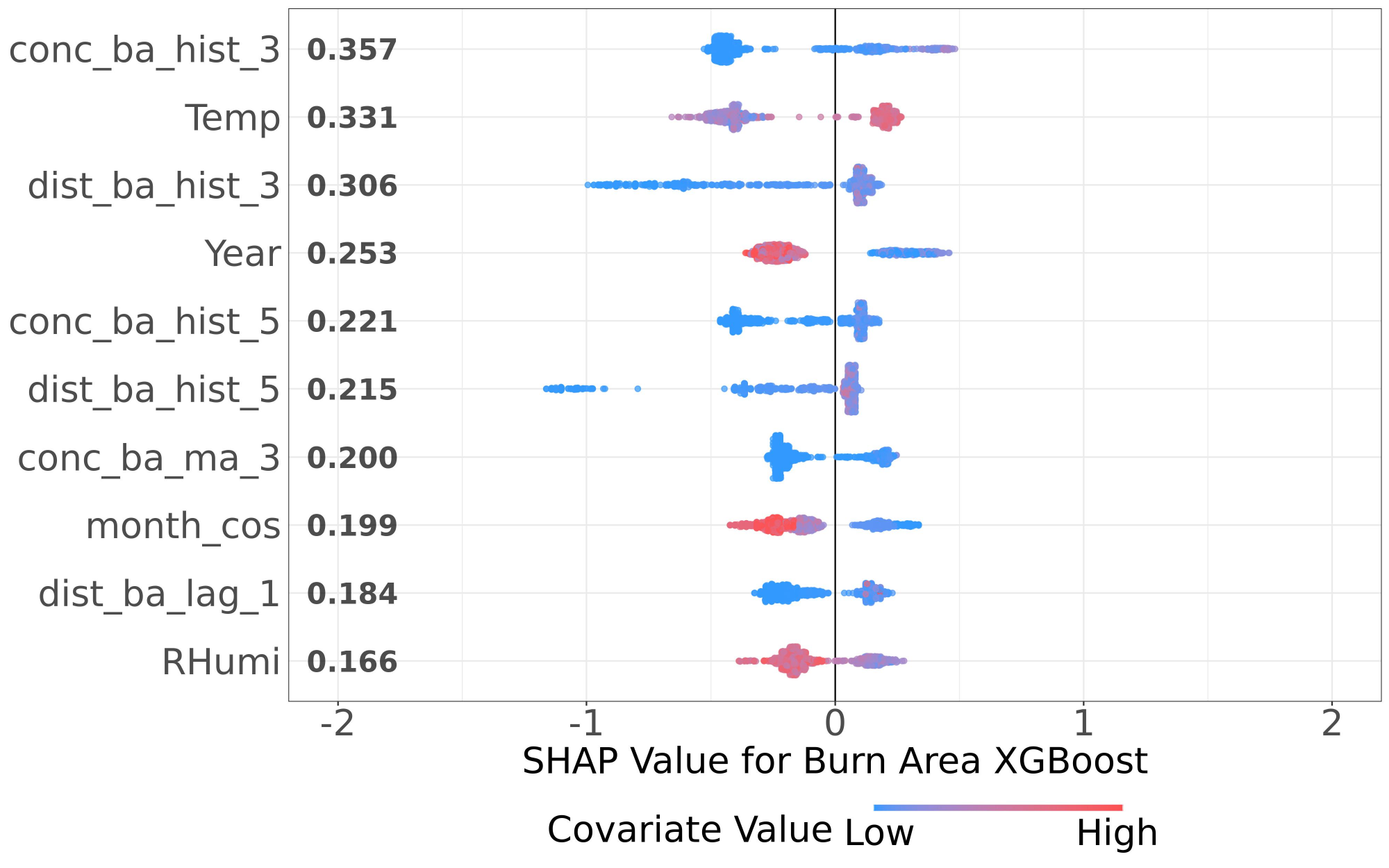}\\
\end{center}
\caption{\footnotesize{SHAP values for the top 10 covariates in the XGBoost models for fire count and burnt area. 
Covariates are ranked by the mean absolute SHAP value (numbers next to the covariate names) across all predictions. Full descriptions of the covariates are provided in Table \ref{tab:covariates_appendix} and \ref{tab: autoregressive_covariates}.}}
\label{fig: SHAP_values}
\end{figure}

Figure \ref{fig: SHAP_values} displays the SHAP values for the ten most influential covariates in the XGBoost models for fire count \eqref{eq: xgb_cnt} and burnt area \eqref{eq: xgb_ba}. 
In both models, autoregressive terms dominate the set of top covariates, suggesting that historical wildfire activity contributes more to predictive accuracy than the environmental variables.
The most influential covariates are the averages of fire count and burnt area of the three months centred at the forecast month over the past three years, aggregated at the council level 
(\texttt{conc\char`_fc\char`_hist\char`_3} and \texttt{conc\char`_ba\char`_hist\char`_3}, respectively).
While high values of these autoregressive covariates do not always lead to large forecasts, they are generally positively correlated with wildfire events.
{Among the environmental covariates, average air temperature (\texttt{Temp}) contributes most strongly to the fire count model, while both \texttt{Temp} and relative humidity (\texttt{RHumi}) are most important for the burnt area model.}
The SHAP values of these variables exhibit an intuitive, non-causal relationship with the wildfire activity. 
For instance, high fire count and burnt area are often associated with elevated temperatures (reflected by large positive SHAP values), whereas low fire activity can occur across a broader range of temperature levels (indicated by the mix of blue and red at the lower end of the SHAP value), consistent with domain expectations.

\subsubsection{Latent Gaussian effect of Year and Covariates}
Figure \ref{fig: Effect_year_scores} shows the posterior estimates of the year-specific effects $T(\cdot)$ and the effects of XGBoost predictions $R(\cdot)$ in $\eta^C$ and $\eta^B$ as specified in \eqref{eq: linear_predictor}.
{While the year effects vary over time, only those for fire presence in 2017 and 2018 are significantly different from zero.
This aligns with historical records: Portugal experienced the highest number of fire ignitions in 2017 in the past decade, followed by a sharp decline in 2018, possibly due to fire regulation adjustments after the catastrophic wildfire activity of 2017.
As the XGBoost model already incorporates temporal information, the absence of significant year effects in most other years suggests that its predictions capture interannual variation effectively.

The lower panels of Figure \ref{fig: Effect_year_scores} illustrate the effects of the XGBoost predictions for fire count and burnt area within the linear predictors $\eta^C$ and $\eta^B$, respectively. 
In both cases, a generally increasing relationship is observed, indicating that higher XGBoost predictions are associated with higher contributions to the linear predictor.
However, the relationship is not strictly linear, especially in the case of large burnt area values. 
This nonlinearity may stem from the differing likelihood assumptions in the two modelling components: the XGBoost model assumes a Gamma distribution (coming from the Tweedie loss) for positive burnt areas, while the latent Gaussian model employs an extended Generalised Pareto (eGP) distribution.

\begin{figure}[ht]
\begin{center}
\includegraphics[width=0.6\textwidth]{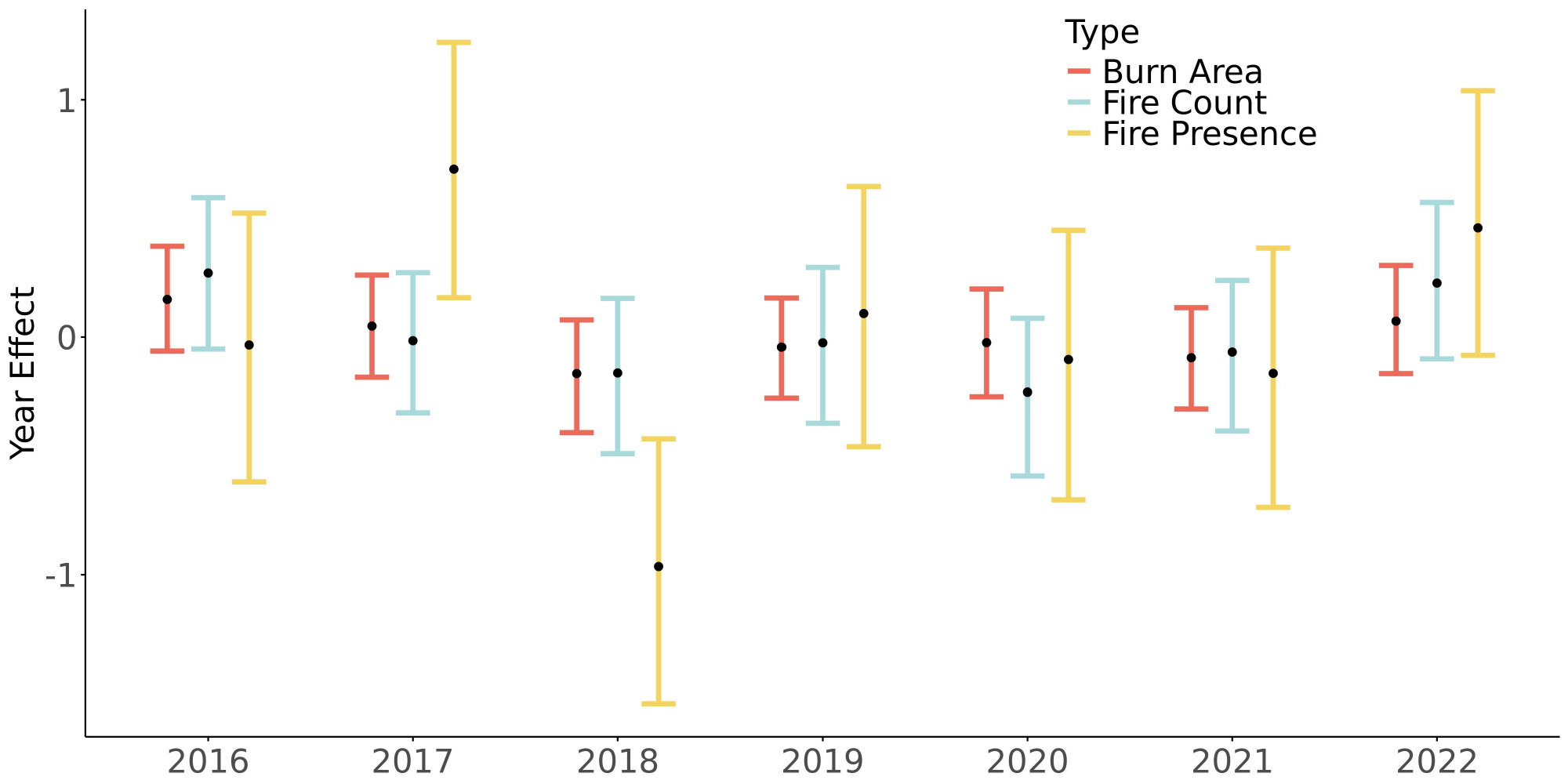}
\includegraphics[width=0.45\textwidth]{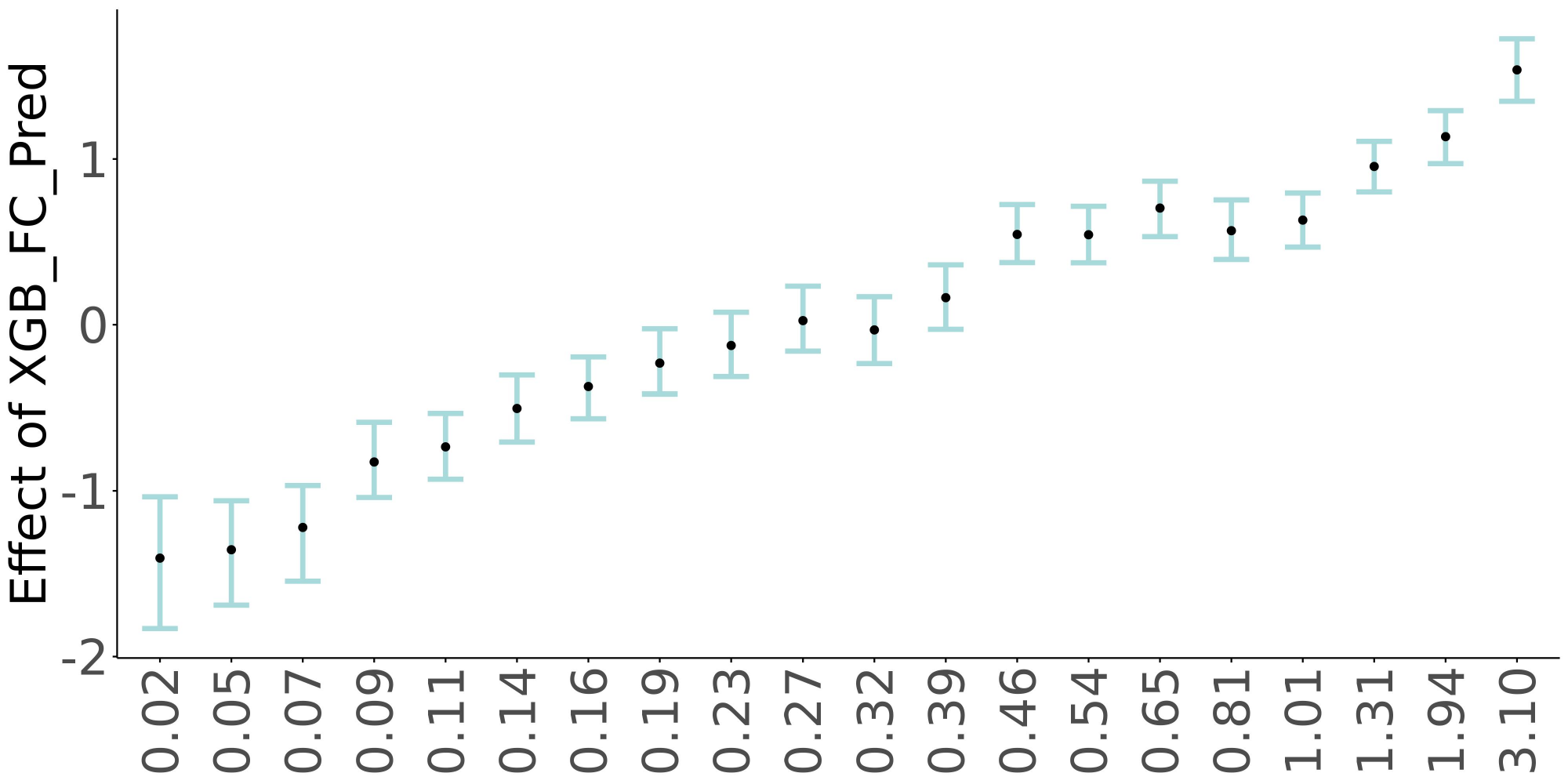}
\includegraphics[width=0.45\textwidth]{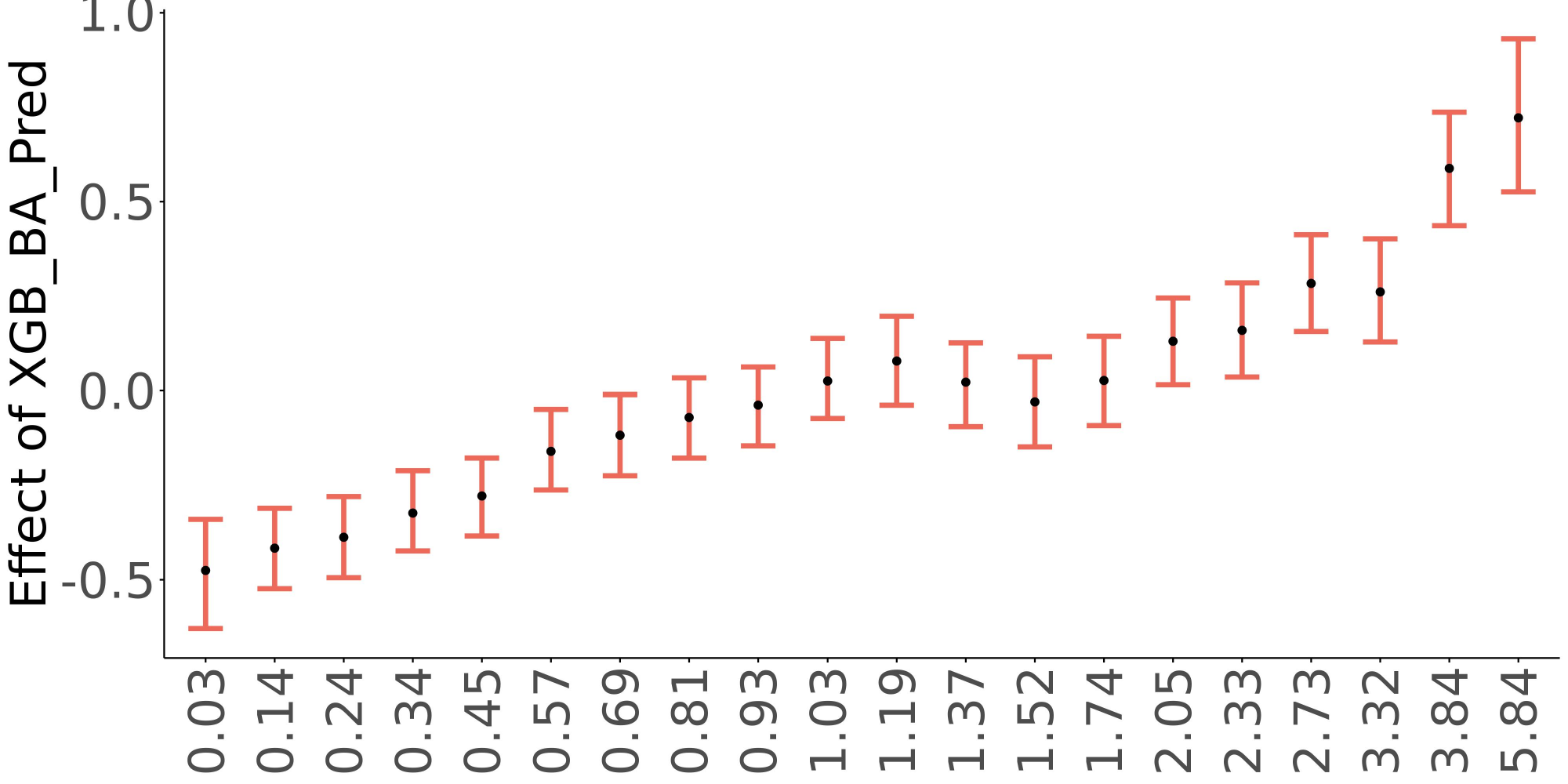}
\end{center}
\caption{\footnotesize{Posterior estimation of year effect $T(\cdot)$ (top) and XGBoost prediction effects $R(\cdot)$ in $\eta^C$ (bottom left) and $\eta^B$ (bottom right).
The values on the x-axis in the bottom two correspond to the raw XGBoost forecasts at each spatio-temporal unit $s,t$.
Black points in the three panels represent posterior means, while the vertical bars show 95\% credible intervals.}}
\label{fig: Effect_year_scores}
\end{figure}

\subsubsection{Shared Spatio-Temporal Effects}
\begin{figure}
\begin{center}
\includegraphics[width=0.39\textwidth]{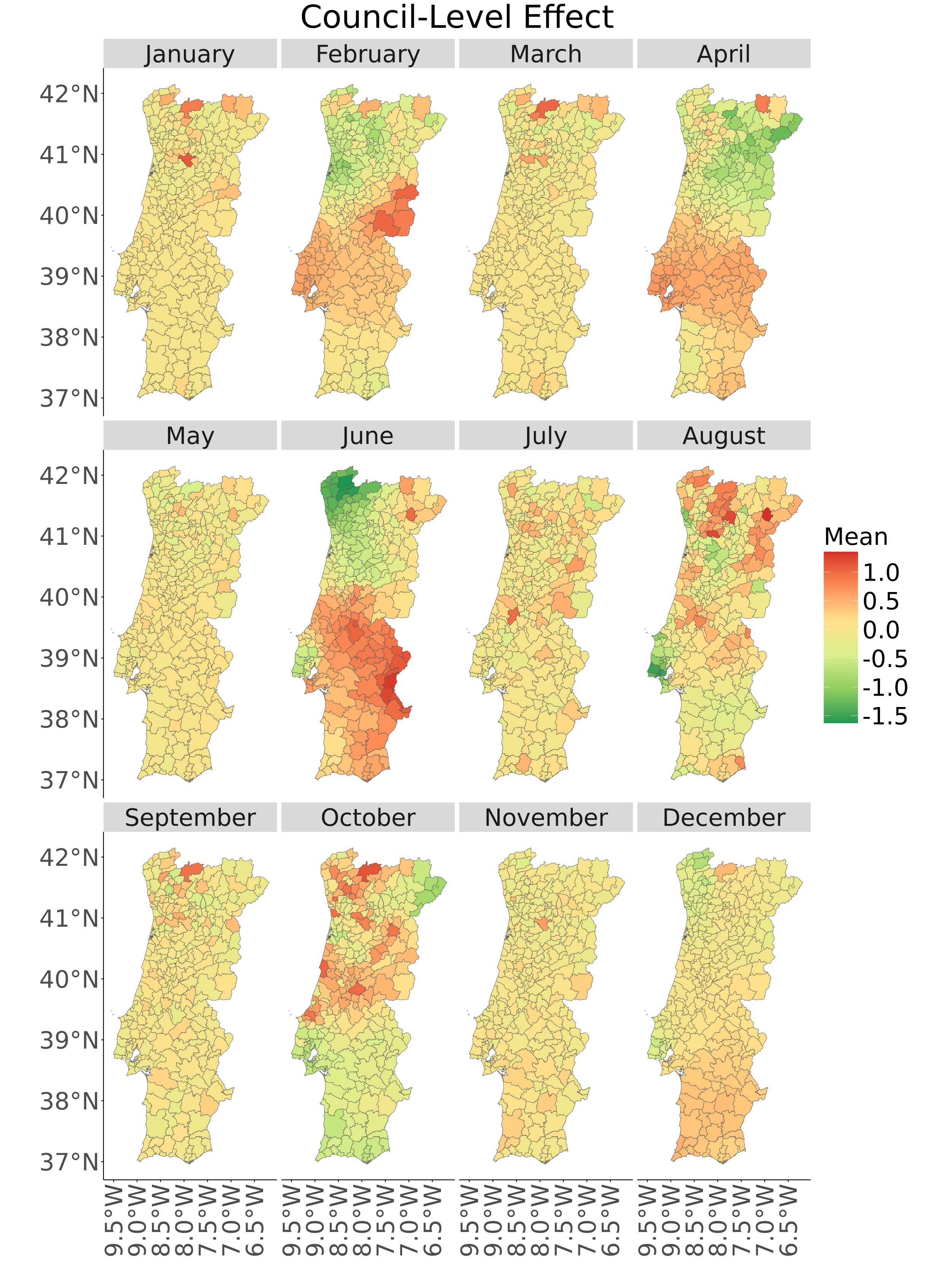}
\includegraphics[width=0.39\textwidth]{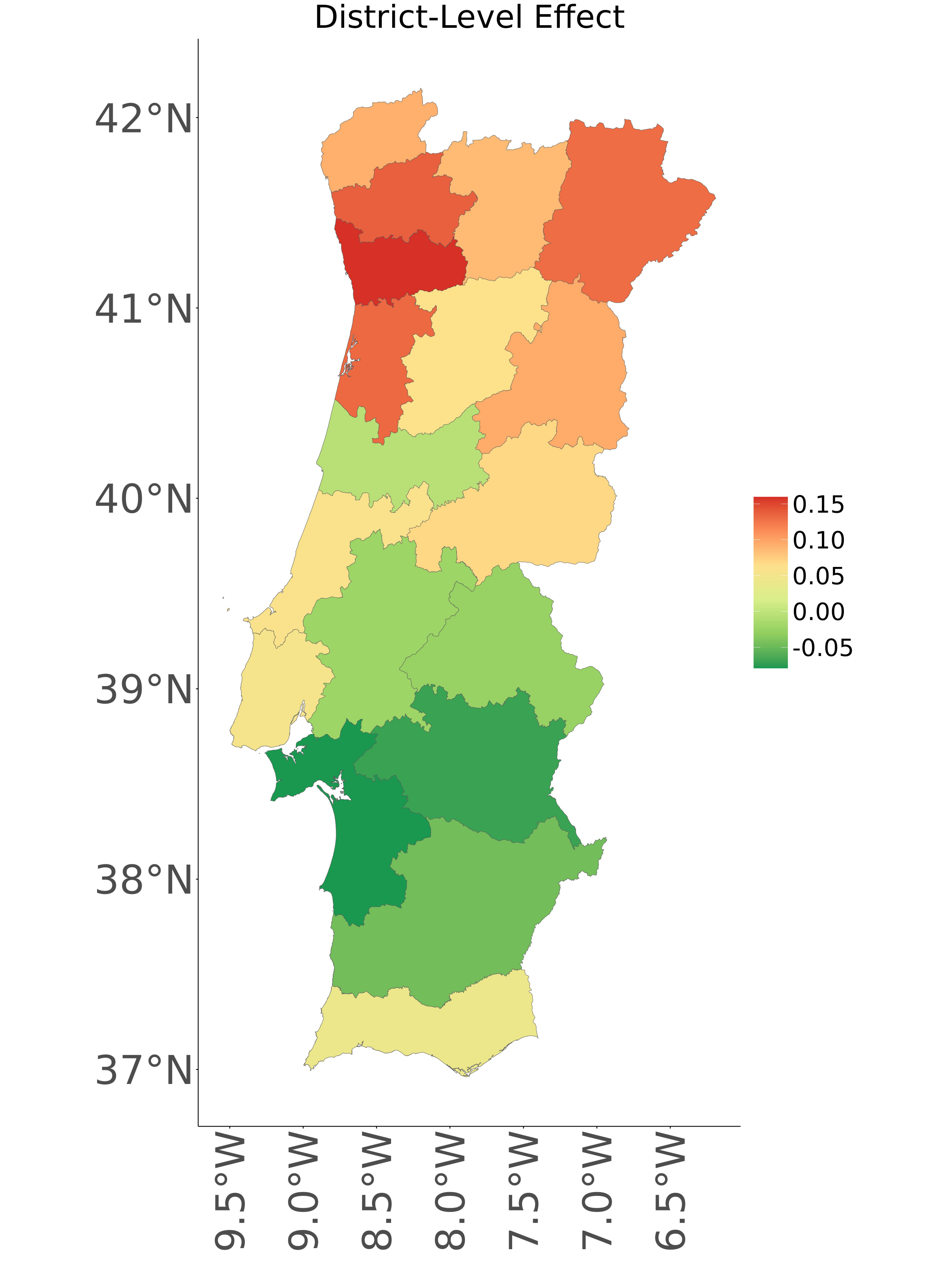}
\end{center}
\caption{\footnotesize{Posterior mean of the council-level spatio-temporal effect $G_c(\cdot)$ by grouped month (left), and average of the posterior mean of the district-level spatio-temporal effect $G_d(\cdot)$ (right), aggregated over all time indices.  }}
\label{fig: Effect_Spatial_Temporal}
\end{figure}

Figure \ref{fig: Effect_Spatial_Temporal} displays the posterior means of the council-level spatio-temporal effect $G_c(\cdot)$, grouped by month, and the average district-level spatio-temporal effect $G_d(\cdot)$ across all time indices.
Their uncertainty, quantified by 0.025 and 0.975 posterior quantiles, is provided in Figure \ref{fig: Effect_Spatial_Temporal_Uncertatinty}.
During the high-risk wildfire season, particularly in July, August, and October, the council-level effects exhibit greater spatial variability, with notable contrasts between neighbouring councils. 
In contrast, during the remaining months, the spatial effects are more homogeneous and show smooth transitions across adjacent regions. 
Noteworthy exceptions include \textit{Montalegre} and \textit{Vinhais} councils: the former shows unusually elevated effects in January and March, while the latter displays pronounced effects in April relative to its surrounding areas.
{At the district level, a general spatial gradient is evident, with higher effect values observed in the northern districts and lower values in the south.
However, the scale of the average district level is relatively small, with the maximum magnitude reaching only about 15\% of that in the council-level.}

{The estimated posterior means of the scaling parameters for the shared effects are  $\widehat{\beta}_1^C=-0.69$, $\widehat{\beta}_1^B=-0.43$, $\widehat{\beta}_2^C=0.77$, $\widehat{\beta}_2^B=0.27$.
The signs of these parameters are not straightforward to interpret, given the dependence between council- and district-level spatio-temporal effects.
Focusing instead on magnitudes within each wildfire quantity, the scaling parameters for fire count are consistently larger than those for burnt area, suggesting that the shared spatio-temporal effects for fire count are more strongly correlated with those for fire presence.
This is because fire count is obtained by aggregating fire presence events, whereas burnt area is a continuous measure conditional on those events.
}

\subsection{Posterior Distributions of eGP Parameters}
We now provide insights into the posterior inference for the parameters $\xi$ and $\kappa$ from the perspective of implementing the eGP likelihood within the INLA framework. 
These parameters are treated as global hyperparameters in the latent Gaussian model, meaning they are shared across all observations.
Figure \ref{fig: Posterior_density} shows the prior and posterior distributions for $\xi$ and $\kappa$.
The posterior of $\xi$, compared to its prior centred around zero, shifts markedly to the right and concentrates near $0.45$.
This suggests that the fitted eGP likelihood possesses a heavy-tailed structure, which may be suitable for capturing the extreme behaviour observed in burnt area data. 
Notably, the mode of the posterior for $\xi$ lies close to the upper bound of its prior, potentially conflicting with the imposed constraint of $(-0.5,0.5)$.
However, we argue that the posterior of $\xi$ is particularly sensitive to the skewness and overall shape of the response distribution, and that its value has only a minor influence on the posterior predictive distribution of burnt area.
This argument is further examined in Section \ref{sec: discussion_similar_perf}.
Therefore, rather than focusing on the interpretability of the posterior of $\xi$, it is more critical to ensure that the fitted eGP likelihood retains desirable properties, such as finite variance, by appropriately constraining the prior.
As for $\kappa$, its posterior distribution deviates substantially from the prior, with mode concentrated around $4.6$, though the posterior of $\kappa$ shows larger dispersion compared to the posterior of $\xi$.
This indicates that the posterior predictive probability $f_{\text{eGP}}(y_{s,t+1} \mid \eta_{s,t+1}^B, \xi, \kappa)$ does not have a singularity at 0, and its shape is approximately bell-like, though potentially skewed (see Figure~\ref{fig: Density_likelihood}).

\begin{figure}[ht]
\begin{center}
\includegraphics[width=0.49\textwidth]{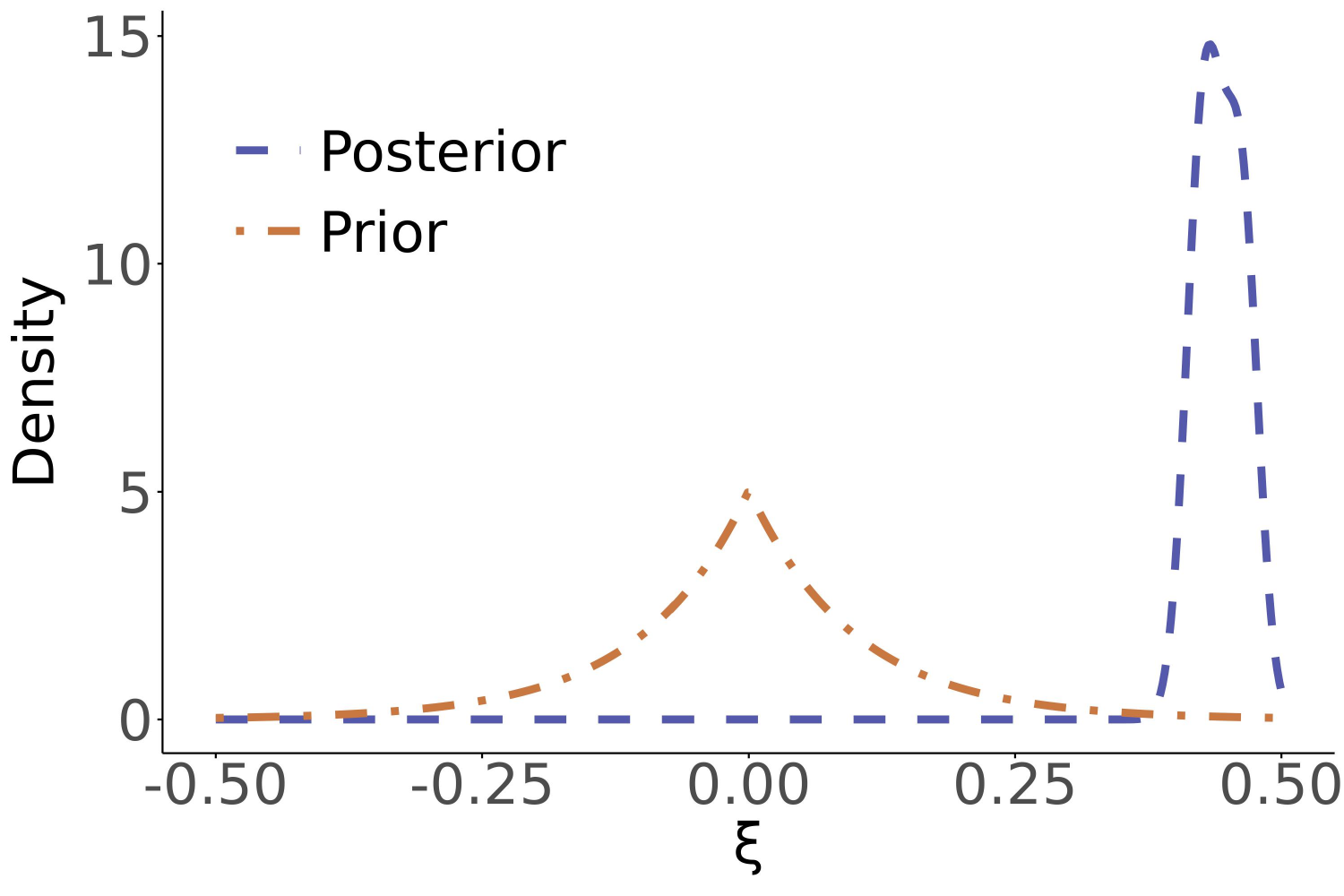}
\includegraphics[width=0.49\textwidth]{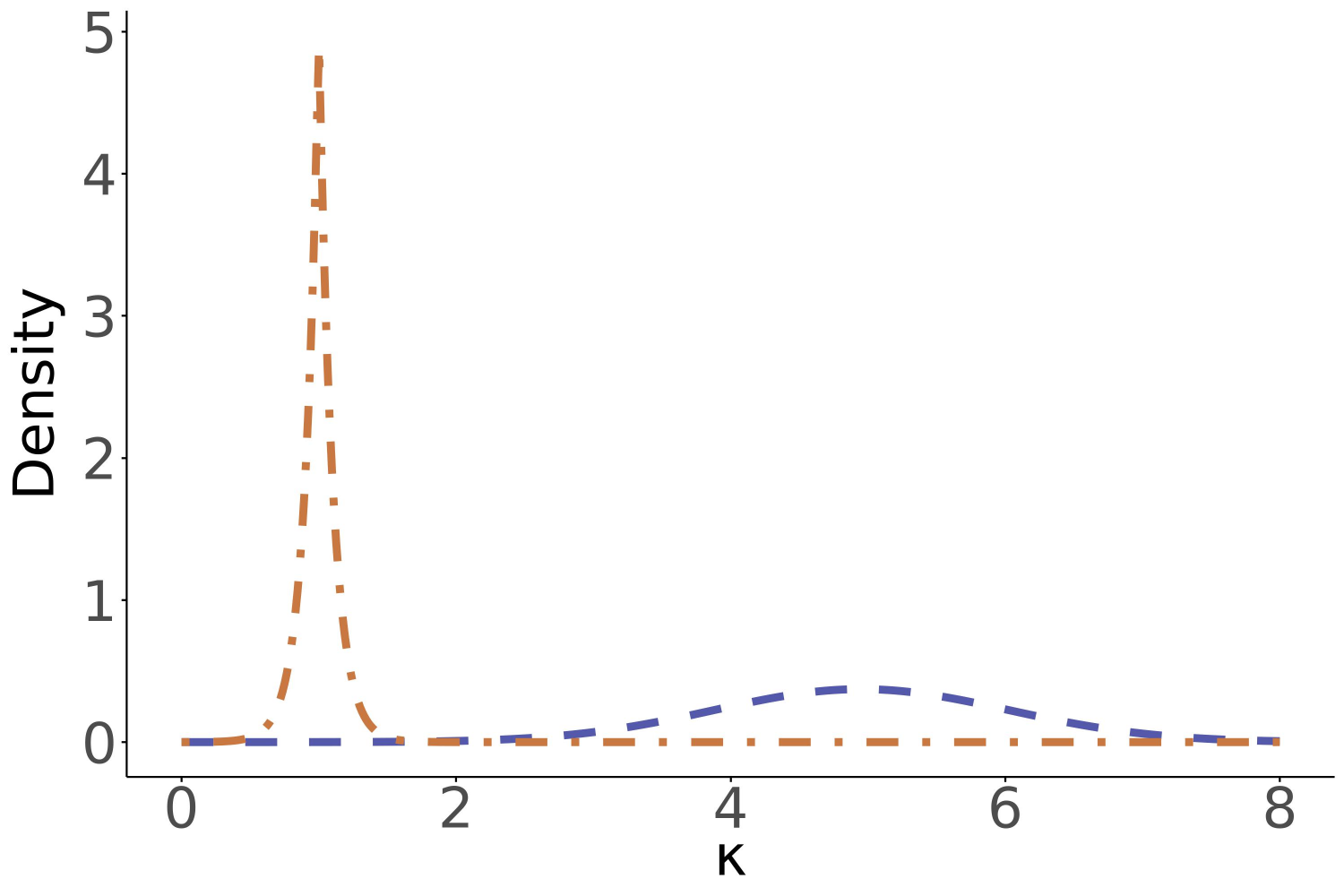}
\end{center}
\vspace{-0.2cm}
\caption{\footnotesize{Prior and posterior distributions of the hyperparameters $\xi$ (left) and $\kappa$ (right) in the eGP likelihood.
The prior for $\xi$ is $\pi(\xi) = 10\exp\{-10|\xi|\}/(2-2\exp\{-5\}), \quad -0.5<\xi<0.5$, and the prior for $\kappa$ is $\pi(\kappa)=    \pi_2(\kappa)\ = {10\exp(-10|\kappa-1|)}/{(2-\exp(-10))}, \quad\kappa>0.$ }}
\label{fig: Posterior_density}
\end{figure}

\section{Discussion }
\label{sec: discussion}
\subsection{Data Transformation and \texorpdfstring{$\xi$}{xi} in eGP}
\label{sec: transformation}
In our framework, the burnt area is modelled on the square root scale, rather than on its original scale or under alternative transformations.
This decision is guided by both theoretical and empirical considerations. The square root transformation reduces the extreme skewness of the burnt area distribution while preserving a meaningful distinction between small and large events. This results in a more stable fit of the extended Generalised Pareto (eGP) model across the full range of the data.
In particular, the eGP tail parameter $\xi$ is sensitive to the shape of the distribution in both the bulk and the tail. When working on the original scale, the strong skewness of burnt area data leads the model to prioritise fitting the bulk, often inflating tail estimates ($\xi > 0.5$), and causing tension with the prior support. On the other hand, aggressive transformations such as the logarithm overly compress the upper tail, causing $\xi$ to collapse toward the lower bound of its prior support ($-0.5$), which in turn can distort inference about extremes.
% This choice is motivated by the need to ensure that the posterior mass of the eGP tail parameter $\xi$ remains well within the interior of the prior support, $(-0.5,0.5)$, rather than accumulating near its boundaries.
% The skewness of the data has a pronounced influence on the estimated value of $\xi$ as this parameter governs the shape of both the bulk and the tail of the eGP distribution.
% When the data are highly skewed, as is the case for burnt area, where the majority of the burnt area is much smaller than upper quantiles, the eGP tends to prioritise fitting the bulk, which can result in an overestimated tail (i.e. $\xi > 0.5$).
% Conversely, if the data are transformed such that the distribution becomes overly symmetric (e.g., using a logarithmic transformation), the posterior of $\xi$ tends to concentrate near the lower bound of the prior, at -0.5, leading to poor posterior inference.

The square root transformation strikes a practical balance, moderating skewness without unduly suppressing large values. Our empirical investigations, which are guided by posterior predictive checks and sensitivity analyses, show that this transformation yields stable and interpretable posteriors for $\xi$, consistent with prior beliefs and with tail behaviour observed in historical burnt area data. 
This choice aligns with modelling choices in recent wildfire literature, such as \citet{cisneros2024deep}. 
While alternative power transformations (e.g., cube root, fourth root, or logarithm) influence the posterior of $\xi$, they result in nearly identical posterior predictive distributions once back-transformed, reinforcing the square root as a pragmatic and robust choice.

% \red{\subsection{PC prior for \texorpdfstring{$\kappa$}{kappa}}\label{sec:pckappa}}
% \red{A PC prior for the parameter $\kappa$ in the eGP distribution would have $\kappa = 1$ as its base model, under which the eGP reduces to the standard GP distribution. However, in order for the PC prior to reflect only the additional shape complexity introduced by deviations in $\kappa$, we must avoid confounding these changes with variations in location or scale.
% To achieve this, the eGP distribution should be reparameterised so that both a fixed $\alpha$-quantile and a fixed measure of spread (such as the interquartile range) are preserved across values of $\kappa$. This will ensure that comparisons to the base model are made on a common scale.
% Since the model is already parametrised in terms of $\eta = \log(q_\alpha)$, this reparameterisation is straightforward. For instance, by fixing $q_\alpha$ at the median (e.g., $\alpha=0.5$) and adjusting $\sigma$ accordingly, we can isolate the effect of $\kappa$ on the shape of the distribution. 
% Ideally, we would also hold constant a measure of spread (such as the interquartile range) across values of $\kappa$ to further disentangle its influence, in the spirit of~\cite{castro2022practical}.
% Under this standardised family of distributions, the Kullback–Leibler divergence used to define the PC prior can be computed; however, no closed-form expression is available. As a result, implementing a proper PC prior for $\kappa$ requires additional numerical integration.}

\subsection{Similar Performance to Gamma and Weibull Likelihoods}
\label{sec: discussion_similar_perf}
% Somewhat unexpectedly, the eGP likelihood does not significantly outperform the Gamma or Weibull likelihoods, despite the fact that the burnt area data are clearly heavy-tailed and the eGP distribution is explicitly designed to accommodate such behaviour. 
{Although the eGP likelihood outperforms the Gamma and Weibull likelihoods in Table \ref{tab: model_comparison}, the differences are modest, particularly relative to the Weibull, which also exhibits heavy-tailed behaviour with a posterior mean shape parameter of approximately 1.35.}
% While the Gamma distribution is light-tailed, the Weibull distribution, given the posterior mean of its shape parameter at approximately 1.39, effectively exhibits heavy-tailed behaviour in this context.
Figure \ref{fig: Density_likelihood} compares the three likelihoods using estimated hyperparameters. 
The linear predictors have been adjusted so that the medians of the resulting distributions align at approximately 0.8, facilitating a shape comparison. 
{While the eGP distribution displays a heavier tail than the other two, the overall shapes of the three densities are broadly similar, with the largest differences appearing near the mode.
% These findings are consistent with Table \ref{tab: model_comparison}, suggesting that tailoring the likelihood to better capture tail behaviour alone does not substantially improve forecast performance.
} 
{These findings, consistent with Table \ref{tab: model_comparison}, indicate that refining the likelihood to capture tail behaviour has limited impact on forecast performance; instead, accuracy is primarily determined by the flexibility of the latent structure and the incorporation of informative covariates into the linear predictor.
Indeed, within the hierarchical structure of our model,}
% These results suggest that extreme value theory does not offer a universally superior solution for modelling extremes.
% One plausible reason lies in the hierarchical structure of the model.
% Within this framework, 
observations are assumed conditionally independent given their respective linear predictors, which are functions of Gaussian latent effects. 
The central tendency and quantiles of the marginal distribution (e.g., its mean or $\alpha$-quantile) are controlled by the linear predictor.
Since hyperparameters such as $\xi$ and $\kappa$ are shared across all observations, their influence on individual predictive densities is limited relative to the localised effect of the linear predictor.
When the latent structure is sufficiently expressive, especially through the inclusion of informative covariates such as the XGBoost predictions, the linear predictor can effectively account for both moderate and extreme observations. 
Consequently, the observations tend to cluster within the high-density region of the fitted marginal distribution, diminishing the role of the tail parameter in shaping the likelihood. 
As a result, the contribution of tail behaviour to overall model performance becomes less critical, which explains the comparable predictive accuracy across the three likelihoods.
\begin{figure}[ht]
\begin{center}
\includegraphics[width=0.8\textwidth]{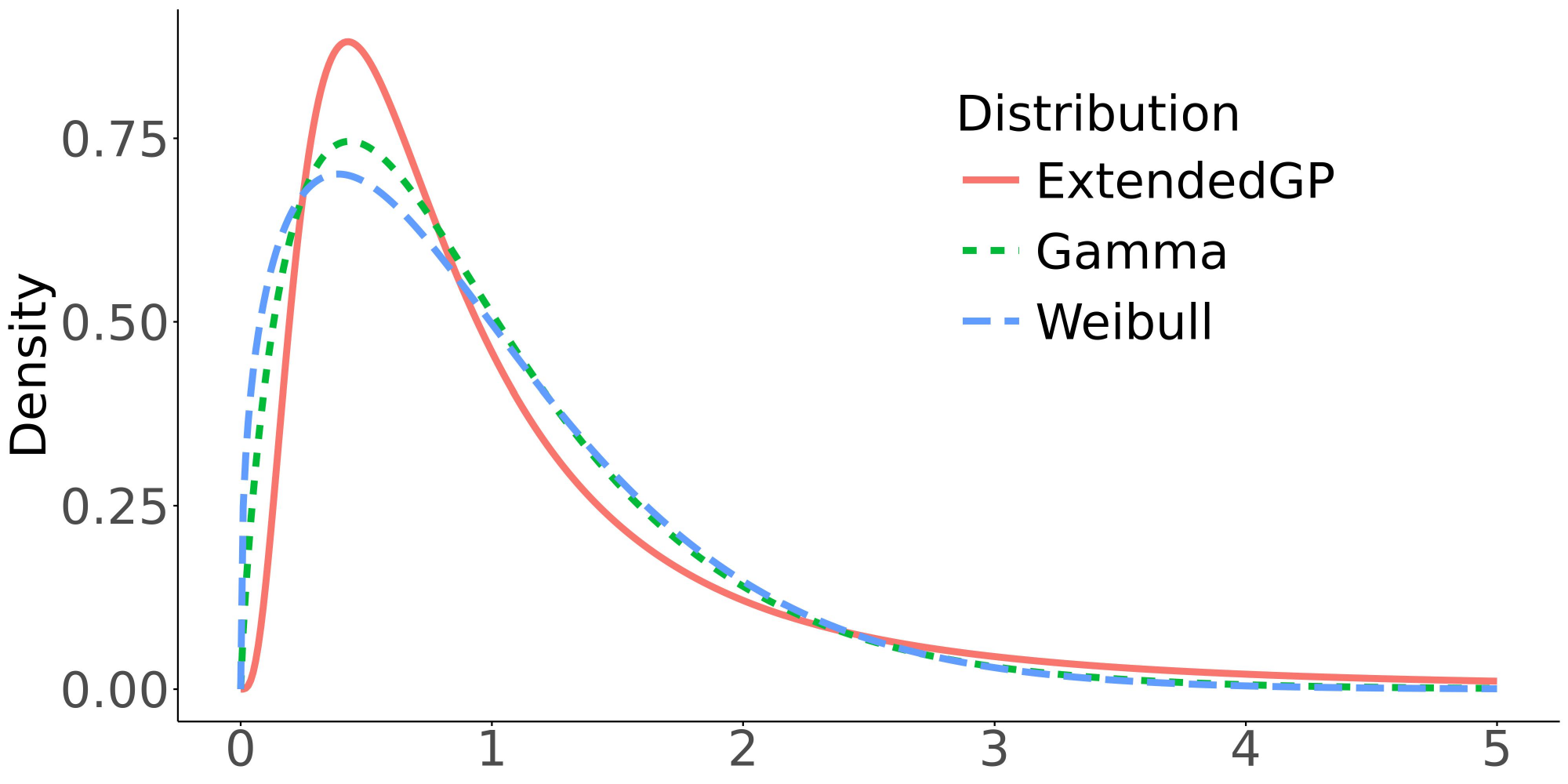}
\end{center}
\caption{\footnotesize{Densities of Gamma (shape = 1.80), Weibull (shape = 1.35) and eGP ($\kappa=4.64,\, \xi=0.45$) likelihoods, based on estimated hyperparameters. 
The densities are scaled via the linear predictor to have approximately equal medians (around 0.8).}}
\label{fig: Density_likelihood}
\end{figure}

\subsection{Longer Forecasting Horizons}
The current forecasting horizon considered in our framework is one month. That is, we generate forecasts $y_{t+1}$ conditional on the past observations $y_{1:t}$ and additional covariates $\widetilde{\boldsymbol{x}}_{t}$. 
Extending this to longer-term forecasts (e.g. forecasting $y_{t+h}$ for $h>1$) presents a key challenge: neither the XGBoost model nor the latent Gaussian model is inherently designed to produce multi-horizon predictions in a unified structure.
As such, it is not straightforward to generate forecasts across multiple future time points using a single model fit.

A practical solution is to apply the full two-stage modelling framework (as illustrated in Figure~\ref{fig: model_framework}) independently for each forecasting horizon $h$. 
While the process has been detailed for $h=1$, the extension to $h=2,3,\ldots$ involves adjusting and retraining the XGBoost model to predict future outcomes based on features lagged by $h$ months.
{We emphasise that this approach does not constitute a fully coherent joint multi-horizon forecasting model, but rather a pragmatic, horizon-specific solution that remains compatible with the INLA framework and preserves interpretability and uncertainty quantification.}
Specifically, the XGBoost models can be modified to produce:
\begin{align*}
\widehat{y}_{s,t+h}^C = \sum_{m}f_m^C(\widetilde{\boldsymbol{x}}_{s,t}^C),\label{eq: xgb_cnt}\\
\sqrt{\widehat{y}_{s,t+h}^B} = \sum_{m}f_m^B(\widetilde{\boldsymbol{x}}_{s,t}^B),
\end{align*}
where $\widetilde{x}_{s,t}^{(\cdot)}$ denotes the feature vectors at space $s$ and time $t$ for fire count ($C$) and burnt area ($B$).
These yield point forecasts of fire count and burnt area at time $t+h$, conditional on information available at time  $t$.
These forecasts then serve as inputs to the latent Gaussian model in the second stage, which is used to generate probabilistic predictions for the target variables at the specified horizon. 
This iterative, horizon-specific approach offers a tractable solution for multi-step forecasting while preserving the interpretability and modularity of the original model structure.

{Since the most influential covariates in the XGBoost models for $h=1$ are derived from historical wildfire activity at the same forecast month (See Figure \ref{fig: SHAP_values} and Table \ref{tab: autoregressive_covariates}), these predictors tend to continue to dominate the feature set when $h>1$. 
Consequently, the predictive accuracy of the XGBoost models does not deteriorate substantially for longer forecast horizons ($h>1$). This stability implies that the proposed multi-horizon forecast method is less susceptible to error propagation over time and can deliver robust uncertainty quantification.
Table~\ref{tab: longer_horizon_forecast} presents a comprehensive comparison of forecast accuracy and uncertainty for model M1 with the eGP likelihood across horizons $h = 1, 2, 3$. In Stage~1, mean squared error (MSE) is used to assess the performance of the XGBoost models. In Stage~2, forecast accuracy is evaluated using the same metrics as in Table~\ref{tab: model_comparison}, while uncertainty is quantified by the average length of the 95\% credible interval derived from INLA. Only minor variations in predictive performance and uncertainty are observed across horizons from one to three months, with no evidence of a systematic degradation in performance.
}

\begin{table}[ht]
\centering \caption{\footnotesize{Comparison of prediction accuracy and uncertainty for model M1 across forecast horizons $h = 1, 2, 3$. Mean squared error (MSE) evaluates the accuracy of the two Stage~1 XGBoost models. Stage~2 prediction metrics are identical to those reported in Table~\ref{tab: model_comparison}. Uncertainty in Stage~2 is quantified by the average length of the 95\% credible interval.}} \vspace{0.3cm} 
\begin{tabular}{@{}llccc@{}} 
\toprule Forecast Horizon & \multicolumn{1}{c}{} & h=1 & h=2 & h=3 \\ \midrule \multirow{2}{*}{Stage 1 Prediction} & MSE for count & \multicolumn{1}{l}{0.34} & \multicolumn{1}{l}{0.36} & \multicolumn{1}{l}{0.30} \\ & MSE for burnt area & \multicolumn{1}{l}{8.26} & \multicolumn{1}{l}{8.27} & \multicolumn{1}{l}{8.24} \\ \midrule \multirow{6}{*}{Stage 2 Prediction} & AUC & 0.862 & 0.857 & 0.864 \\ & CRPS & 4.75 & 4.61 & 4.61 \\ & Unweighted $r^C$ & 362 & 367 & 345 \\ & Weighted $r^C$ & 4.84 & 5.11 & 4.83 \\ & Unweighted $r^B$ & 565 & 585 & 599 \\ & Weighted $r^B$ & 14.2 & 14.8 & 14.4 \\ \midrule \multirow{2}{*}{Stage 2 Uncertatinty} & Avg. CI for count & \multicolumn{1}{l}{3.38} & \multicolumn{1}{l}{3.49} & \multicolumn{1}{l}{3.50} \\ & Avg. CI for burnt area & \multicolumn{1}{l}{24.33} & \multicolumn{1}{l}{24.28} & \multicolumn{1}{l}{23.92} \\ \bottomrule 
\end{tabular} 
\label{tab: longer_horizon_forecast} 
\end{table}

\subsection{Choices of the Model in Stage 1}
{In this study, XGBoost is adopted in Stage~1 to generate one-month-ahead wildfire forecasts at the council level. Since Stage~1 only requires point forecasts, XGBoost is not the only viable choice. Alternative models, such as ensemble methods (e.g., Random Forest; \citealp{breiman2001random}) or deep learning approaches for time series forecasting (e.g., long short-term memory networks; \citealp{hochreiter1997long}, and attention-based models; \citealp{lim2021temporal,zhou2021informer}), could serve a similar role.
In principle, multiple models targeting the same response could be fitted in Stage~1, with their outputs jointly incorporated as covariates in Stage~2. This strategy is not pursued here because the number of hyperparameters associated with existing model components is already close to 20, which is near the recommended upper limit for INLA. When selecting a single model, numerous studies have demonstrated that XGBoost achieves state-of-the-art performance in both time series forecasting \citep{elsayed2021reallyneeddeeplearning,januschowski2022forecasting,santoro2024comparison} and tabular data modelling \citep{gorishniy2021revisiting,grinsztajn2022tree}. For these reasons, XGBoost is chosen as the Stage~1 model in our framework.
}

\section{Conclusion}
\label{sec: conclusion}
In this study, we propose a two-stage ensemble modelling framework that addresses the challenge of incorporating multiple future covariates in spatio-temporal forecasting using INLA. 
Our approach integrates window-based XGBoost predictions as proxy covariates for future fire count and burnt area, and couples them with a latent Gaussian model to produce calibrated posterior forecasts.
{By compressing complex spatio-temporal dynamics and dozens of environmental predictors into two informative proxy covariates, the framework enables INLA to exploit information that would otherwise be difficult to accommodate, thereby improving predictive performance.}
Furthermore, we introduce and implement the novel sub-asymptotic eGP likelihood within the INLA framework and its companion R-INLA library, enabling joint modelling of both moderate and extreme wildfire events.

By comparing posterior predictions under the eGP, Weibull, and Gamma likelihoods while keeping the remainder of the model structure fixed, we observe only a marginal improvement in predictive performance with the eGP likelihood.
We attribute this to the conditional independence assumption and the dominance of the linear predictor in shaping the marginal likelihoods. 
Similar findings are reported in \citet{yadav2023joint}, who also reported minimal sensitivity to likelihood choice in Bayesian hierarchical models for landslide size.

We also discuss a strategy for extending the framework to longer forecasting horizons by replicating the two-stage procedure for each horizon separately.
While this does not offer a unified multi-horizon forecast, it provides a practical path forward within the current model constraints.
The result shows that our proposed framework can provide a stable forecast and uncertainty for longer horizons.

One remaining limitation of the present work is the absence of explicit covariates that capture human activity, which is known to play a critical role in wildfire ignition and propagation. 
Incorporating such information, e.g., data on population density, land use, or proximity to infrastructure, could further enhance the predictive capacity of the model and is a promising direction for future research.

\section{Statements}
\subsection{Data Availability Statement}
The code for implementing the two-stage model and reproducing the results in this paper is available at {\url{https://github.com/hcl516926907/Portugal_Wildfire}}. 
The wildfires data from ICNF (\url{https://www.icnf.pt/florestas/gfr/gfrgestaoinformacao/estatisticas}) and ERA5 data (\url{https://cds.climate.copernicus.eu/datasets/reanalysis-era5-land?tab=overview}) are publicly available online. Due to the size of the data, they are not shared in the above GitHub repository.

\subsection{Funding Statement}
\begin{itemize}
    \item RB work is partially funded by national funds through the FCT – Fundação para a Ciência e a Tecnologia, I.P., under the scope of the projects UIDB/00297/2020 (\url{https://doi.org/10.54499/UIDB/00297/2020}) and UIDP/00297/2020 (\url{https://doi.org/10.54499/UIDP/00297/2020}) (Centre for Mathematics and Applications)
    \item  CC was partially funded by the Portuguese Fundação para a Ciência e a Tecnologia (FCT) I.P./MCTES through national funds (PIDDAC) – UID/50019/2025, LA/P/0068/2020 
\end{itemize}

\subsection{Conflict of Interest Statement}
We declare that they have no known competing financial interests or personal relationships that could have appeared to influence the work reported in this paper.

  \appendix
  \begin{appendices}

\section{ Supplementary material}
\renewcommand{\theequation}{A.\arabic{equation}}
\setcounter{equation}{0}

\renewcommand{\thefigure}{A.\arabic{figure}}
\setcounter{figure}{0}

\renewcommand{\thetable}{A.\arabic{table}}
\setcounter{table}{0}

% \subsection{Code and Data}
% The code for implementing the two-stage model and reproducing the results in this paper is available at {\url{https://github.com/hcl516926907/Portugal_Wildfire}}. 
% The wildfires and ERA5 data are publicly available online. Due to the size of the data, they are not shared in the above GitHub repository.

\subsection{Derivation of the PC prior for \texorpdfstring{$\kappa > 0$}{kappa}} \label{sec:appendix_pckappa}

We construct a penalised complexity (PC) prior for the parameter $\kappa$ in the extended generalised Pareto (eGP) distribution. The base model corresponds to $\kappa = 1$, under which the eGP reduces to the standard Generalised Pareto distribution (GPD).

Let $f_\kappa(y) \equiv f_{\text{eGP}}(y; \kappa)$ and $f_{\kappa_1}(y) \equiv f_{\text{eGP}}(y; \kappa = 1)$ denote the eGP densities for general $\kappa > 0$ and for the base model, respectively. The PC prior is defined by assigning an exponential distribution to the Kullback--Leibler-based distance between $f_\kappa$ and the base model:
\[
d(\kappa) = \sqrt{2\,\text{KLD}(f_\kappa \| f_{\kappa_1})}.
\]
This distance quantifies the additional complexity introduced by allowing $\kappa \neq 1$. The PC prior is then given by:
\[
\pi(\kappa) = \lambda \exp\{-\lambda d(\kappa)\} \left| \frac{\partial d(\kappa)}{\partial \kappa} \right|,
\]
where $\lambda > 0$ is a user-defined rate parameter.

\paragraph{Case \boldmath$\xi \neq 0$.}
The Kullback--Leibler divergence from $f_{\kappa_1}$ to $f_\kappa$ is defined as:
\[
\text{KLD}(f_\kappa \| f_{\kappa_1}) = \int_0^\infty f_\kappa(y) \log \left( \frac{f_\kappa(y)}{f_{\kappa_1}(y)} \right) \, \text{d}y.
\]
Using the expression for the eGP density when $\xi \neq 0$, we have:
\begin{align*}
f_\kappa(y) &= \kappa \left[1 - \left(1 + \xi \frac{y}{\sigma} \right)^{-\frac{1}{\xi}} \right]^{\kappa - 1} \cdot \frac{\xi}{\sigma} \left(1 + \xi \frac{y}{\sigma} \right)^{-\frac{1}{\xi} - 1},\\
f_{\kappa_1}(y) &= \frac{\xi}{\sigma} \left(1 + \xi \frac{y}{\sigma} \right)^{-\frac{1}{\xi} - 1}.
\end{align*}

Hence,
\[
\log\left( \frac{f_\kappa(y)}{f_{\kappa_1}(y)} \right) = \log(\kappa) + (\kappa - 1) \log\left(1 - \left(1 + \xi \frac{y}{\sigma} \right)^{-\frac{1}{\xi}} \right),
\]
and the KLD becomes:
\begin{equation}\label{eq:kdl}
    \int_0^\infty f_\kappa(y) \cdot \log\left(\frac{f_\kappa(y)}{f_{\kappa_1}(y)}\right) \, \text{d}y = \log(\kappa) \int_0^\infty f_\kappa(y) \, \text{d}y + (\kappa - 1) \int_0^\infty f_\kappa(y) \log\left(1 - \left(1 + \xi \frac{y}{\sigma}\right)^{-\frac{1}{\xi}}\right) \, \text{d}y.
\end{equation}
The first term in \eqref{eq:kdl} simplifies directly in $\log \kappa$, since $\int_0^\infty f_\kappa(y) \text{d}y=1$.
% \[
% \log(\kappa) \int_0^\infty f_\kappa(y) \, \text{d}y = \log\kappa.
% \]
{For the second term, applying the change of variable $u=1 - (1 + \xi {y}/{\sigma})^{-1/\xi}$, yielding:
\begin{align*}
    &(\kappa - 1) \int_0^\infty f_\kappa(y) \log\left(1 - \left(1 + \xi \frac{y}{\sigma}\right)^{-\frac{1}{\xi}}\right) \, \text{d}y = (\kappa - 1) \int_0^1 \kappa u^{\kappa-1}\log u \; \text{d}u = -\frac{\kappa-1}{\kappa} 
\end{align*}
% Let \( \mathbb{E}_\kappa \) denote the expectation under \( f_\kappa(y) \). Then
% \[
% (\kappa - 1) \int_0^\infty f_\kappa(y) \log\left(1 - \left(1 + \xi \frac{y}{\sigma}\right)^{-\frac{1}{\xi}}\right) \, \text{d}y = 
% (\kappa - 1) \mathbb{E}_\kappa \left[\log\left(1 - \left(1 + \xi \frac{y}{\sigma}\right)^{-\frac{1}{\xi}}\right)\right].
% \]
Therefore,
\begin{equation}
\text{KLD}(f_\kappa \| f_{\kappa_1}) = \log\kappa - \frac{\kappa - 1}{\kappa}. 
\label{eq: kld_kappa}
\end{equation}
Using the exact KLD in \eqref{eq: kld_kappa}, the corresponding PC prior is:
\begin{equation*}
    \pi(\kappa) = \begin{cases}
        \frac{\lambda |\kappa-1|}{2\kappa^2\sqrt{2\log \kappa -2(\kappa-1)/\kappa}}\exp\left\{-\lambda \left( \sqrt{2 \log \kappa -{2(\kappa-1)}/{\kappa}} \right) \right\}, \qquad &\kappa>0, \;\kappa \neq1,\\
        \lambda/2, & \kappa=1.
    \end{cases}
\end{equation*}
}
{Alternatively, approximating the KLD around $\kappa=1$ via a second-order Taylor expansion yields:
\begin{equation*}
\text{KLD}(f_\kappa \| f_{\kappa_1}) = \frac{1}{2}(\kappa-1)^2 + o((\kappa-1)^3), \qquad \text{as }\kappa \rightarrow1 . 
\end{equation*}
This leads to a locally symmetric PC prior:
\begin{equation*}
    \pi(\kappa)\ = \frac{\lambda\exp(-\lambda|\kappa-1|)}{2-\exp(-\lambda)}, \qquad\kappa>0.
\end{equation*}
}
\paragraph{Case \boldmath$\xi = 0$.}
When $\xi = 0$, the eGP reduces to a power transformation of the exponential distribution:
\begin{align*}
f_\kappa(y) &= \kappa \left(1 - \exp\left\{-\frac{y}{\sigma}\right\} \right)^{\kappa - 1} \cdot \frac{1}{\sigma} \exp\left\{- \frac{y}{\sigma}\right\},\\
f_{\kappa_1}(y) &= \frac{1}{\sigma} \exp\left\{-\frac{y}{\sigma}\right\}.
\end{align*}
Using the change of variable $v=1-\exp\{-y/\sigma\}$, one can derive the same KLD expression as in the case $\xi \neq 0$, thereby recovering the same PC prior $\pi(\kappa)$.
\subsection{Covariates in XGBoost}
Table \ref{tab:covariates_appendix} provides a description of the environmental covariates derived from the ERA5 dataset, while Table \ref{tab: autoregressive_covariates} summarises the feature-engineered autoregressive covariates constructed from historical wildfire records.

\begin{table}[H]
\footnotesize
\centering
\caption{\footnotesize{Environmental Covariates used in the XGBoost model.}}
\vspace{0.5cm}
\begin{adjustbox}{width=0.9\textwidth}
\begin{tabular}{@{}M{2cm} M{2.5cm} M{2.5cm} M{2cm} L{6.5cm}@{}}
\toprule
\textbf{Name} & \textbf{Source} & \textbf{Spatial \newline Resolution}    & \textbf{Temporal Resolution} & \multicolumn{1}{c}{\textbf{Description}}                                                                                      \\ \midrule
Pricp         & ERA5-Land       & $0.1^\circ \times 0.1^\circ$   & Hourly                       & Accumulated   liquid and frozen water, including rain and snow, that falls to the Earth's   surface.                           \\ \midrule
Temp          & ERA5-Land       & $0.1^\circ \times 0.1^\circ$   & Hourly                       & Temperature of air at 2m above the surface of land.                                                                          \\ \midrule
Ucomp         & ERA5-Land       & $0.1^\circ \times 0.1^\circ$   & Hourly                       & Eastward component of the 10m wind.                                                                                          \\ \midrule
Vcomp         & ERA5-Land       & $0.1^\circ \times 0.1^\circ$   & Hourly                       & Northward component of the 10m wind.                                                                                         \\ \midrule
DewPoint      & ERA5-Land       & $0.1^\circ \times 0.1^\circ$   & Hourly                       & Temperature to which the air, at   2 metres above the surface of the Earth, would have to be cooled for saturation to occur. \\ \midrule
HVegLAI       & ERA5-Land       & $0.1^\circ \times 0.1^\circ$   & Hourly                       & One-half of the total green leaf area per unit horizontal ground surface area for high vegetation type.                   \\ \midrule
LVegLAI       & ERA5-Land       & $0.1^\circ \times 0.1^\circ$   & Hourly                       & One-half of the total green leaf area per unit. horizontal ground surface area for low vegetation type.                      \\ \midrule
HVegCov       & ERA5            & $0.25^\circ \times 0.25^\circ$ & Constant                     & The   fraction of the grid box that is covered with vegetation that is classified   as ``high".                                \\ \midrule
LVegCov       & ERA5            & $0.25^\circ \times 0.25^\circ$ & Constant                     & The fraction of the grid box that is covered with vegetation that is classified as ``low".                                   \\ \midrule
HVegTyp       & ERA5            & $0.25^\circ \times 0.25^\circ$ & Constant                     & Indicator of the 6 types of high vegetation recognised by the ECMWF Integrated Forecasting   System.                         \\ \midrule
LVegTyp       & ERA5            & $0.25^\circ \times 0.25^\circ$ & Constant                     & Indicator of the 10 types of low vegetation recognised by the ECMWF Integrated   Forecasting System.                         \\ \midrule
FWI           & Derived              & $0.1^\circ \times 0.1^\circ$   & Hourly                       & Fire Weather Index                                                                                                            \\ \midrule
RHumi         & Derived             & $0.1^\circ \times 0.1^\circ$   & Hourly                       & Relative Humidity                                                                                                             \\ \bottomrule
\end{tabular}
\end{adjustbox}
\phantomsection
\label{tab:covariates_appendix}
\end{table}

\begin{table}[H]
\caption{\footnotesize{Feature-engineered autoregressive covariates for fire count and burnt area. 
In this table, $t$ denotes the forecasting time point, {$h$ is the forecast horizon}, $X$ indicates the spatial scale, taking values ``dist" (district level) or ``conc" (council level); and $Y$ specifies the source variable, with ``fc" for fire count and ``ba" for burnt area.
}}
\vspace{0.5cm}
\footnotesize
\centering
\begin{adjustbox}{width=0.9\textwidth}
\begin{tabular}{@{}M{2cm} M{3cm} M{5cm} L{6.5cm} @{}}
\toprule
\textbf{Name} & \textbf{Index Range} & \textbf{Formula}                                            & \multicolumn{1}{c}{\textbf{Description}}                                                   \\ \midrule
X\_Y\_lag\_j  & $j=1,2,\cdots,9$     & $y_{s,t-j}$                                                 & Lag $j$ of monthly fire count/burnt area at council/district level                                 \\ \midrule
X\_Y\_ma\_j   & $j=3,6,9,12,24,36$   & $\frac{\sum_{i=1}^j y_{s,t-i}}{j}$                          & Moving average of past $j$ months of fire count/burnt area at council/district level        \\ \midrule
X\_Y\_hist\_j & $j=1,3,5$            & $\frac{\sum_{k=1}^3\sum_{i=-(j-1)/2}^{(j-1)/2} y_{s,t+h-12k+i}}{3j}$ & Average fire count/burnt area  over centred $j$ month around month $t+h$ in the past 3 years \\ \midrule
month\_sin    & NA                   & $\sin(2\pi t/12)$                                           & Angular representation of the month                                                        \\ \midrule
month\_cos    & NA                   & $\cos(2\pi t/12)$                                           & Angular representation of the month                                                        \\ \midrule
Year          & NA                   & NA                                                          & Year of the wildfire occurrence                                                            \\ \midrule
Lon           & NA                   & NA                                                          & Longitude of the centroid of the council                                                   \\ \midrule
Lat           & NA                   & NA                                                          & Latitude of the centroid of the council                                                    \\ \bottomrule
\end{tabular}
\end{adjustbox}
\phantomsection
\label{tab: autoregressive_covariates}
\end{table}

\subsection{Additional Diagnostic Plots}
We use the ACF plots in Figure \ref{fig: ACF_plots} to determine the extent of long-term temporal dependence, specifically, the number of lags or amount of historical data to include in constructing the autoregressive covariates.
For each month, we first calculate the average fire count and burnt area across all councils, and then compute the autocorrelation using these monthly averages following the standard ACF formula.
\begin{figure}[H]
% \begin{center}
\centering
\includegraphics[width=0.49\textwidth]{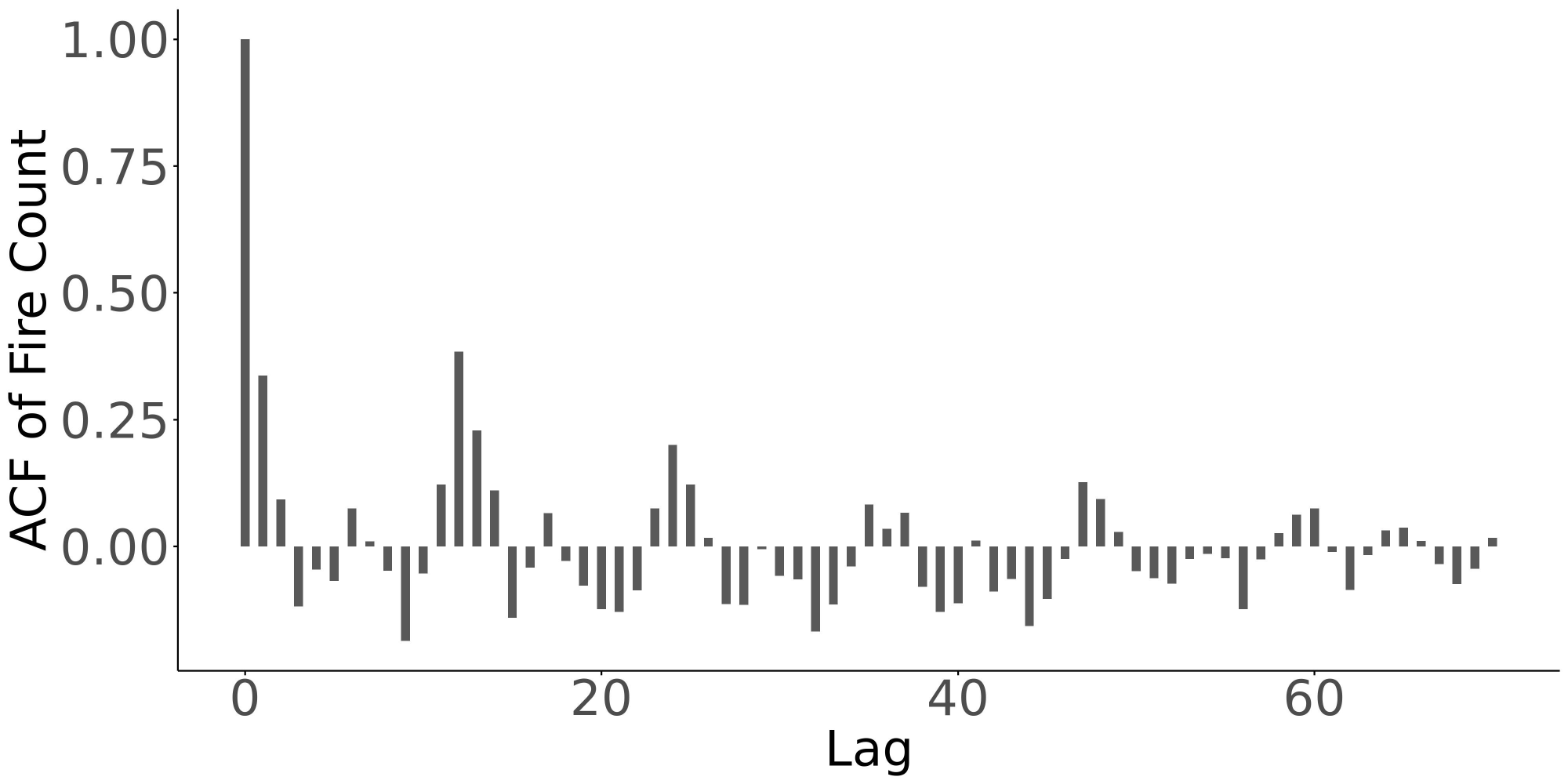}
\includegraphics[width=0.49\textwidth]{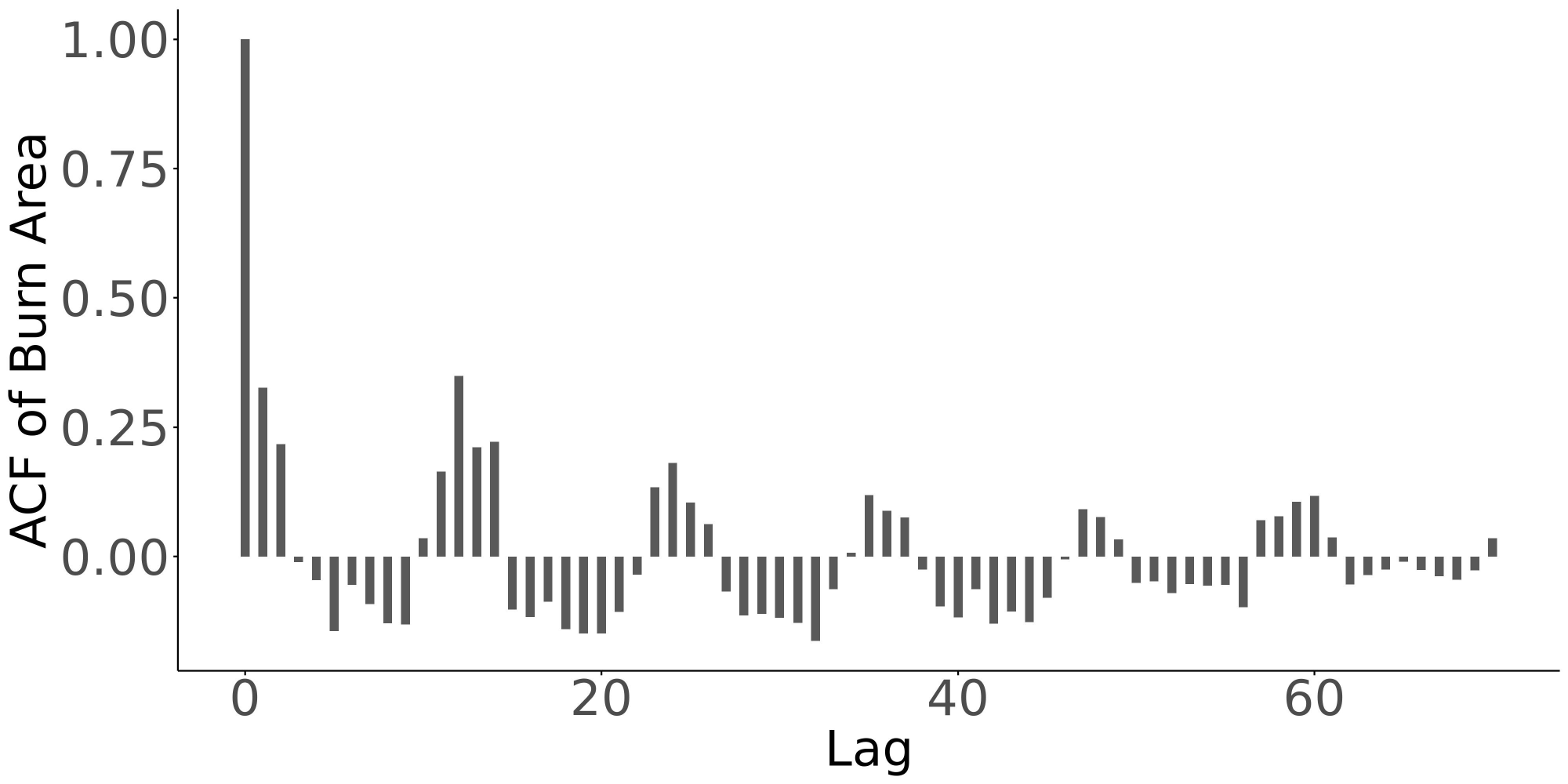}
% \end{center}
\caption{Autocorrelation Function plots of average fire count and burnt area at council-monthly level.}
\phantomsection
\label{fig: ACF_plots}
\end{figure}

The uncertainty associated with the council-level spatio-temporal effect $G_c(\cdot)$ and the district-level spatio-temporal effect $G_d(\cdot)$ is illustrated below. Uncertainty is quantified using the 0.025 and 0.975 posterior quantiles of the corresponding random effects. 
\begin{figure}[H]
\begin{center}
\includegraphics[width=0.4\textwidth]{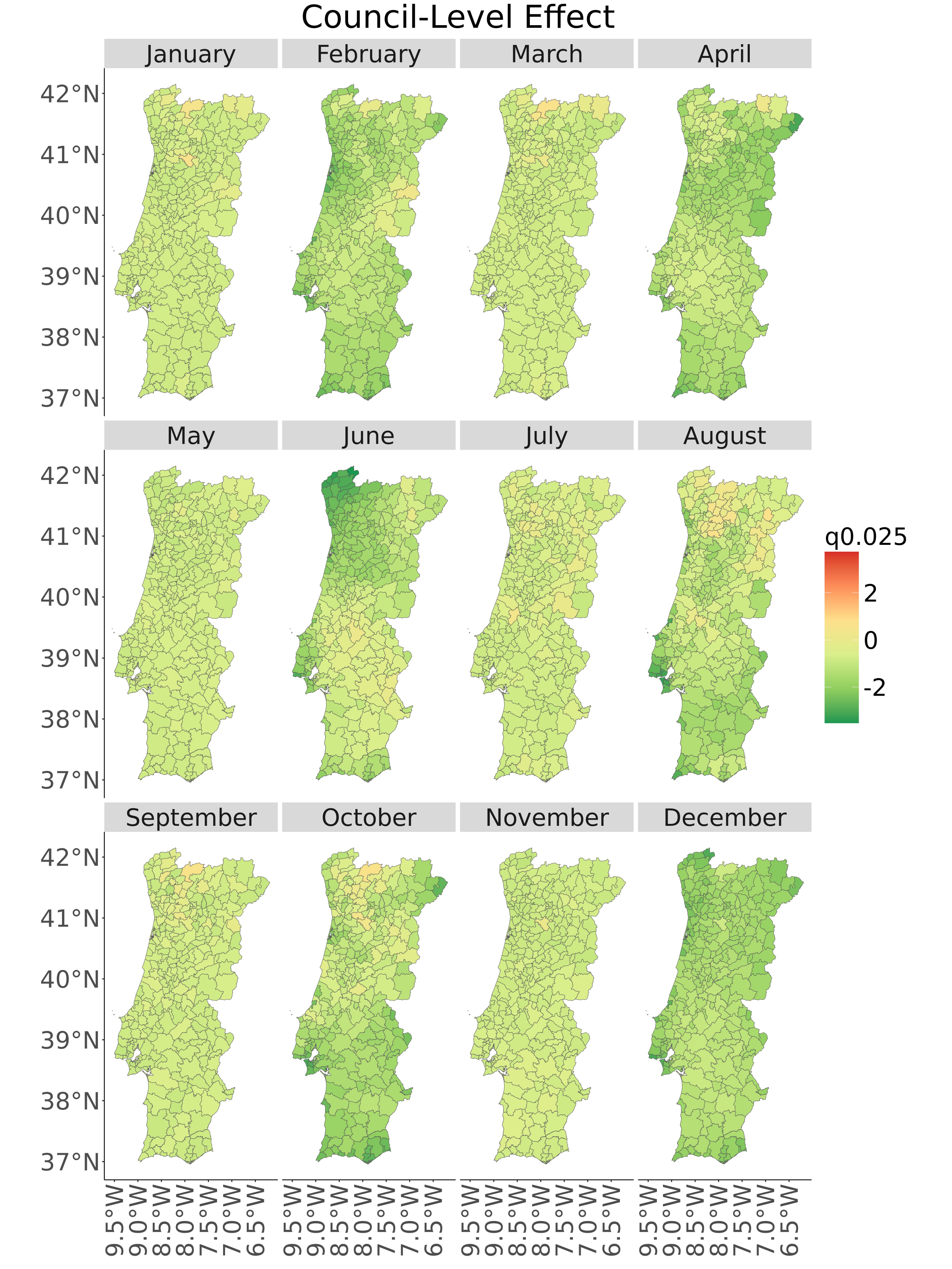}
\includegraphics[width=0.4\textwidth]{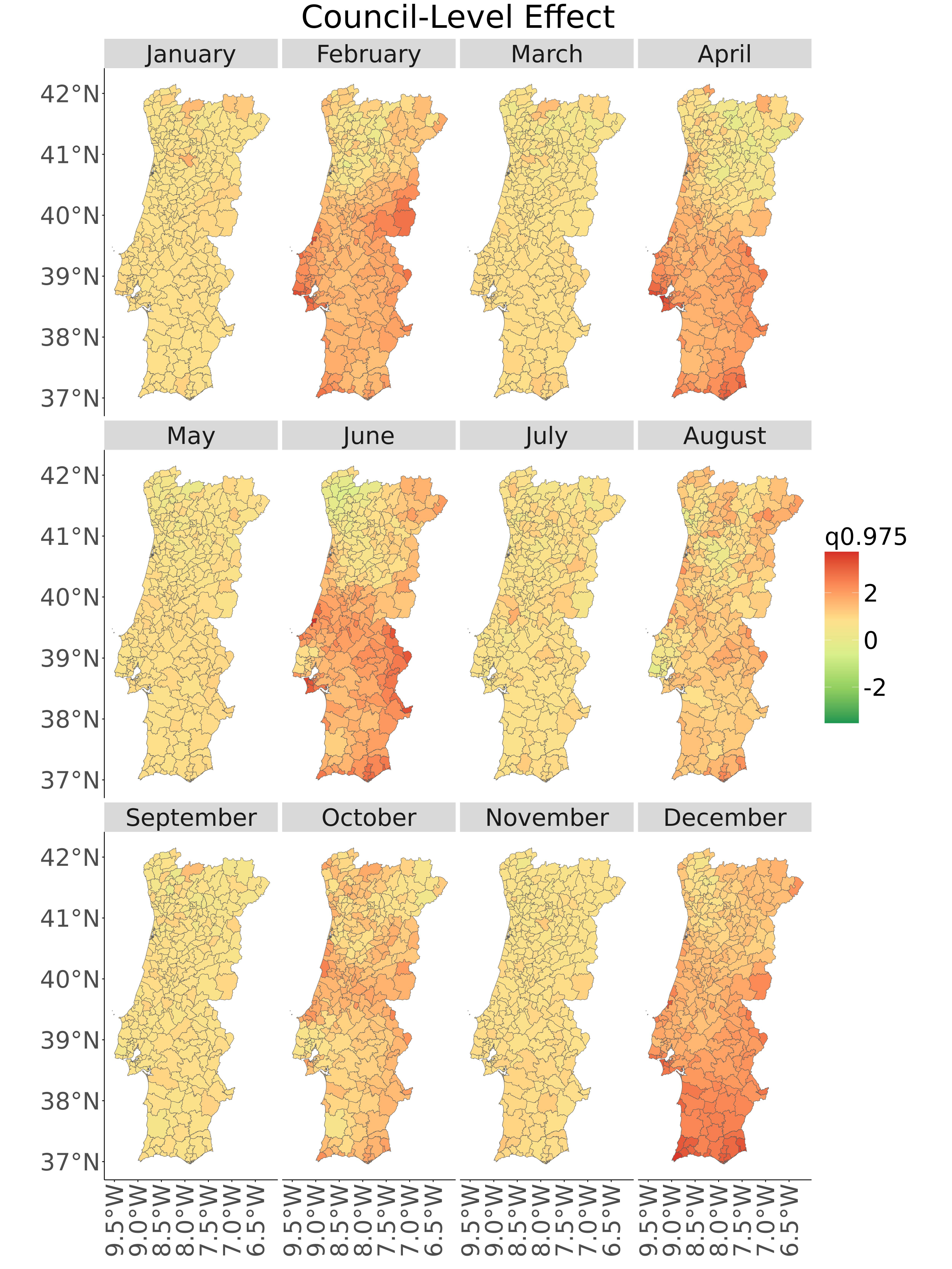}
\includegraphics[width=0.4\textwidth]{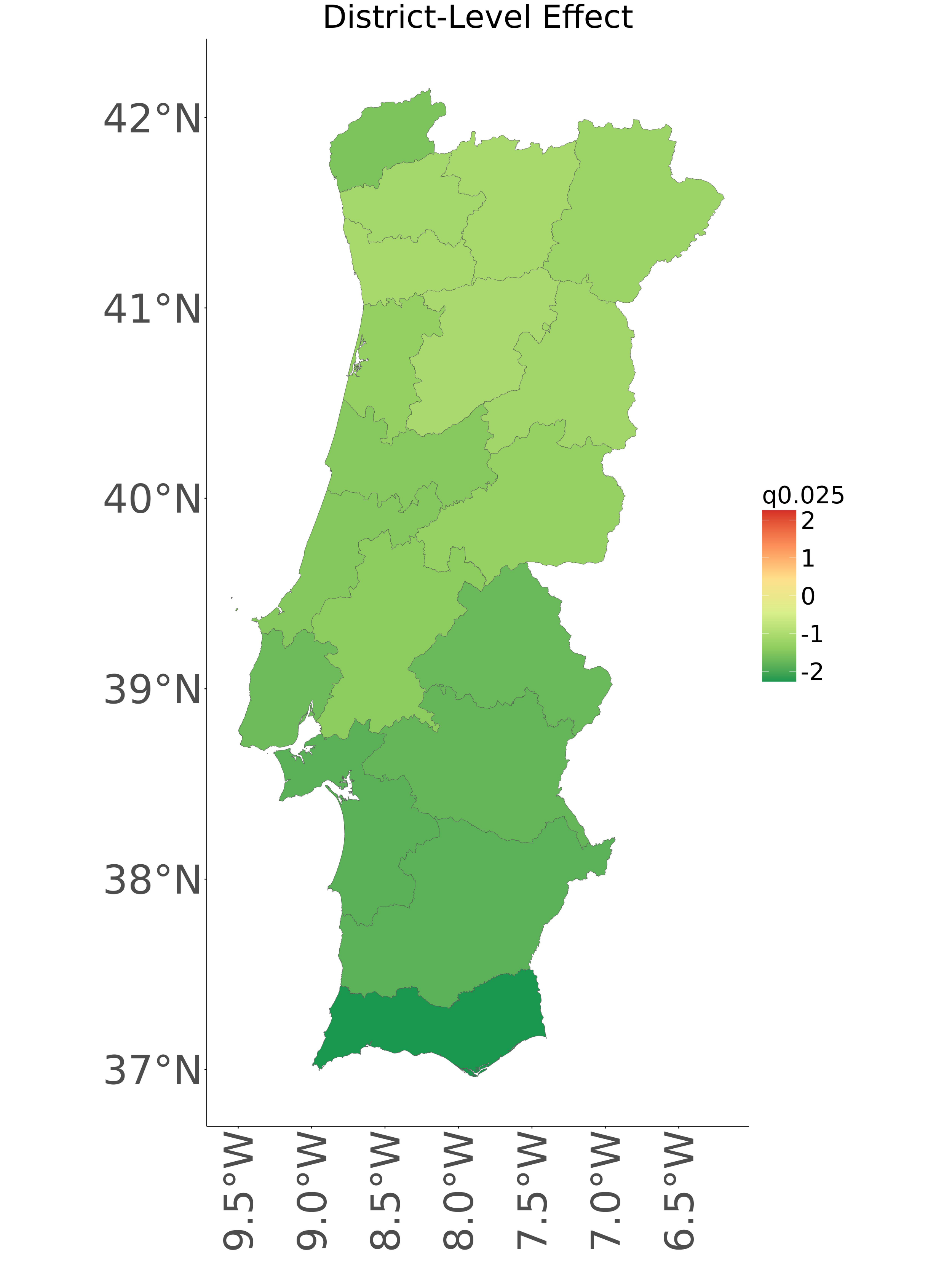}
\includegraphics[width=0.4\textwidth]{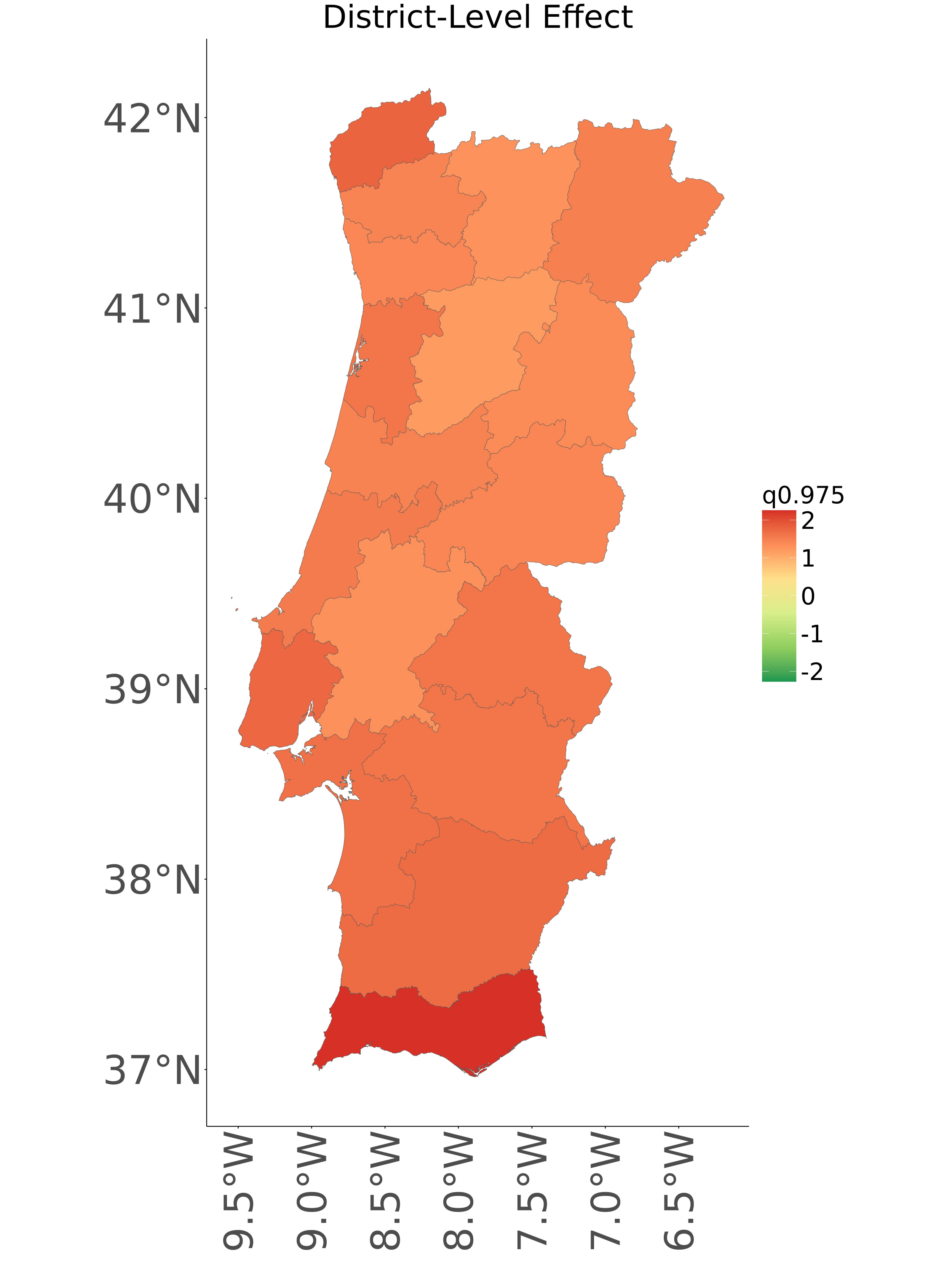}
\end{center}
\caption{\footnotesize{Posterior uncertainty of the spatio-temporal effects. 
The top panel shows the 0.025 and 0.975 posterior quantiles of the council-level spatio-temporal effect $G_c(\cdot)$ grouped by month, while the bottom panel displays the averages of the 0.025 and 0.975 posterior quantiles of the district-level spatio-temporal effect $G_d(\cdot)$ aggregated over all time indices.}}
\phantomsection
\label{fig: Effect_Spatial_Temporal_Uncertatinty}
\end{figure}

  \end{appendices}
  \FloatBarrier
  \bibliographystyle{apalike_Capital}
\bibliography{My_Library}
\end{document}